\documentclass[
twocolumn,
]{aastex701}

\usepackage{amsmath}
\usepackage{graphicx}
\usepackage{xcolor}
\usepackage{xparse}
\usepackage{bm}
\usepackage[
capitalize,
noabbrev
]{cleveref}
\usepackage{tikz}
\usepackage[normalem]{ulem}
\usepackage{paralist}
\usepackage{multirow}

\newcommand{\Bb}{\ensuremath{\bm{b}}}
\newcommand{\Bv}{\ensuremath{\bm{v}}}
\newcommand{\Bz}{\ensuremath{\bm{z}}}

\newcommand{\BB}{\ensuremath{\bm{B}}}

\NewDocumentCommand{\curl}{d{_}{_} o}{\ensuremath{\IfNoValueTF{#1}{\nabla}{\nabla_{#1}} \times #2}}
\NewDocumentCommand{\grad}{d{_}{_} o}{\ensuremath{\IfNoValueTF{#1}{\nabla}{\nabla_{#1}} #2}}
\NewDocumentCommand{\diverg}{d{_}{_} o}{\ensuremath{\IfNoValueTF{#1}{\nabla}{\nabla_{#1}} \cdot #2}}

\NewDocumentCommand{\nT}{o}{\ensuremath{
\IfNoValueF{#1}{#1 \,}
\mathrm{nT}
}}

\NewDocumentCommand{\km}{o}{\ensuremath{
\IfNoValueF{#1}{#1 \,}
\mathrm{km}
}}

\NewDocumentCommand{\au}{o}{\ensuremath{
\IfNoValueF{#1}{#1 \,}
\mathrm{AU}
}}

\NewDocumentCommand{\pct}{o}{\ensuremath{
\IfNoValueF{#1}{#1 \;}
\%
}}

\NewDocumentCommand{\Rs}{o}{\ensuremath{
\IfNoValueF{#1}{#1 \;}
\mathrm{R_S}
}}

\NewDocumentCommand{\kms}{o}{\ensuremath{
\IfNoValueF{#1}{#1 \;}
\mathrm{km \, s^{-1}}
}}

\NewDocumentCommand{\cc}{o}{\ensuremath{
\IfNoValueF{#1}{#1 \;}
\mathrm{cm^{-3}}
}}

\NewDocumentCommand{\eV}{o}{\ensuremath{
\IfNoValueF{#1}{#1 \;}
\mathrm{eV}
}}

\NewDocumentCommand{\keV}{o}{\ensuremath{
\IfNoValueF{#1}{#1 \;}
\mathrm{keV}
}}

\NewDocumentCommand{\MeV}{o}{\ensuremath{
\IfNoValueF{#1}{#1 \;}
\mathrm{MeV}
}}

\NewDocumentCommand{\nucleon}{s o}{\ensuremath{
\IfNoValueF{#2}{#2 \;}
\IfBooleanTF{#1}{\mathrm{nucleon}}{\mathrm{nuc}}
}}

\NewDocumentCommand{\MeVnuc}{s o}{\ensuremath{
\IfNoValueF{#2}{#2 \;}
\MeV \! /\IfBooleanTF{#1}{\nucleon*}{\nucleon}
}}

\NewDocumentCommand{\keVe}{o}{\ensuremath{
\IfNoValueF{#1}{#1 \;}
\mathrm{keV/e}
}}

\NewDocumentCommand{\Element}{m}{\ensuremath{\mathrm{#1}}}
\NewDocumentCommand{\QState}{m m}{\ensuremath{\mathrm{#1}^{#2+}}}

\NewDocumentCommand{\Hy}{o}{\IfNoValueTF{#1}{\Element{H}}{\QState{H}{#1}}}
\NewDocumentCommand{\He}{o}{\IfNoValueTF{#1}{\Element{He}}{\QState{He}{#1}}}

\NewDocumentCommand{\FIP}{s o O{=}}{\ensuremath{\mathrm{FIP}
\IfNoValueF{#2}{
#3
\IfBooleanTF{#1}{#2}{\eV[#2]}}
}}

\NewDocumentCommand{\AbSEP}{O{X} O{\Ox}}{\ensuremath{#1/#2}}

\NewDocumentCommand{\PLawExp}{s o}{\ensuremath{b
\IfNoValueF{#2}{\IfBooleanTF{#1}{\approx}{=} #2}}}

\newcommand{\he}{\Element{He}}

\NewDocumentCommand{\MpQ}{o}{\ensuremath{
\IfNoValueTF{#1}{\mathrm{M/Q}}{(\mathrm{M/Q})_{#1}}}}

\NewDocumentCommand{\n}{o o O{=}}{\ensuremath{n
\IfNoValueF{#1}{_{#1}}
\IfNoValueF{#2}{
\IfNoValueTF{#3}{=}{#3} \cc[#2]
}
}}

\NewDocumentCommand{\dn}{o o O{=}}{\ensuremath{\delta n
\IfNoValueF{#1}{_{#1}}
\IfNoValueF{#2}{
\IfNoValueTF{#3}{=}{#3} \cc[#2]
}
}}

\NewDocumentCommand{\dnn}{s D{_}{_}{\Hy} o O{=}}{\ensuremath{
\IfBooleanTF{#1}{\dn[#2] / \n[#2]}{\abs{\dn[#2] / \n[#2]}}
\IfNoValueF{#3}{
\IfNoValueTF{#4}{=}{#4} {#3}
}
}}

\NewDocumentCommand{\Wk}{o o O{=}}{\ensuremath{W
\IfNoValueTF{#1}{_k}{_{k,#1}}
\IfNoValueF{#2}{
\IfNoValueTF{#3}{=}{#3} \mWmsq[#2]
}
}}

\NewDocumentCommand{\rc}{d<> o O{=}}{\ensuremath{r_{c\IfNoValueF{#1}{;{#1}}}
\IfNoValueF{#2}{#3 \Rs[#2]}
}}

\NewDocumentCommand{\rA}{d<> o O{=}}{\ensuremath{r_{A\IfNoValueF{#1}{;{#1}}}
\IfNoValueF{#2}{#3 \Rs[#2]}
}}

\NewDocumentCommand{\vsw}{s o O{=}}{\ensuremath{v_\sw
\IfNoValueF{#2}{
\IfNoValueTF{#3}{=}{#3}
\IfBooleanTF{#1}{#2}{\kms[#2]}}
}}

\NewDocumentCommand{\vs}{s o O{=}}{\ensuremath{v_s
\IfNoValueF{#2}{
\IfNoValueTF{#3}{=}{#3}
\IfBooleanTF{#1}{#2}{\kms[#2]}}
}}

\NewDocumentCommand{\vheavy}{s o O{=}}{\ensuremath{v_\mathrm{heavy}
\IfNoValueF{#2}{
\IfNoValueTF{#3}{=}{#3}
\IfBooleanTF{#1}{#2}{\kms[#2]}}
}}

\NewDocumentCommand{\vslow}{s o O{=}}{\ensuremath{v_\mathrm{slow}
\IfNoValueF{#2}{
\IfNoValueTF{#3}{=}{#3}
\IfBooleanTF{#1}{#2}{\kms[#2]}}
}}

\NewDocumentCommand{\vfast}{s o O{=}}{\ensuremath{v_\mathrm{fast}
\IfNoValueF{#2}{
\IfNoValueTF{#3}{=}{#3}
\IfBooleanTF{#1}{#2}{\kms[#2]}}
}}

\NewDocumentCommand{\vau}{s o O{=}}{\ensuremath{v_\mathrm{\au[1]}
\IfNoValueF{#2}{
\IfNoValueTF{#3}{=}{#3}
\IfBooleanTF{#1}{#2}{\kms[#2]}}
}}

\NewDocumentCommand{\vK}{s o O{=}}{\ensuremath{v_K
\IfNoValueF{#2}{
\IfNoValueTF{#3}{=}{#3}
\IfBooleanTF{#1}{#2}{\kms[#2]}}
}}

\NewDocumentCommand{\vi}{s o O{=}}{\ensuremath{v_i
\IfNoValueF{#2}{
\IfNoValueTF{#3}{=}{#3}
\IfBooleanTF{#1}{#2}{\kms[#2]}}
}}

\NewDocumentCommand{\vv}{s o O{=}}{\ensuremath{v_0
\IfNoValueF{#2}{
\IfNoValueTF{#3}{=}{#3}
\IfBooleanTF{#1}{#2}{\kms[#2]}}
}}

\NewDocumentCommand{\vn}{s o O{=}}{\ensuremath{v_n
\IfNoValueF{#2}{
\IfNoValueTF{#3}{=}{#3}
\IfBooleanTF{#1}{#2}{\kms[#2]}}
}}

\NewDocumentCommand{\vsigma}{s o O{=}}{\ensuremath{v_\sigma
\IfNoValueF{#2}{
\IfNoValueTF{#3}{=}{#3}
\IfBooleanTF{#1}{#2}{\kms[#2]}}
}}

\NewDocumentCommand{\vdn}{s o O{=}}{\ensuremath{v_{\delta n}
\IfNoValueF{#2}{
\IfNoValueTF{#3}{=}{#3}
\IfBooleanTF{#1}{#2}{\kms[#2]}}
}}

\NewDocumentCommand{\vdnp}{s o O{=}}{\ensuremath{v_{\delta n'}
\IfNoValueF{#2}{
\IfNoValueTF{#3}{=}{#3}
\IfBooleanTF{#1}{#2}{\kms[#2]}}
}}

\NewDocumentCommand{\vIP}{s o O{=}}{\ensuremath{v_{IP}
\IfNoValueF{#2}{
\IfNoValueTF{#3}{=}{#3}
\IfBooleanTF{#1}{#2}{\kms[#2]}}
}}

\NewDocumentCommand{\vIPW}{s o O{=}}{\ensuremath{v_{IPW}
\IfNoValueF{#2}{
\IfNoValueTF{#3}{=}{#3}
\IfBooleanTF{#1}{#2}{\kms[#2]}}
}}

\NewDocumentCommand{\As}{s o O{=}}{\ensuremath{A_s
\IfNoValueF{#2}{
\IfNoValueTF{#3}{=}{#3}
\IfBooleanTF{#1}{#2}{#2 \%}}
}}

\NewDocumentCommand{\mfast}{s o O{=}}{\ensuremath{m_s
\IfNoValueF{#2}{
\IfNoValueTF{#3}{=}{#3}
\IfBooleanTF{#1}{#2}{\verify{\pten{-3}[#2] \km^{-1} \, s}}}
}}

\NewDocumentCommand{\grate}{o o}{\ensuremath{
\gamma\IfNoValueF{#1}{/\Omega_{#1}}
\IfNoValueF{#2}{= 10^{{#2}}}
}}

\NewDocumentCommand{\gmax}{o}{\ensuremath{
\gamma_\mathrm{max}\IfNoValueF{#1}{/\Omega_{#1}}
}}

\NewDocumentCommand{\kvec}{o}{\ensuremath{
\vec{k} \rho\IfNoValueF{#1}{{_{#1}}}
}}

\NewDocumentCommand{\kpar}{o}{\ensuremath{
{k_\parallel} \rho\IfNoValueF{#1}{{_{#1}}}
}}

\NewDocumentCommand{\kper}{o}{\ensuremath{
{k_\perp} \rho\IfNoValueF{#1}{{_{#1}}}
}}

\NewDocumentCommand{\ani}{s o}{\ensuremath{
R\IfNoValueF{#2}{_{#2}}
\IfBooleanT{#1}{\, [\perp\!/\!\parallel]}
}}

\NewDocumentCommand{\Temp}{o}{\ensuremath{T{\IfNoValueF{#1}{_{#1}}}}}

\NewDocumentCommand{\Trat}{s m m o}{\ensuremath{
T_{\IfNoValueF{#4}{{#4};}#2}/T_{\IfNoValueF{#4}{{#4};}#3}
 \IfBooleanT{#1}{\, [\#]}
}}

\NewDocumentCommand{\pbeta}{s o}{\ensuremath{
\beta\IfNoValueF{#2}{_{#2}}
 \IfBooleanT{#1}{\, [\#]}
}}

\NewDocumentCommand{\pbetaR}{o}{\ensuremath{
(\pbeta[\parallel
\IfNoValueF{#1}{;#1}], \ani[#1])
}}

\NewDocumentCommand{\dv}{o}{\ensuremath{\Delta v\IfNoValueF{#1}{_{#1}}}}
\NewDocumentCommand{\ca}{o}{\ensuremath{C_{A\IfNoValueF{#1}{;#1}}}}
\NewDocumentCommand{\dvca}{o o}{\ensuremath{\dv[#1]/\ca[#2]}}

\NewDocumentCommand{\nuc}{o}{\ensuremath{\nu_{c\IfNoValueF{#1}{;#1}}}}
\NewDocumentCommand{\Nc}{o}{\ensuremath{N_{c\IfNoValueF{#1}{;#1}}}}
\NewDocumentCommand{\Ac}{o}{\ensuremath{A_{c\IfNoValueF{#1}{;#1}}}}

\NewDocumentCommand{\tauEXP}{o}{\ensuremath{
\tau_{\mathrm{exp}\IfNoValueF{#1}{;#1}
}}}

\NewDocumentCommand{\tauCC}{o}{\ensuremath{
\tau_{\mathrm{C}\IfNoValueF{#1}{;#1}
}}}

\NewDocumentCommand{\SSN}{o}{\ensuremath{\mathrm{SSN}
\IfNoValueF{#1}{#1}}}
\NewDocumentCommand{\NSSN}{o}{\ensuremath{\mathrm{NSSN}
\IfNoValueF{#1}{#1}}}

\newcommand{\sw}{\ensuremath{\mathrm{sw}}}

\NewDocumentCommand{\qpar}{o}{\ensuremath{
q_{\parallel
\IfNoValueF{#1}{;#1}
}}}

\NewDocumentCommand{\edv}{o}{\ensuremath{
\tilde{E}_{\dv[#1]
}}}

\NewDocumentCommand{\ndays}{o}{
\ensuremath{N_\mathrm{days}{\IfNoValueF{#1}{= {#1}}}}
}

\NewDocumentCommand{\se}{o}{\ensuremath{
S{\IfNoValueF{#1}{_{#1}}}
}}

\NewDocumentCommand{\ab}{o}{\ensuremath{
A{\IfNoValueF{#1}{_{#1}}}
}}

\NewDocumentCommand{\ahe}{s o O{=}}{\ensuremath{\ab[\he]
\IfNoValueF{#2}{
\IfNoValueTF{#3}{=}{#3}
\IfBooleanTF{#1}{#2}{#2 \%}}
}}

\NewDocumentCommand{\corr}{o}{\ensuremath{
\rho
\IfNoValueF{#1}{(#1)}
}}

\NewDocumentCommand{\xhel}{s o O{=} o}{\ensuremath{
\IfBooleanTF{#1}{\sigma_{c\IfNoValueF{#4}{,#4}}}{\abs{\sigma_{c\IfNoValueF{#4}{,#4}}}}
\IfNoValueF{#2}{
\IfNoValueTF{#3}{=}{#3}
#2}
}}

\NewDocumentCommand{\SpecInd}{o}{\ensuremath{\gamma
\IfNoValueF{#1}{_{#1}}}}
\NewDocumentCommand{\QT}{o}{\ensuremath{\mathrm{QT}
\IfNoValueF{#1}{= #1}}}

\NewDocumentCommand{\pten}{o m}{\ensuremath{
\IfNoValueF{#1}{#1 \times} 10^{#2}
}}

\NewDocumentCommand{\abs}{m}{\ensuremath{\left| #1 \right|}}

\NewDocumentCommand{\func}{m m o O{=}}{\ensuremath{#1\left({#2}\right)\IfNoValueF{#3}{\! #4 \! #3}}}

\definecolor{q}{HTML}{228B22}
\definecolor{wc}{HTML}{FF8C00}
\definecolor{dnc}{HTML}{FF00FF}
\definecolor{todo}{HTML}{e13748}
\definecolor{ben}{HTML}{e13748}
\definecolor{bob}{HTML}{0080FF}

\NewDocumentCommand{\todo}{s o m}{\IfBooleanF{#1}{\textcolor{todo}{\textbf{TODO}\IfNoValueF{#2}{ (#2)}: \textit{#3}}}}
\NewDocumentCommand{\verify}{s o m}{\IfBooleanTF{#1}{#3}{\textcolor{todo}{\textbf{VERIFY}\IfNoValueF{#2}{ (#2)}: \textit{#3}}}}
\NewDocumentCommand{\goal}{s o m}{\IfBooleanTF{#1}{#3}{\textcolor{todo}{\textbf{GOAL}\IfNoValueF{#2}{ (#2)}: \textit{#3}}}}

\NewDocumentCommand{\sect}{o m}{Section~\ref{sec:#2}\IfNoValueF{#1}{ #1}}

\usepackage{tablefootnote}

\newcommand{\gridfigscale}{1}
\newcommand{\gridfigbasepath}{}

\NewDocumentCommand{\plotVswHist}{s}{
\IfBooleanTF{#1}{\begin{figure*}}{\begin{figure}}
\includegraphics[width=\linewidth]{vsw-hist-with-indicators-Ahe_dnn}
\caption{\label{fig:vsw-hist}
The probability density function (PDF) of the solar wind speed observed by the Wind Faraday cups at \au[1] during solar minima 23 and 24 is plotted in black.
The colored segments indicate speed ranges derived in this paper, \citetalias{\BibOne}, and \citet{Wind:SWE:Wk}.
The \vsw\ range corresponding to each segment is indicated by a horizontal, labeled line of the same color below the PDF.
The vertical location of these horizontal bars is chosen so that they do not overlap, but has no other significance.
The speed ranges are summarized in \cref{tbl:speeds}.
}
\IfBooleanTF{#1}{\end{figure*}}{\end{figure}}
}

\RenewDocumentCommand{\plotVswHist}{s}{
\IfBooleanTF{#1}{\begin{figure*}}{\begin{figure}}
\includegraphics[width=\linewidth]{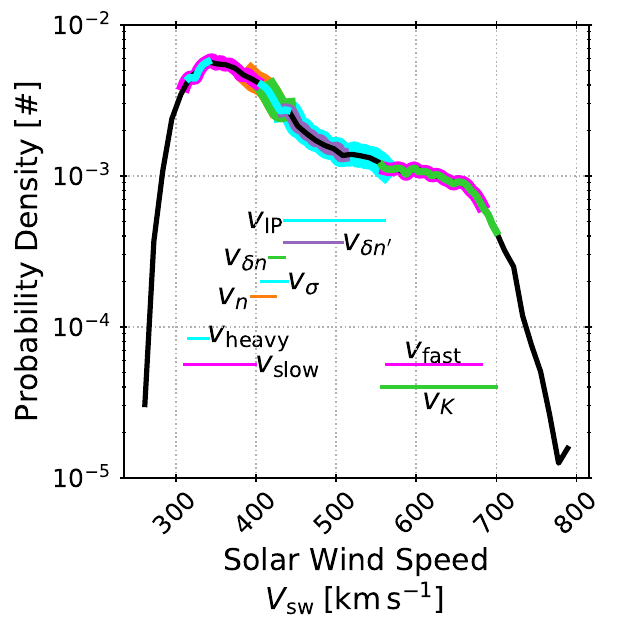}
\caption{\label{fig:vsw-hist}
The probability density function (PDF) of the solar wind speed observed by the Wind Faraday cups at \au[1] during solar minima 23 and 24 is plotted in black.
The colored segments indicate speed ranges derived in this paper, \citetalias{\BibOne}, \citet{ACE:SWICS:FStransition}, and \citet{Wind:SWE:Wk}.
The \vsw\ range corresponding to each segment is plotted in the same color below the PDF and labeled.
The vertical location of these \vsw\ ranges is chosen to avoid overlap.
The speed ranges are summarized in \cref{tbl:speeds}.
}
\IfBooleanTF{#1}{\end{figure*}}{\end{figure}}
}

\NewDocumentCommand{\plotAheVsw}{s}{
\IfBooleanTF{#1}{\begin{figure*}}{\begin{figure}}
\includegraphics[width=\linewidth]{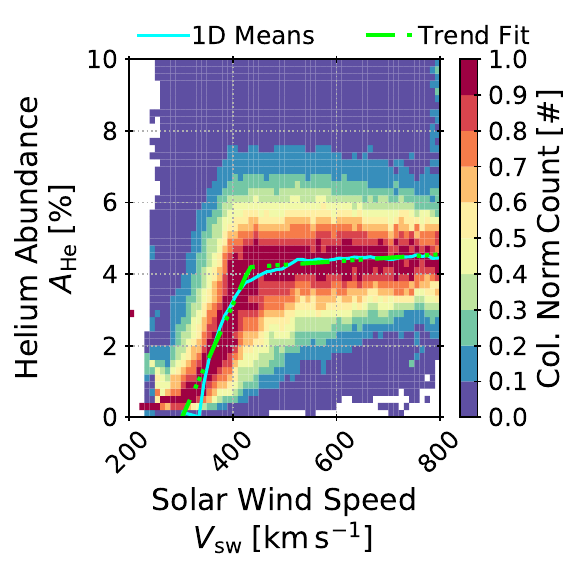}
\caption{\label{fig:vsw-ahe}
The helium abundance as a function of solar wind speed.
\ahe\ has been normalized to its maximum value in each column.
The blue line indicates the mean value of \ahe\ in each column.
The dash-dotted green line is a bilinear fit to these means using \cref{eq:two-line}.
The helium abundance monotonically increases from $0\%$ to $4.19\%$ in slow wind and saturates to this \ahe[4.19] in fast solar wind for which \vsw[433][>].
}
\IfBooleanTF{#1}{\end{figure*}}{\end{figure}}
}

\NewDocumentCommand{\plotCategorizationCartoon}{s}{
\IfBooleanTF{#1}{\begin{figure*}}{\begin{figure}}
\includegraphics[width=\linewidth]{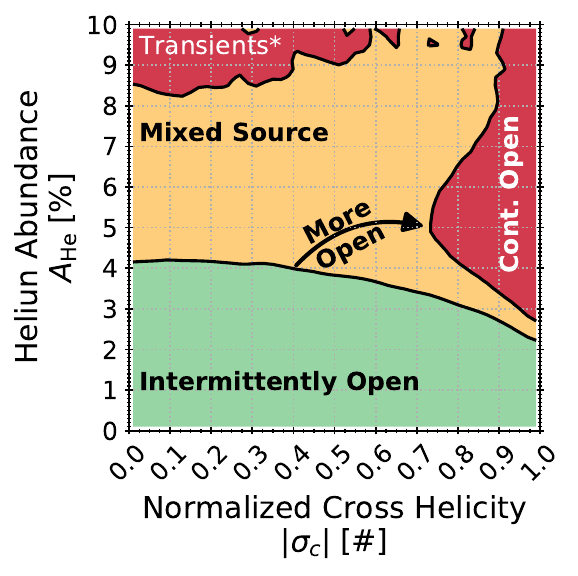}
\caption{\label{fig:xhel-ahe-vsw:cartoon}
A plot of \vsw\ as a function of \xhel\ and \ahe\ with contours at \vsw*[300] and \kms[460].
\citetOne\ considers helium-poor solar wind with \ahe*[\As][<] to originate in intermittently open source regions and Alfvénic solar wind with \xhel[0.7][>] in the red region to originate in continuously open source regions.
That work hypothesizes that the speed enhancement of non-Alfvénic, helium rich solar wind in the top left corner of the plot is due to transients.
}
\IfBooleanTF{#1}{\end{figure*}}{\end{figure}}
}

\NewDocumentCommand{\plotVswDnn}{s}{
\IfBooleanTF{#1}{\begin{figure*}}{\begin{figure}}
\includegraphics[width=\linewidth]{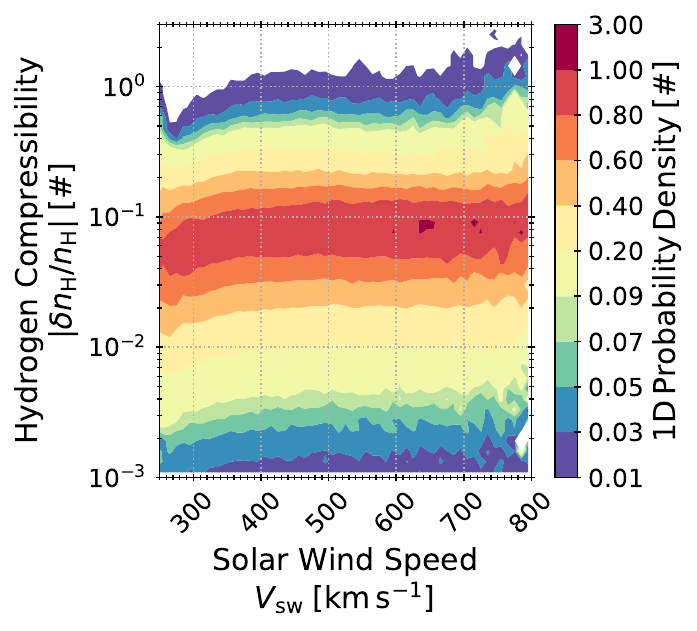}
\caption{\label{fig:vsw:dnn}
A contour plot of the PDF of \func{\dnn}{\vsw}.
In the underlying 2D histogram, the frequency of observing \dnn\ in each column is normalized to its maximum value.
The contours are smoothed with a $1\sigma$ Gaussian kernel for visual clarity.
}
\IfBooleanTF{#1}{\end{figure*}}{\end{figure}}
}

\NewDocumentCommand{\plotVswXhel}{s}{
\IfBooleanTF{#1}{\begin{figure*}}{\begin{figure}}
\includegraphics[width=\linewidth]{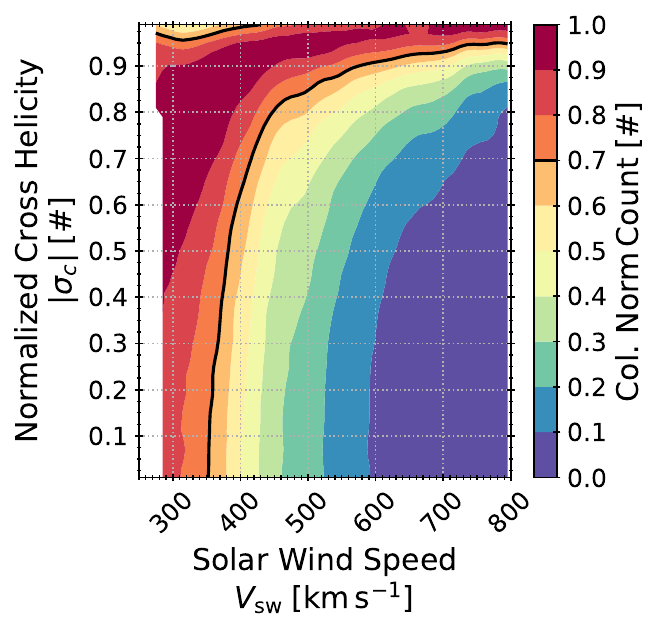}
\caption{\label{fig:vsw:xhel}
A contour plot of the PDF of \xhel.
The columns in the underlying 2D histogram have been normalized to their maximum value.
The contour at 0.7 is indicated in black.
}
\IfBooleanTF{#1}{\end{figure*}}{\end{figure}}
}

\NewDocumentCommand{\plotVswXhelDnn}{s}{
\IfBooleanTF{#1}{\begin{figure*}}{\begin{figure}}
\includegraphics[width=\linewidth]{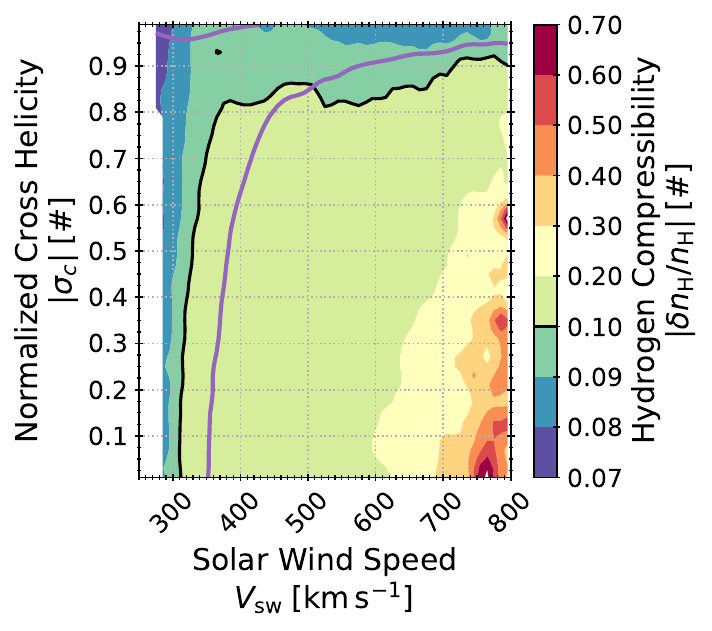}
\caption{\label{fig:vsw:xhel:dnn}
A contour plot of the PDF of \func{\dnn}{\vsw,\xhel}.
Average \dnn\ is calculated as the logarithmic mean.
Contours are smoothed with a $1\sigma$ Gaussian kernel for visual clarity.
The purple line is the 0.7 contour from \cref{fig:vsw:xhel}.
}
\IfBooleanTF{#1}{\end{figure*}}{\end{figure}}
}

\NewDocumentCommand{\plotVswAheXhel}{s}{
\IfBooleanTF{#1}{\begin{figure*}}{\begin{figure}}
\begin{centering}
\includegraphics[width=\linewidth]{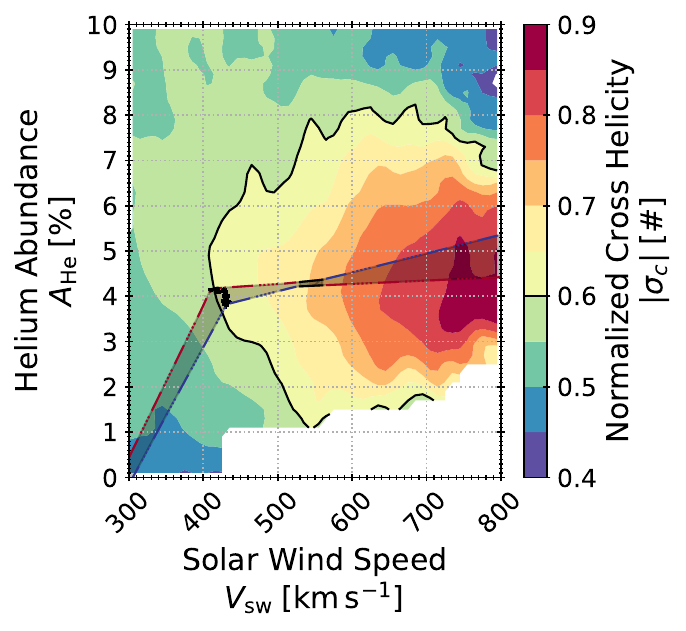}
\end{centering}
\caption{\label{fig:vsw-ahe-xhel}
A contour plot of the normalized cross helicity as a function of solar wind speed and helium abundance.
The contours are smoothed with a $1\sigma$ gaussian filter for clarity.
The contour at \xhel[0.6] is highlighted in black.
The black points mark the saturation point \satpoint.
The saturation points occur at \xhel[0.6][>].
The shaded black region corresponds to fits to \func{\ahe}{\vsw} for each \xhel\ quantile: the red dash-dotted line is the high \xhel\ edge and the blue dash-dotted line is the low \xhel\ edge.
The region from 530 to \kms[560] is where the maximum and minimum \xhel\ fits to \func{\ahe}{\xhel} cross.
This region is bounded by solid black lines and crosses the \xhel[0.7] contour.
}
\IfBooleanTF{#1}{\end{figure*}}{\end{figure}}
}

\NewDocumentCommand{\plotVswAheDnn}{s}{
\IfBooleanTF{#1}{\begin{figure*}}{\begin{figure}}
\begin{centering}
\includegraphics[width=\linewidth]{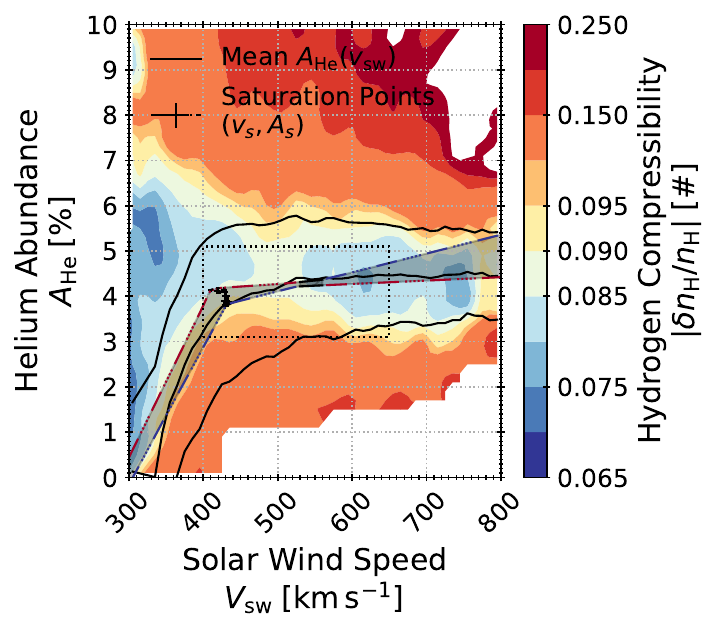}
\end{centering}
\caption{\label{fig:vsw-ahe-dnn}
A contour plot of the compressibility as a function of solar wind speed and helium abundance.
Contours are smoothed with a $1\sigma$ gaussian filter for clarity.
Solid black lines are the mean and standard deviations from 1D fits plotted in \citetalias[Fig.~2]{\BibOne}.
The shaded black region bounded by red and blue dash-dotted lines matches \cref{fig:vsw-ahe-xhel}.
The region encircled by the box highlights the location of the saturation points \satpoint\ and 
is plotted in \cref{fig:vsw-ahe-dnn:zoom}.
}
\IfBooleanTF{#1}{\end{figure*}}{\end{figure}}
}

\NewDocumentCommand{\plotVswAheDnnZoom}{s}{
\IfBooleanTF{#1}{\begin{figure*}}{\begin{figure}}
\begin{centering}
\includegraphics[width=\linewidth]{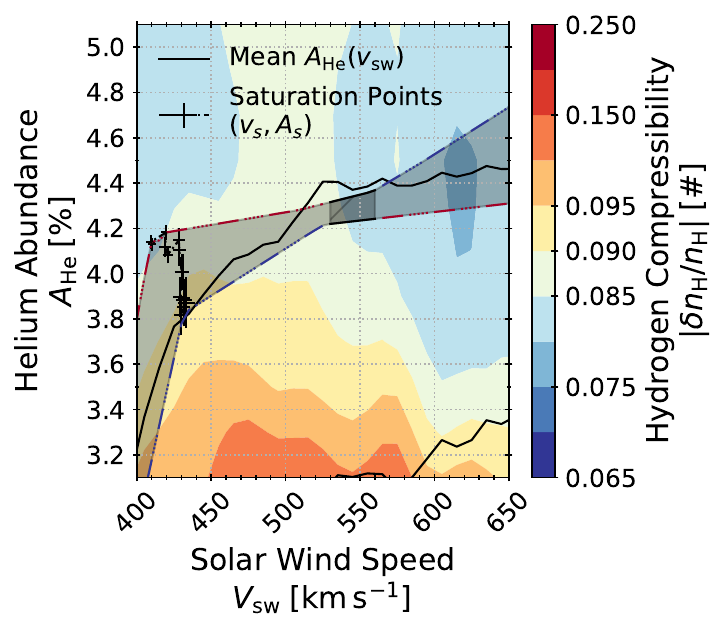}
\end{centering}
\caption{\label{fig:vsw-ahe-dnn:zoom}
A zoom in of the region indicated in \cref{fig:vsw-ahe-dnn}.
The gradient of the contour for \dnn_\Hy_[0.09] changes across the saturation points.
The fits  \func{\ahe}{\vsw} cross over the speed range \vsw*[530][\approx] to \kms[560], which is bounded by solid black lines instead of red and blue dash-dotted lines.
This region where the \func{\ahe}{\vsw} fits at low \xhel\ become larger than the fits at high \xhel\ crosses the \dnn_\Hy_[0.085] contour.
}
\IfBooleanTF{#1}{\end{figure*}}{\end{figure}}
}

\NewDocumentCommand{\plotSaturationFits}{s}{
\IfBooleanTF{#1}{\begin{figure*}}{\begin{figure}}
\begin{centering}
\includegraphics[width=\linewidth]{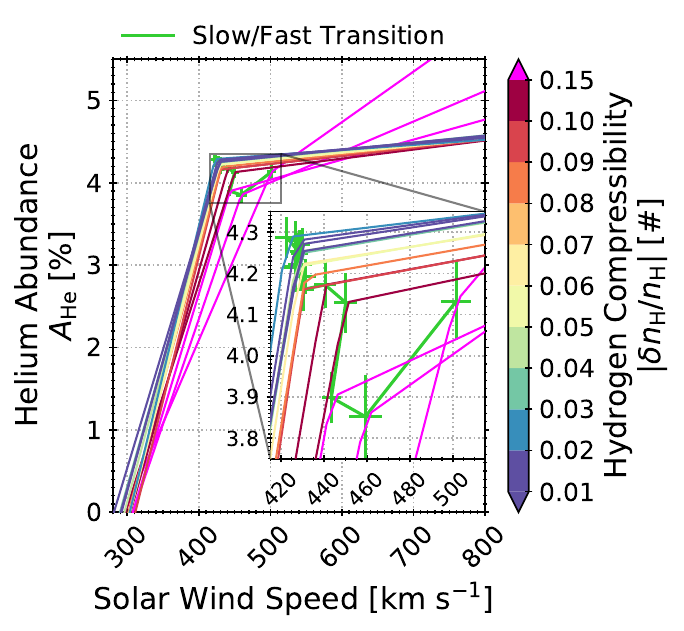}
\end{centering}
\caption{\label{fig:sat-fits}
Fits of \func{\ahe}{\vsw} for 15 quantiles of hydrogen compressibility (\dnn).
The color bar identifies \dnn.
Quantiles with \dnn_\Hy_[0.15][>] are indicated in pink.
The green plots indicate the saturation points \satpoint.
}
\IfBooleanTF{#1}{\end{figure*}}{\end{figure}}
}

\NewDocumentCommand{\plotSaturationFitsScaled}{s}{
\IfBooleanTF{#1}{\begin{figure*}}{\begin{figure}}
\begin{centering}
\includegraphics[width=\linewidth]{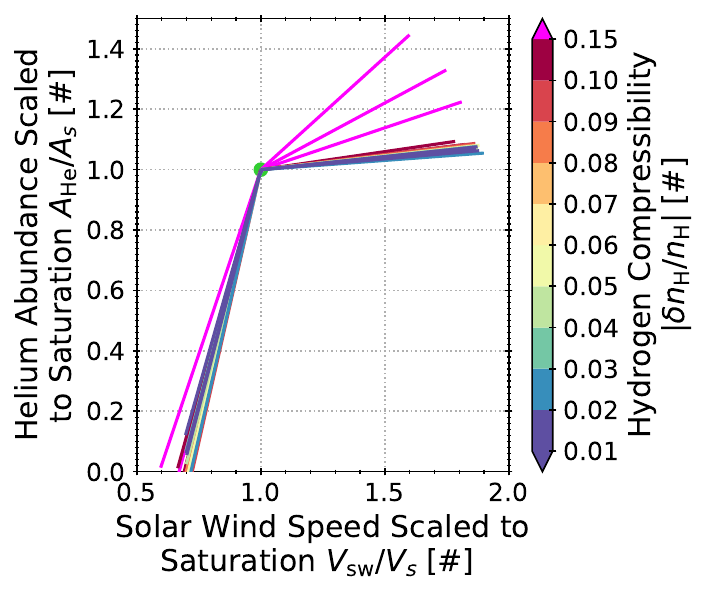}
\end{centering}
\caption{\label{fig:sat-fits:scaled}
Fits to \func{\ahe}{\vsw} in 15 \dnn\ quantiles.
Each fit is scaled to its situation point \satpoint, which is indicated at $\left(1, 1\right)$ in green.
As in \cref{fig:sat-fits}, the quantiles are indicated by the color bar.
}
\IfBooleanTF{#1}{\end{figure*}}{\end{figure}}
}

\NewDocumentCommand{\plotAfastAs}{s}{
\IfBooleanTF{#1}{\begin{figure*}}{\begin{figure}}
\includegraphics[page=111, width=\linewidth]{FastAhe_Scaled_to_As}
\caption{\label{fig:ahe-enh}
Helium abundance at \kms[795] derived from the fits plotted in \cref{fig:sat-fits} scaled to the saturation abundance \As.
Error bars and markers are colored by \xhel\ to facilitate comparison with prior figures.
The dash-dotted line is a fit to the trend.
\todo{Describe figure inline}
\todo{Make same plot as function of \dnn[p]}
}
\IfBooleanTF{#1}{\end{figure*}}{\end{figure}}
}

\RenewDocumentCommand{\plotAfastAs}{s}{
\IfBooleanTF{#1}{\begin{figure*}}{\begin{figure}}
\includegraphics[width=\linewidth]{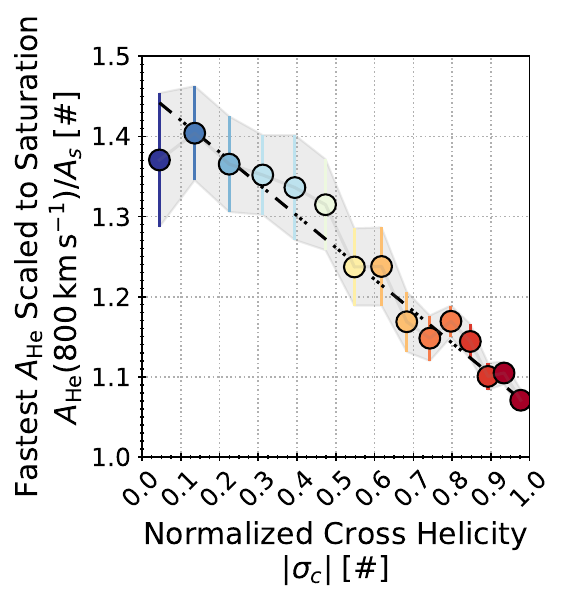}
\caption{\label{fig:ahe-enh}
Helium abundance at \kms[800] derived from the fits of \func{\ahe}{\vsw} across \xhel\ quantiles in \citetOne, each scaled to the saturation abundance \As\ in that paper.
Error bars and markers are colored by \xhel\ to facilitate comparison with prior figures.
The dash-dotted line is a fit to the trend.
}
\IfBooleanTF{#1}{\end{figure*}}{\end{figure}}
}

\NewDocumentCommand{\plotAfastAsEnhCube}{s}{
\IfBooleanTF{#1}{\begin{figure*}}{\begin{figure}}
\includegraphics[width=\linewidth]{vsw-xhel-AheEnh}
\caption{\label{fig:ahe-enh:3d}
The helium abundance enhancement above the saturation abundance \As\ as a function of solar wind speed and cross helicity.
\todo{Describe figure inline}
}
\IfBooleanTF{#1}{\end{figure*}}{\end{figure}}
}

\NewDocumentCommand{\plotAbundanceEnhSlopes}{s}{
\IfBooleanTF{#1}{\begin{figure*}}{\begin{figure}}
\includegraphics[page=2,width=\linewidth]{Abundance-Enh-Over-Saturation}
FastAhe_Scaled_to_As-with-fit
\caption{\label{fig:ahe-enh:slopes}
The gradient of the helium abundance enhancement with respect to the saturation abundance \As\ as a function of \xhel\ for increasing \vsw.
In other words, the gradient along the \xhel\ axis in \cref{fig:ahe-enh:3d}.
\todo{Describe figure inline}
}
\IfBooleanTF{#1}{\end{figure*}}{\end{figure}}
}

\NewDocumentCommand{\plotXhelAheDnn}{s}{
\renewcommand{\gridfigscale}{1}
\renewcommand{\gridfigbasepath}{HiddenVar-xhel-ahe-vsw}
\IfBooleanTF{#1}{\begin{figure*}}{\begin{figure}}
\begin{centering}
\includegraphics[width=\gridfigscale\linewidth]{\gridfigbasepath/abs_xhel-Ahe-dn_p_abs-dn_p_Contours}
\end{centering}
\caption{\label{fig:xhel-ahe-dnn}
Contour plots of the compressibility \dnn\ as a function of the cross helicity and helium abundance.
Contours are smoothed with a $1\sigma$ gaussian filter for visual clarity.
The compressibility decreases with increasing \xhel\ and the range of \ahe\ corresponding to a given level of compressibility decreases as \xhel\ increased and \dnn_\Hy_ decreases.
Contours at the \dnn_\Hy_[0.09] and 0.085 levels are highlighted.
}
\IfBooleanTF{#1}{\end{figure*}}{\end{figure}}
}

\NewDocumentCommand{\plotXhelAheDnnContours}{s}{
\renewcommand{\gridfigscale}{1}
\renewcommand{\gridfigbasepath}{HiddenVar-xhel-ahe-vsw}
\IfBooleanTF{#1}{\begin{figure*}}{\begin{figure}}
\begin{centering}
\includegraphics[width=\gridfigscale\linewidth]{\gridfigbasepath/abs_xhel-Ahe-dn_p_abs-vsw_Contours}
\end{centering}
\caption{\label{fig:xhel-ahe-dnn:contours}
The hydrogen compressibility as a function of the helium abundance and normalized cross helicity.
Contours at the \dnn_\Hy_[0.085] and 0.090 levels are highlighted.
Labeled contours indicate the slow wind peak (\vslow), saturation speed (\vs), and speed at Gaussian distributions fit to the slow and fast wind peak during solar minima intersect (\vi). 
These three speeds are derived in \citetalias{\BibOne}.
}
\IfBooleanTF{#1}{\end{figure*}}{\end{figure}}
}

\NewDocumentCommand{\plotXhelAheDnnMultiPanel}{s}{
\renewcommand{\gridfigscale}{1}
\IfBooleanTF{#1}{\begin{figure*}}{\begin{figure}}
\begin{centering}
\includegraphics[width=\gridfigscale\linewidth,page=2]{xhel_ahe_dnn_quantiles}\\
\includegraphics[width=\gridfigscale\linewidth,page=1]{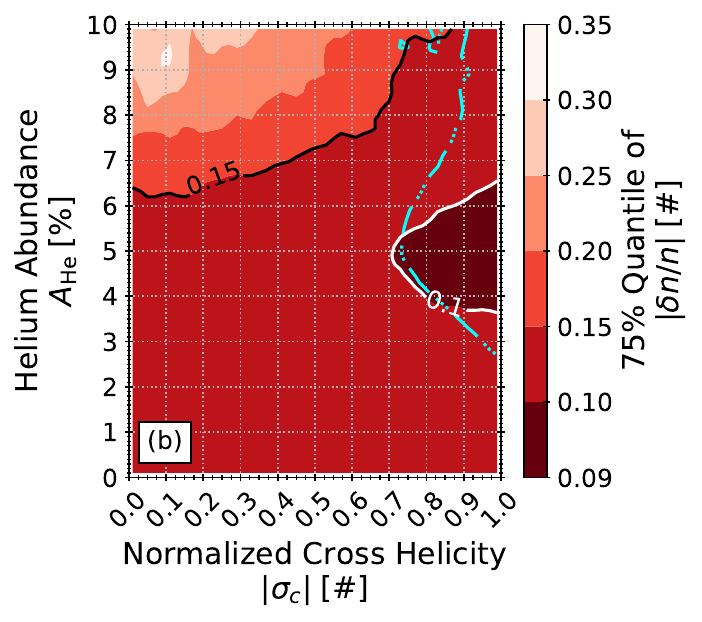}
\end{centering}
\caption{\label{fig:xhel-ahe-dnn}
Contour plots of \textbf{(a)} mean and \textbf{(b)} 75\% quantile \dnn\ as a function of \xhel\ and \ahe.
Contours are smoothed with a $1\sigma$ Gaussian kernel for visual clarity.
Solid black contours in panel (a) indicate \dnn_\Hy_[0.085], 0.1, and 0.15.
The dash-dotted black contour indicates where the 75\% quantile of \dnn_\Hy_[0.15] in panel (b).
Solid contours in panel (b) indicate where the 75\% quantile of \dnn_\Hy_[0.1] and 0.15.
The region of the plane with \xhel\ greater than that indicated by the dash-dotted blue line corresponds to solar wind originating in continuously open source regions as defined in \citetOne.
}
\IfBooleanTF{#1}{\end{figure*}}{\end{figure}}
}

\NewDocumentCommand{\plotXhelDnnAhe}{s}{
\renewcommand{\gridfigscale}{0.4}
\IfBooleanTF{#1}{\begin{figure*}}{\begin{figure}}
\begin{centering}
\includegraphics[width=\gridfigscale\linewidth, page=3, trim={0 0 0 0.75cm}, clip]{xhel_dnn_ahe_or_vsw_quantiles}\\
\includegraphics[width=\gridfigscale\linewidth, page=2, trim={0 0 0 0.75cm}, clip]{xhel_dnn_ahe_or_vsw_quantiles}
\includegraphics[width=\gridfigscale\linewidth, page=1, trim={0 0 0 0.75cm}, clip]{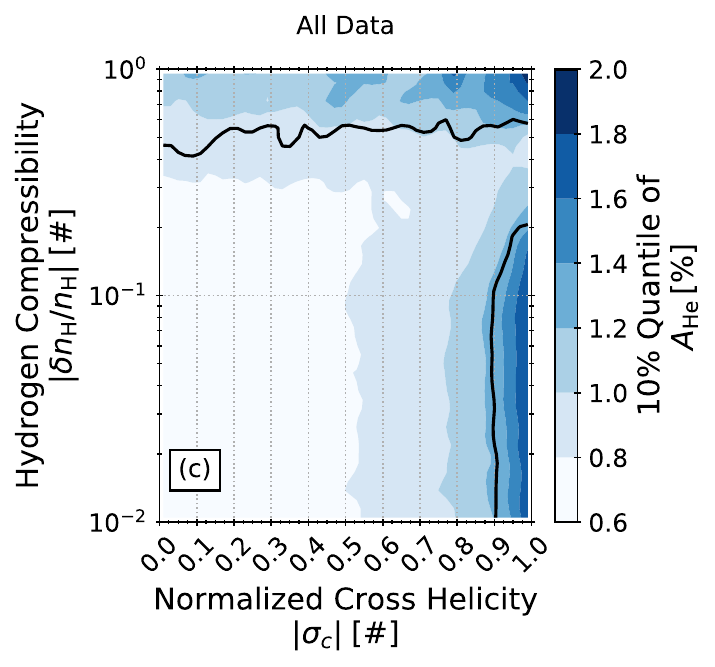}\\
\includegraphics[width=\gridfigscale\linewidth, page=4]{xhel_dnn_ahe_or_vsw_quantiles}\\
\end{centering}
\caption{\label{fig:xhel-dnn-ahe}Contour plots of the \textbf{(a)} mean, \textbf{(b)} 90\% quantile, and \textbf{(c)} 10\% quantile of the helium abundances as a function of cross helicity and solar wind compressibility.
Panel \textbf{(d)} plots the variability of the helium abundance.
The black contours in all panels indicate \ahe[3.82] from panel (a), which  is the smallest \As\ derived in \citetOne\ and this work.
All contours are smoothed with a $1\sigma$ Guassian kernel for visual clarity.
While the ranges for the color scales in panels (a) to (c) are different, the color scales themselves are chosen to provide visual continuity such that the 90\% quantiles in panel (b) are plotted in red to match colors at the high \ahe\ range of panel (a) and the 10\% quantiles in panel (c) are plotted in blue to match the low \ahe\ range in panel (a).
}
\IfBooleanTF{#1}{\end{figure*}}{\end{figure}}
}

\NewDocumentCommand{\plotXhelDnnVsw}{s}{
\renewcommand{\gridfigscale}{1}
\IfBooleanTF{#1}{\begin{figure*}}{\begin{figure}}
\begin{centering}
\includegraphics[width=\gridfigscale\linewidth, page=13, trim={0 0 0 0.75cm}, clip]{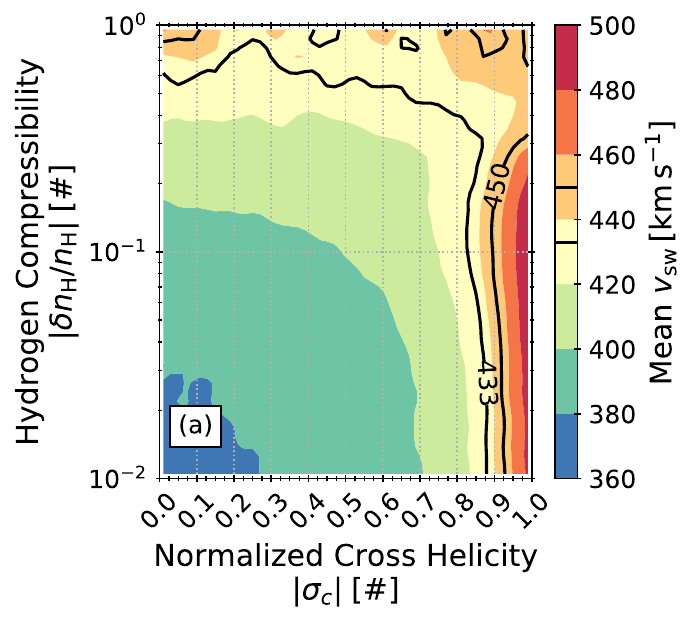}\\
\includegraphics[width=\gridfigscale\linewidth, page=14]{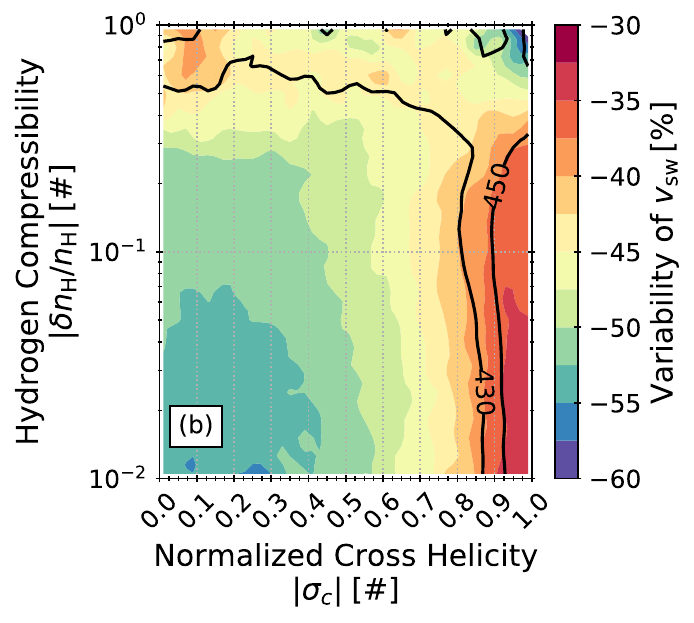}\\
\end{centering}
\caption{\label{fig:xhel-dnn-vsw}\textbf{(a)} The solar wind speed as a function of cross helicity and compressibility.
The black contour indicates \vsw*[433] and \kms[450], the fastest \vs\ derived in \citetOne\ and the fastest \vs\ derived for incompressible solar wind (\dnn_\Hy_[0.15][<]) in this work.
\textbf{(b)} The variability of \vsw\ calculated in the same manner as in \cref{fig:xhel-dnn-ahe}.
}
\IfBooleanTF{#1}{\end{figure*}}{\end{figure}}
}

\NewDocumentCommand{\plotLinesVswDnnXhel}{s}{
\IfBooleanTF{#1}{\begin{figure*}}{\begin{figure}}
\begin{centering}
\includegraphics[width=\linewidth]{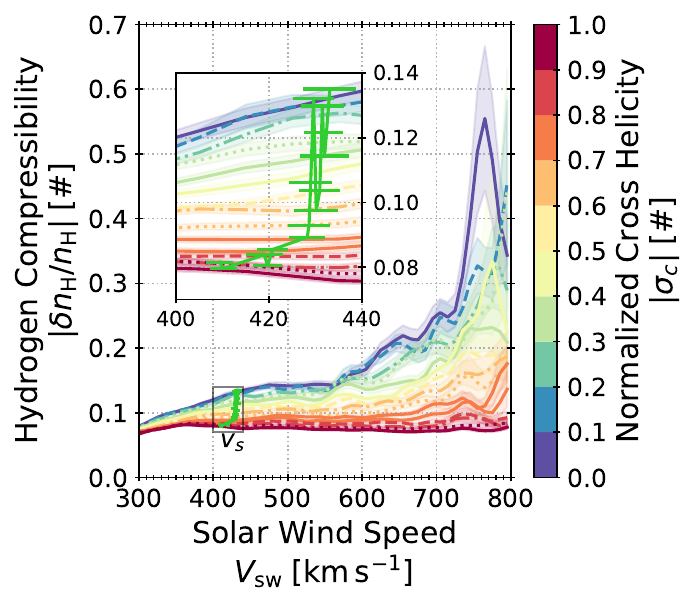}
\caption{\label{fig:lines:vsw-dnn-xhel}
The hydrogen compressibility \dnn_\Hy_ as a function of the solar wind speed \vsw\ in quantiles of \xhel, which is given by the colorbar.
The green line labeled \vs\ indicates \dnn_\Hy_ in each \xhel\ quantile at the corresponding saturation abundance.
}
\end{centering}
\IfBooleanTF{#1}{\end{figure*}}{\end{figure}}
}

\NewDocumentCommand{\plotLinesAheDnnXhel}{s}{
\IfBooleanTF{#1}{\begin{figure*}}{\begin{figure}}
\includegraphics[width=\linewidth]{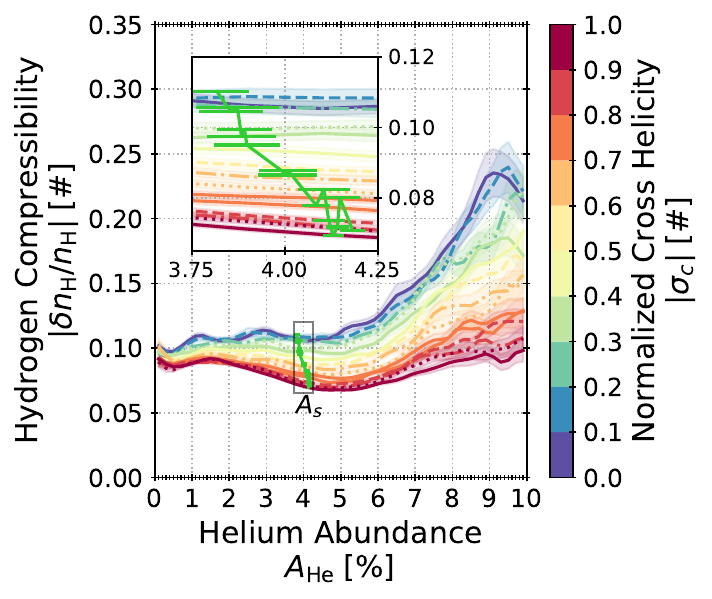}
\caption{\label{fig:lines:ahe-dnn-xhel}
The hydrogen compressibility \dnn_\Hy_ as a function of the helium abundance \ahe\ in quantiles of \xhel, which is given by the colorbar.
The green line labeled \As\ indicates \dnn_\Hy_ in each \xhel\ quantile at the corresponding saturation abundance.
}
\IfBooleanTF{#1}{\end{figure*}}{\end{figure}}
}

\NewDocumentCommand{\plotLinesAheXhelDnn}{s}{
\IfBooleanTF{#1}{\begin{figure*}}{\begin{figure}}
\includegraphics[page=1, width=\linewidth]{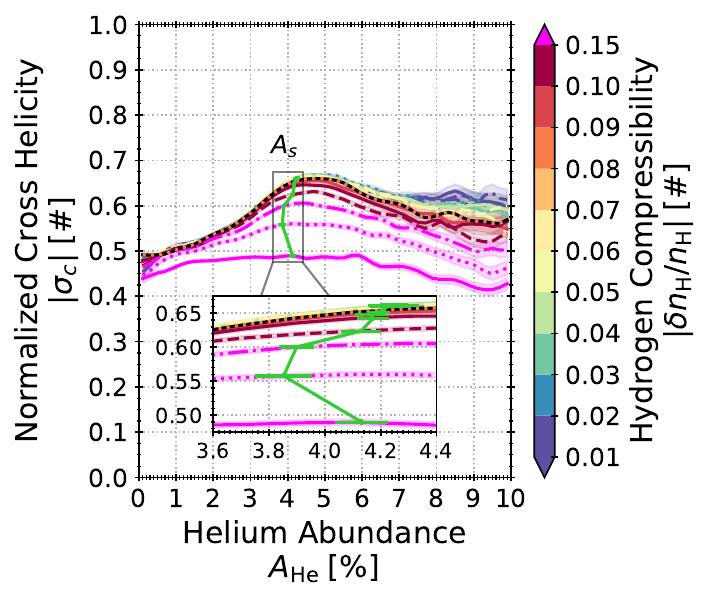}\\
\caption{\label{fig:lines:ahe-xhel-dnn}
The average normalized cross helicity (\xhel) as a function of the helium abundance (\ahe) in each of 15 compressibility (\dnn) quantiles.
Shaded regions indicate the standard error of the mean.
The color of each \dnn\ quantile is given by the color bar.
Lines are smoothed with a $1\sigma$ Gaussian filter for visual clarity.
Compressibilities \dnn[0.15][>] are plotted in pink.
The quantile corresponding to \dnn[0.085] is highlighted by plotting the dotted orange line on top of a black solid line.
The green line plots \func{\As}{\xhel} in each \dnn\ quantile.
}
\IfBooleanTF{#1}{\end{figure*}}{\end{figure}}
}

\NewDocumentCommand{\plotLinesAheXhelDnnZoom}{s}{
\IfBooleanTF{#1}{\begin{figure*}}{\begin{figure}}
\includegraphics[page=2, width=\linewidth]{sat_on_other}\\
\caption{\label{fig:lines:ahe-xhel-dnn:zoom}
A zoom into the region surrounding \As\ in \cref{fig:lines:ahe-xhel-dnn}.
The green markers are connected in \dnn\ order and the error bars indicate the \verify{uncertainty}.
}
\IfBooleanTF{#1}{\end{figure*}}{\end{figure}}
}

\NewDocumentCommand{\plotLinesAheVswDnn}{s}{
\IfBooleanTF{#1}{\begin{figure*}}{\begin{figure}}
\includegraphics[page=2, width=\linewidth]{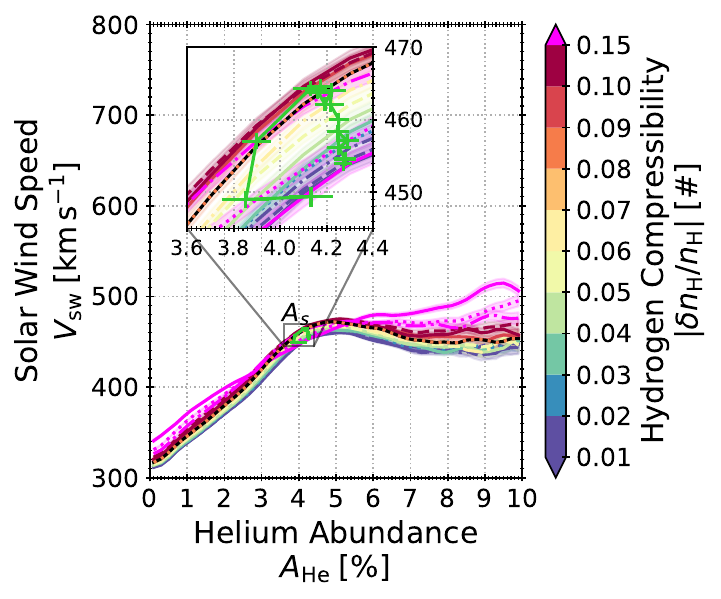}\\
\caption{\label{fig:lines:ahe-vsw-dnn}
The solar wind speed as a function of \ahe\ for each \dnn\ quantile.
The style matches \cref{fig:lines:ahe-xhel-dnn}.
}
\IfBooleanTF{#1}{\end{figure*}}{\end{figure}}
}

\NewDocumentCommand{\plotLinesAheVswDnnZoom}{s}{
\IfBooleanTF{#1}{\begin{figure*}}{\begin{figure}}
\includegraphics[page=4, width=\linewidth]{sat_on_other}\\
\caption{\label{fig:lines:ahe-vsw-dnn:zoom}
A zoom in on the area surrounding \As\ in \cref{fig:lines:ahe-vsw-dnn:zoom}.
}
\IfBooleanTF{#1}{\end{figure*}}{\end{figure}}
}

\NewDocumentCommand{\plotLinesVswXhelDnn}{s}{
\IfBooleanTF{#1}{\begin{figure*}}{\begin{figure}}
\includegraphics[page=3, width=\linewidth]{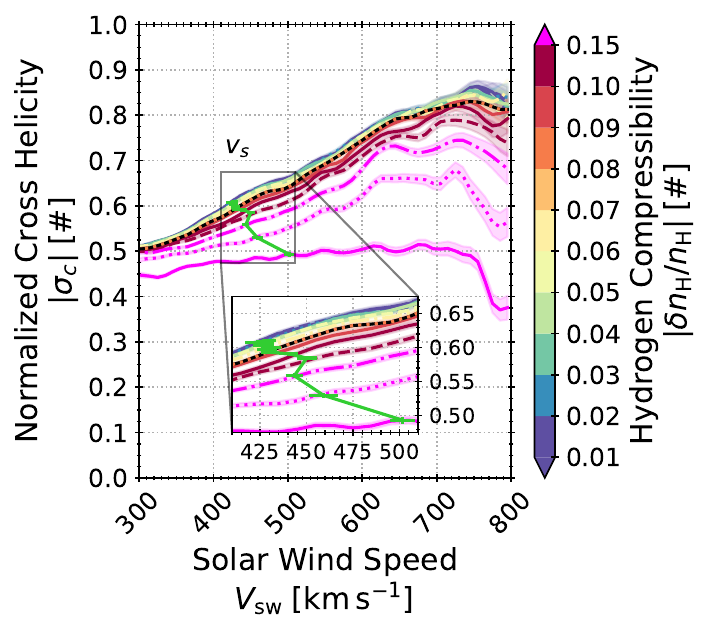}\\
\caption{\label{fig:lines:vsw-xhel-dnn}
The normalized cross helicity as a function of \vsw\ in the 15 \dnn\ quantiles.
Style matches \cref{fig:lines:ahe-xhel-dnn}, with the green line indicating \vs\ instead of \As.
}
\IfBooleanTF{#1}{\end{figure*}}{\end{figure}}
}

\NewDocumentCommand{\plotLinesVswXhelDnnZoom}{s}{
\IfBooleanTF{#1}{\begin{figure*}}{\begin{figure}}
\includegraphics[page=6, width=\linewidth]{sat_on_other}\\
\caption{\label{fig:lines:vsw-xhel-dnn:zoom}
A zoom in on the region surrounding \vs\ in \cref{fig:lines:ahe-vsw-dnn}.
The style follows \cref{fig:lines:ahe-vsw-dnn:zoom}.
In particular, green points for \vs\ are connected in \dnn\ order.
}
\IfBooleanTF{#1}{\end{figure*}}{\end{figure}}
}

\NewDocumentCommand{\plotLinesVswAheDnn}{s}{
\IfBooleanTF{#1}{\begin{figure*}}{\begin{figure}}
\includegraphics[page=4, width=\linewidth]{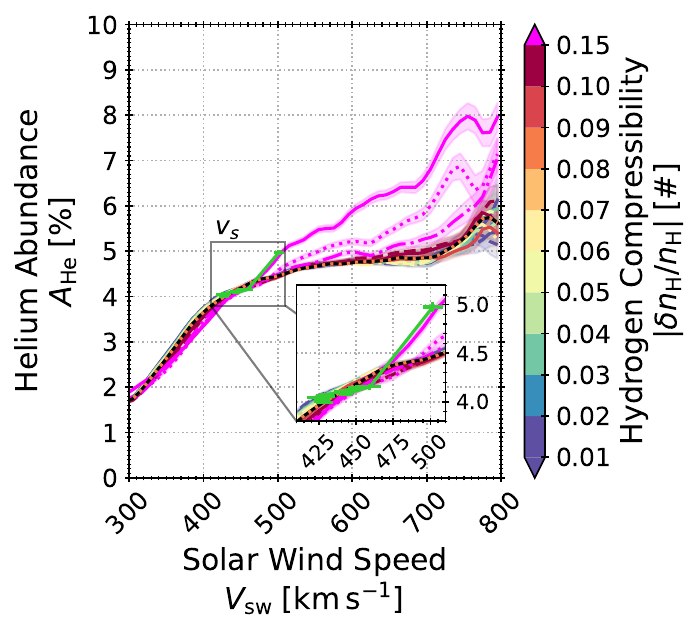}\\
\caption{\label{fig:lines:vsw-ahe-dnn}
The helium abundance (\ahe) as a function of solar wind speed (\vsw) in the 15 \dnn\ quantiles.
Again, style matches \cref{fig:lines:ahe-vsw-dnn}, with the green markers \verify{indicating the saturation points}.
}
\IfBooleanTF{#1}{\end{figure*}}{\end{figure}}
}

\NewDocumentCommand{\plotLinesVswAheDnnZoom}{s}{
\IfBooleanTF{#1}{\begin{figure*}}{\begin{figure}}
\includegraphics[page=8, width=\linewidth]{sat_on_other}\\
\caption{\label{fig:lines:vsw-ahe-dnn:zoom}
A zoom in on the regions surrounding \satpoint\ in \cref{fig:lines:ahe-vsw-dnn}.
}
\IfBooleanTF{#1}{\end{figure*}}{\end{figure}}
}

\NewDocumentCommand{\plotSaturationSlope}{s}{
\IfBooleanTF{#1}{\begin{figure*}}{\begin{figure}}
\includegraphics[width=\linewidth]{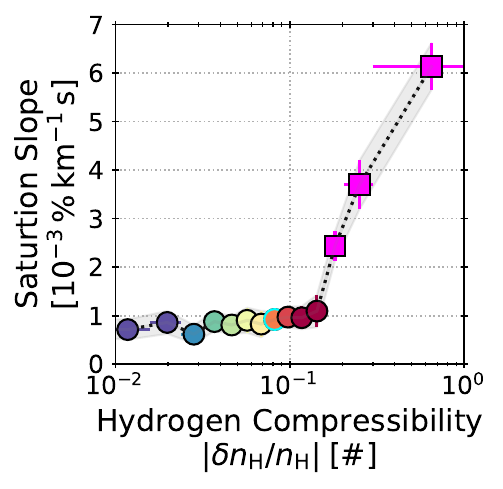}\\
\caption{\label{fig:sat-slopes}
The slopes of \func{\ahe}{\vsw} for speeds \vsw*[\vs][>] in each \dnn\ quantile.
Nominally, this is the slope for the fast solar wind nominally originating in open solar wind regions and we refer to them as ``saturation slopes'' in this work.
Quantiles for which \dnn_\Hy_[0.15][>] are plotted in pink and indicated with squares instead of circles.
The marker indicating the quantile with \dnn_\Hy_[0.085] is blue, while all other quantiles have a black marker edge.
}
\IfBooleanTF{#1}{\end{figure*}}{\end{figure}}
}

\NewDocumentCommand{\plotSaturationSlopeMapped}{s}{
\IfBooleanTF{#1}{\begin{figure*}}{\begin{figure}}
\includegraphics[width=\linewidth]{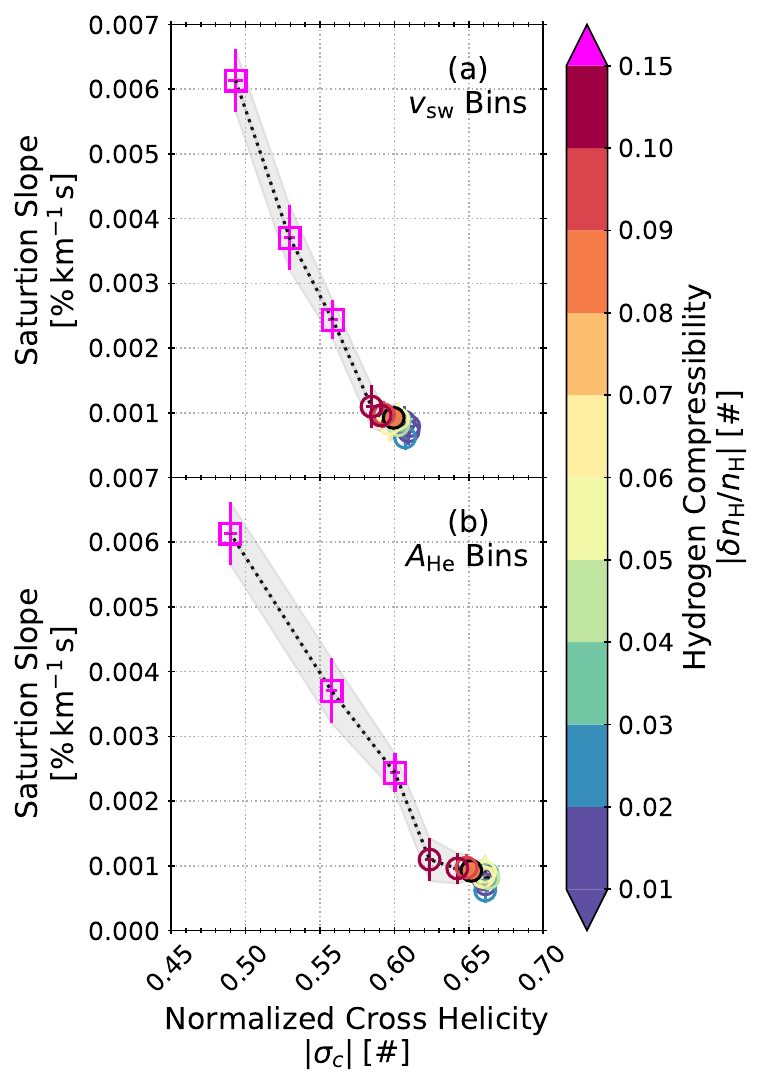}\\
\caption{\label{fig:sat-slopes:mapped}
The saturation slopes from \cref{fig:sat-slopes} as a function of \xhel.
In panel (a), \dnn\ in \cref{fig:sat-slopes} is mapped to \xhel\ using the trends in \cref{fig:lines:vsw-xhel-dnn} by calculating the \xhel\ corresponding to \vs\ in each \dnn\ quantile.
We then plot the saturations slope in that \dnn\ quantile as a function of this \xhel.
In panel (b), we use the same method and determining the \xhel\ corresponding to \As\ in each \dnn\ quantile from \cref{fig:lines:ahe-xhel-dnn}.
Because markers corresponding to \dnn_\Hy_[0.085][\leq] overlap, markers are unfilled.
The exception are markers corresponding to the \dnn_\Hy_[0.085] level, which are filled.
}
\IfBooleanTF{#1}{\end{figure*}}{\end{figure}}
}

\NewDocumentCommand{\plotSaturationSlopeOne}{s}{
\IfBooleanTF{#1}{\begin{figure*}}{\begin{figure}}
\includegraphics[width=\linewidth]{sat-slope-Paper1}\\
\caption{\label{fig:sat-slopes:one}
The slopes of \func{\ahe}{\vsw} for speeds \vsw*[\vs][>] in each \xhel\ quantile as derived in \citetOne.
}
\IfBooleanTF{#1}{\end{figure*}}{\end{figure}}
}

\NewDocumentCommand{\plotSaturationCompressibilityMappedOld}{s}{
\IfBooleanTF{#1}{\begin{figure*}}{\begin{figure}}
\includegraphics[width=\linewidth]{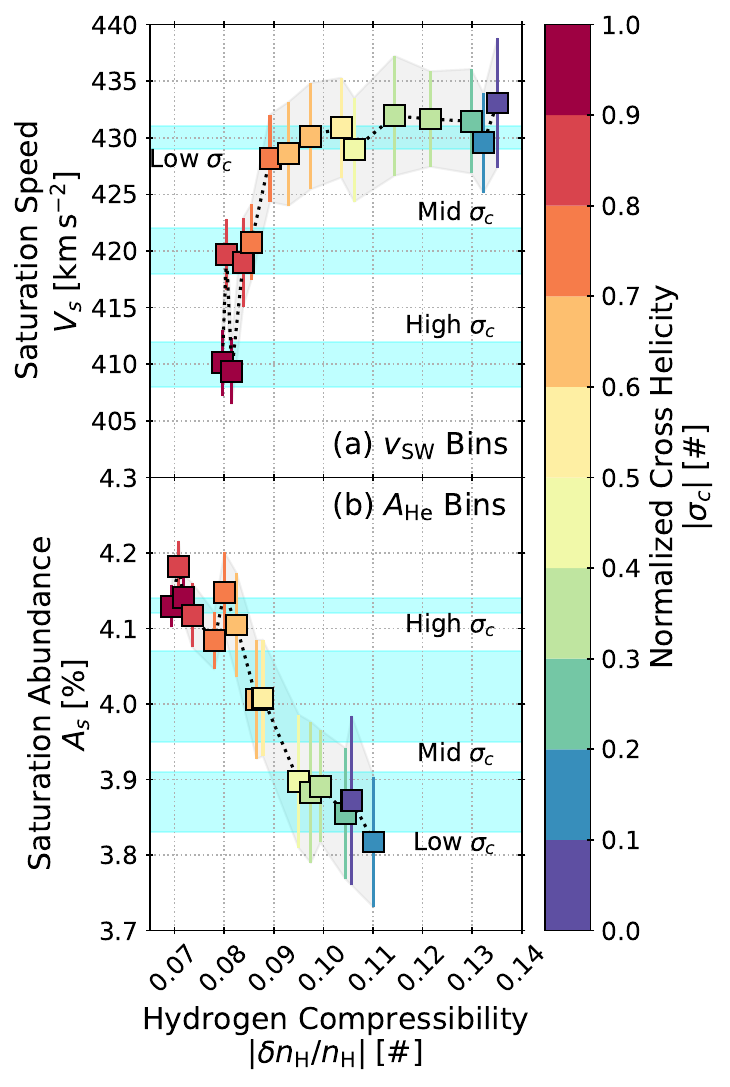}\\
\caption{\label{fig:sat-dn:old}
The saturation \textbf{(a)} speed \vs\ and \textbf{(b)} abundance \As\ as a function of compressibility \dnn_\Hy_ for each \xhel\ quantile, which is given by the color bar.
Points are connected by a dotted line to aid the eye.
Uncertainties are plotted as error bars with the same color as the markers and a semi-transparent gray envelope traces out the envelope surrounding the error bars.
This figure is similar to \cref{fig:sat-dn:mapped} in that it maps \xhel\ for each saturation point derived in \citetalias{\BibOne} to \dnn\ in its \xhel\ quantile.
In effect, it is the analagous mapping in \cref{fig:sat-dn:mapped} for \citetalias[Figure 6]{\BibOne}
}
\IfBooleanTF{#1}{\end{figure*}}{\end{figure}}
}

\NewDocumentCommand{\plotSaturationCompressibilityMapped}{s}{
\IfBooleanTF{#1}{\begin{figure*}}{\begin{figure}}
\includegraphics[width=\linewidth]{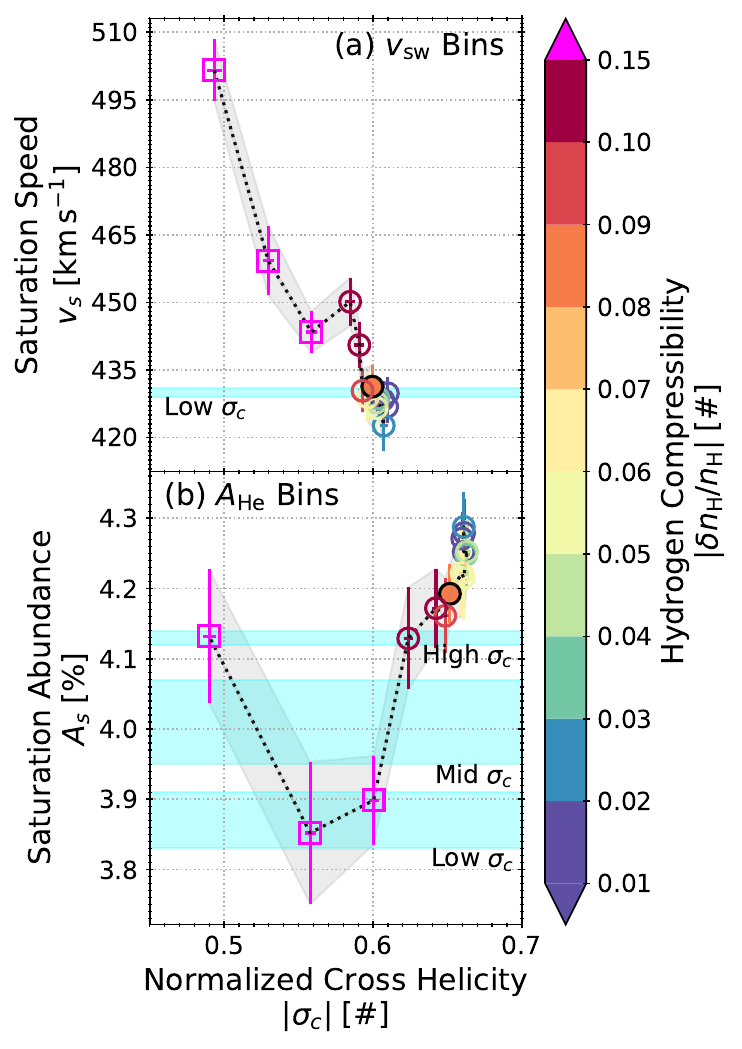}\\
\caption{\label{fig:sat-dn:mapped}
The saturation point \satpoint\ in \cref{fig:sat-dn} plotted as a function of \xhel.
Points are connected by a dotted line to aid the eye.
Uncertainties are plotted as error bars with the same color as the markers and a semi-transparent gray envelope traces out the envelope surrounding the error bars.
Each \dnn\ quantiles is mapped to \xhel\ using 
\begin{inparaenum}[(a)]
\item \cref{fig:lines:vsw-xhel-dnn} and 
\item \cref{fig:lines:ahe-xhel-dnn}.
\end{inparaenum}
Markers excluding the quantile \dnn_\Hy_[0.085] are unfilled.
The \dnn_\Hy_[0.085] quantile marker is filled and has a black edge.
As in \cref{fig:sat-dn}, the blue horizontal regions correspond to average value of (a) \vs\ and (b) \ahe\ for key ranges of \xhel\ derived in \citetalias{\BibOne}.
}
\IfBooleanTF{#1}{\end{figure*}}{\end{figure}}
}

\NewDocumentCommand{\plotSaturationCompressibility}{s}{
\IfBooleanTF{#1}{\begin{figure*}}{\begin{figure}}
\includegraphics[width=\linewidth]{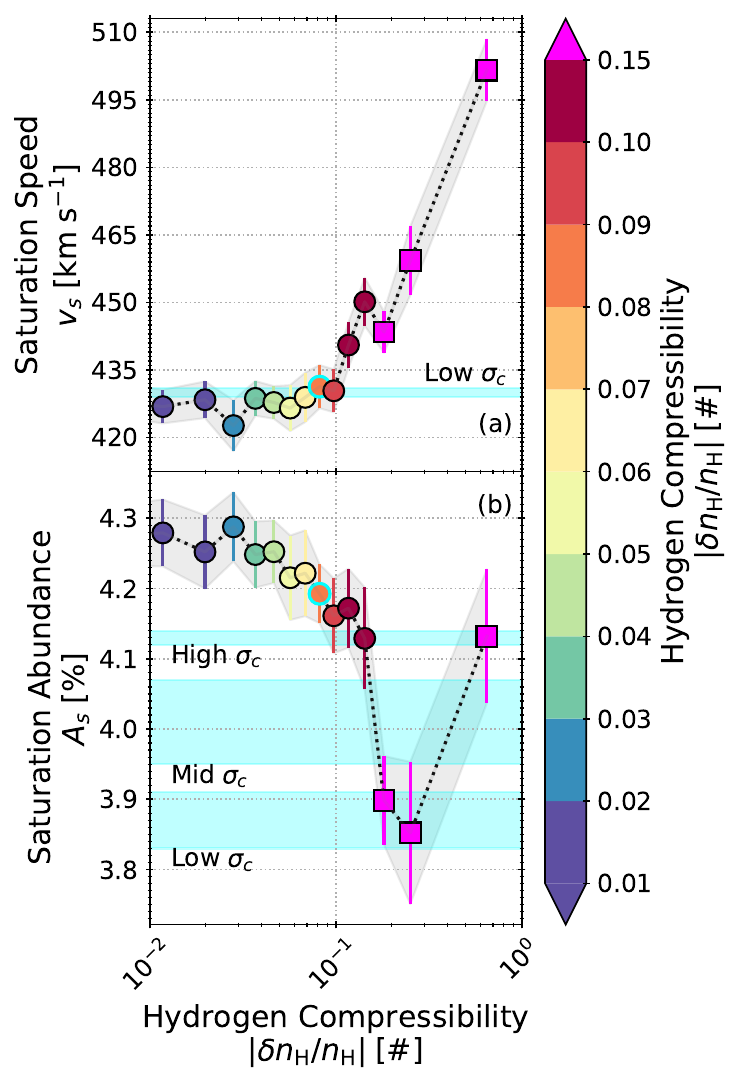}\\
\caption{\label{fig:sat-dn}
The saturation point \satpoint\ as a function of \dnn.
Points are connected by a dotted line to aid the eye.
Uncertainties are plotted as error bars with the same color as the markers and a semi-transparent gray envelope traces out the envelope surrounding the error bars.
Markers are colored by \dnn\ to facilitate comparison with other figures.
As above, quantiles with \dnn_\Hy_[0.15][>] are pink and square.
The edge of the marker corresponding to the quantile for which \dnn_\Hy_[0.085] is blue.
The blue horizontal regions correspond to average value of (a) \vs\ and (b) \ahe\ for key ranges of \xhel\ derived in \citetalias{\BibOne}.
}
\IfBooleanTF{#1}{\end{figure*}}{\end{figure}}
}

\NewDocumentCommand{\plotCompareThresholds}{s}{
\IfBooleanTF{#1}{\begin{figure*}}{\begin{figure}}
\includegraphics[width=\linewidth]{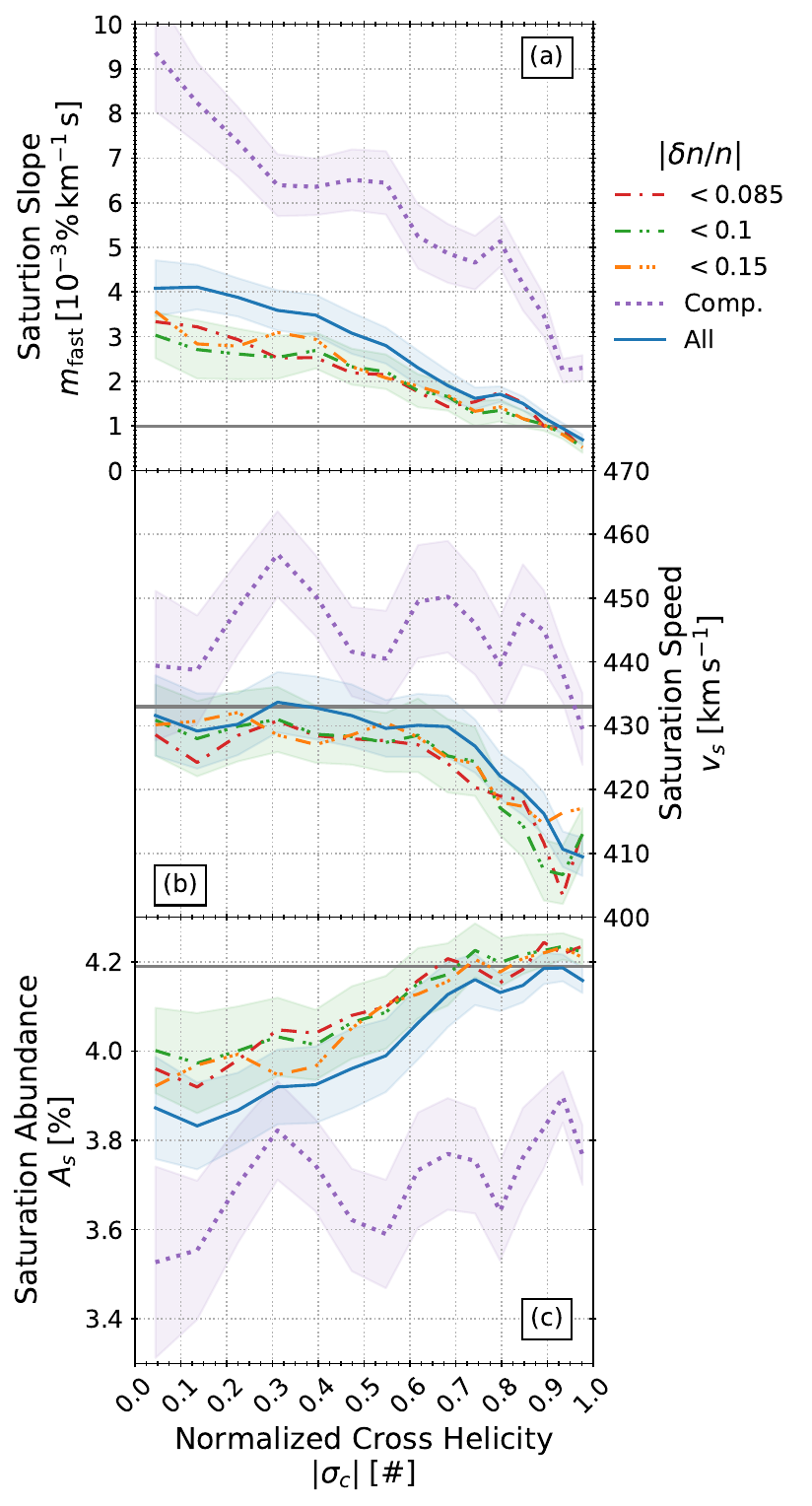}
\caption{\label{fig:compare-thresholds}
Saturation fit parameters for different maximum \dnn_\Hy_, given by the legend, as a function of \xhel\ along with the data, irrespective of \dnn, from \citetOne.
The panels are saturation \textbf{(a)} slope (\mfast),
\textbf{(b)} speed (\vs), and 
\textbf{(c)} abundance (\As).
Gray lines are the fit parameters across all observations in \cref{fig:vsw-ahe}.
The \emph{Comp.} subset is compressible solar wind with \dnn_\Hy_[0.15][>].
}
\IfBooleanTF{#1}{\end{figure*}}{\end{figure}}
}  

\NewDocumentCommand{\eqAhe}{s}{
\begin{equation}\label{eq:ahe}
\ahe = 100 \times \frac{\n[\He]}{\n[\Hy]}\IfBooleanF{#1}{.}
\end{equation}
}

\NewDocumentCommand{\eqXhel}{s}{
\begin{equation}\label{eq:xhel}
\xhel* = \frac{e^+ - e^-}{e^+ + e^-}\IfBooleanF{#1}{.}
\end{equation}
}

\NewDocumentCommand{\eqdnn}{s}{
\begin{equation}\label{eq:dnn}
\abs{\frac{\delta \n[\Hy]}{\n[\Hy]}} = \abs{\frac{\n[\Hy] - \langle \n[\Hy] \rangle}{\n[\Hy]}}
\end{equation}
}

\newcommand{\twolinefcn}{
\begin{equation} \label{eq:two-line}
A(v) = \mathrm{min}\left[m_1(v - v_1), m_2 (v - v_2)\right]
\end{equation}
}

\newcommand{\eqVsat}{
\begin{equation}\label{eq:vs}
v_s = \frac{m_1 v_1 - m_2 v_2}{m_1 - m_2}.
\end{equation}
}

\newcommand{\TblDnn}{
\begin{table*}
\centering
\begin{tabular}{ccccccc}
\hline\hline
&     Saturation Speed    &     Saturation Abundance &    Saturation Slope &         Vanishing Speed\\
&     \vs    &     \As &    \mfast &         \vv \\
\dnn &        $\left[\kms\right]$ &            $\left[\%\right]$ &     $\left[\pten{-3} \, \% \mathrm{km^{-1} \, s}\right]$ &         $\left[\kms\right]$ \\
\hline
0.003 &  $430 \pm 4$ &  $4.27 \pm 0.04$ &   $0.8 \pm 0.2$ &   $281 \pm 8$ \\
0.012 &  $427 \pm 4$ &  $4.28 \pm 0.05$ &   $0.7 \pm 0.2$ &   $292 \pm 4$ \\
0.020 &  $428 \pm 4$ &  $4.25 \pm 0.05$ &   $0.9 \pm 0.2$ &   $291 \pm 4$ \\
0.028 &  $423 \pm 6$ &  $4.29 \pm 0.05$ &   $0.6 \pm 0.2$ &   $303 \pm 9$ \\
0.037 &  $429 \pm 4$ &  $4.25 \pm 0.05$ &   $0.9 \pm 0.2$ &   $290 \pm 4$ \\
0.046 &  $428 \pm 4$ &  $4.25 \pm 0.04$ &   $0.8 \pm 0.2$ &   $296 \pm 4$ \\
0.057 &  $427 \pm 5$ &  $4.22 \pm 0.06$ &   $0.9 \pm 0.3$ &  $303 \pm 10$ \\
0.068 &  $429 \pm 5$ &  $4.22 \pm 0.06$ &   $0.8 \pm 0.3$ &  $304 \pm 10$ \\
0.082 &  $431 \pm 5$ &  $4.19 \pm 0.04$ &   $0.9 \pm 0.2$ &   $302 \pm 8$ \\
0.098 &  $430 \pm 5$ &  $4.16 \pm 0.05$ &   $1.0 \pm 0.2$ &   $311 \pm 6$ \\
0.117 &  $441 \pm 5$ &  $4.17 \pm 0.06$ &   $1.0 \pm 0.2$ &   $304 \pm 8$ \\
0.143 &  $450 \pm 5$ &  $4.13 \pm 0.07$ &   $1.1 \pm 0.3$ &   $298 \pm 5$ \\
0.182 &  $443 \pm 5$ &   $3.9 \pm 0.06$ &   $2.4 \pm 0.3$ &   $308 \pm 5$ \\
0.252 &  $459 \pm 8$ &   $3.85 \pm 0.1$ &   $3.7 \pm 0.5$ &  $308 \pm 12$ \\
0.649 &  $502 \pm 7$ &   $4.13 \pm 0.1$ &   $6.1 \pm 0.5$ &   $296 \pm 5$ \\
\hline\hline
0.085 &  $428 \pm 1 $ &  $4.25 \pm 0.02$ &  $0.81 \pm 0.07$ &  $293 \pm 2 $ \\
0.1 &  $428 \pm 1 $ &  $4.24 \pm 0.02$ &  $0.82 \pm 0.06$ &  $295 \pm 2 $ \\
0.15 &  $430 \pm 1 $ &  $4.23 \pm 0.01$ &  $0.84 \pm 0.06$ &  $296 \pm 2 $ \\
Comp. &  $458 \pm 3 $ &   $4.00 \pm 0.04$ &    $2.7 \pm 0.2$ &  $301 \pm 3 $ \\
\hline\hline
\end{tabular}
\caption{\label{tbl:dnn}
Saturation points and slopes in \cref{fig:sat-dn,fig:sat-slopes}.
We also include the speed \vv\ at which the slow wind portion of the fit (\vsw*[\vs][<]) intersects the x-axis to completeness.
From left to right, first four columns are 
\begin{inparaenum}[]
\item \vs\ (\cref{fig:sat-dn} (a)), 
\item \As\ (\cref{fig:sat-dn} (b)),
\item saturation slope (\cref{fig:sat-slopes}), and
\item \vv\ in \dnn\ quantiles.
\end{inparaenum}
The bottom four rows correspond to weighted averages taken across the incompressible subsets \dnn_\Hy_[0.085][<], 0.1, and 0.15 along with the compressible subset \dnn_\Hy_[0.15][>] under \emph{Comp}.
}
\end{table*}
}

\newcommand{\TblDnnMapping}{
\begin{table}
\centering
\begin{tabular}{ccc}
\hline\hline
&       \func{\xhel}{\As,\dnn} & \func{\xhel}{\vs,\dnn} \\
\dnn & $\left[\#\right]$ & $\left[\#\right]$ \\
\hline
0.003 &  $0.660 \pm 0.003$ &       $0.610 \pm 0.003$ \\
0.012 &   $0.661 \pm 0.003$ &       $0.610 \pm 0.003$ \\
0.020 &  $0.661 \pm 0.003$ &      $0.607 \pm 0.003$ \\
0.028 &    $0.661 \pm 0.003$ &      $0.607 \pm 0.003$ \\
0.037 &     $0.663 \pm 0.003$ &      $0.604 \pm 0.003$ \\
0.046 &     $0.663 \pm 0.003$ &      $0.603 \pm 0.003$ \\
0.057 &     $0.661 \pm 0.003$ &      $0.603 \pm 0.003$ \\
0.068 &    $0.660 \pm 0.003$ &      $0.597 \pm 0.003$ \\
0.082 &    $0.652 \pm 0.003$ &        $0.600 \pm 0.003$ \\
0.098 &   $0.649 \pm 0.003$ &      $0.593 \pm 0.003$ \\
0.117 &    $0.642 \pm 0.003$ &      $0.591 \pm 0.003$ \\
0.143 &    $0.624 \pm 0.003$ &      $0.585 \pm 0.004$ \\
0.182 &     $0.600 \pm 0.003$ &      $0.559 \pm 0.003$ \\
0.252 &     $0.558 \pm 0.004$ &       $0.530 \pm 0.004$ \\
0.649 &    $0.490 \pm 0.004$ &      $0.493 \pm 0.005$ \\
\hline\hline
\end{tabular}
\caption{\label{tbl:dnn:mapping}
The mappings from \dnn\ to \xhel.
The first column is \xhel\ corresponding to \vs\ in a given \dnn\ quantile (Figures \ref{fig:lines:vsw-xhel-dnn} and \ref{fig:sat-dn:mapped} (a)).
The second columns is \xhel\ corresponding to \As\ in a given \dnn\ quantile (Figures \ref{fig:lines:ahe-xhel-dnn} and \ref{fig:sat-dn:mapped} (b)).
}
\end{table}
}

\NewDocumentCommand{\TblXhel}{s}{
\IfBooleanTF{#1}{\begin{table*}}{\begin{table}}
\centering
\begin{tabular}{c cc}
\hline\hline
{} & $\func{\dnn}{\As,\xhel}$ & $\func{\dnn}{\vsw,\xhel}$ \\
\xhel\ & $\left[\#\right]$ & $\left[\#\right]$ \\
\hline
0.98 &        $0.069 \pm 0.001$ &          $0.08 \pm 0.001$ \\
0.89 &        $0.071 \pm 0.001$ &          $0.08 \pm 0.001$ \\
0.93 &        $0.072 \pm 0.001$ &         $0.082 \pm 0.001$ \\
0.85 &        $0.074 \pm 0.001$ &         $0.084 \pm 0.001$ \\
0.80 &        $0.078 \pm 0.001$ &         $0.085 \pm 0.001$ \\
0.74 &         $0.08 \pm 0.001$ &         $0.089 \pm 0.001$ \\
0.68 &        $0.082 \pm 0.001$ &         $0.093 \pm 0.001$ \\
0.62 &        $0.087 \pm 0.002$ &         $0.097 \pm 0.001$ \\
0.55 &        $0.088 \pm 0.001$ &         $0.104 \pm 0.002$ \\
0.47 &        $0.095 \pm 0.002$ &         $0.106 \pm 0.002$ \\
0.39 &        $0.097 \pm 0.002$ &         $0.114 \pm 0.002$ \\
0.31 &        $0.099 \pm 0.002$ &         $0.122 \pm 0.002$ \\
0.22 &        $0.104 \pm 0.002$ &          $0.13 \pm 0.003$ \\
0.05 &        $0.106 \pm 0.002$ &         $0.135 \pm 0.003$ \\
0.14 &         $0.11 \pm 0.003$ &         $0.132 \pm 0.004$ \\
\hline
\end{tabular}\caption{\label{tbl:xhel}
Quantiles of \xhel\ derived in \citetOne\ mapped to \dnn\ using \cref{fig:lines:vsw-dnn-xhel,fig:lines:ahe-dnn-xhel}.
\citetalias[Table 1]{\BibOne} gives \vs, \As, \mfast, and \vv\ derived in \xhel\ quantiles.
}
\IfBooleanTF{#1}{\end{table*}}{\end{table}}
} 

\newcommand{\SpeedTable}{
\begin{table}
\centering
\begin{tabular}{cc}
\hline
\hline
Speed & Typical Value $\left[\mathrm{km \, s^{-1}}\right]$\\
\hline
$v_K$ & $557$ to $700$ \\
$v_\mathrm{fast}$ & $622 \pm 59$ \\
$v_\mathrm{IP} $  & $434$ to $554$\\
$v_i$ & $484 \pm 34$ \\
\vdnp & $435$ to $508$ \\
\vdn & $417$ to $436$ \\
\vsigma & $407$ to $439$ \\
$v_n$ & $409 \pm 15$ \\
$v_\mathrm{slow}$ & $355 \pm 44$ \\
$v_\mathrm{heavy}$ & $316$ to $340$\\
\hline
\hline
\end{tabular}
\caption{\label{tbl:speeds}
Key speeds highlighted in \cref{fig:vsw-hist}.
These are the speeds of 
the slow (\vslow) and fast (\vfast) wind peaks during solar minima, 
the intersection of the Gaussians fit to these peaks (\vi), 
the saturation speed (\vsigma) and 
the range of speeds corresponding to the peak of \n[\He] as a function of \vsw\ both derived in \citetOne, 
the range of incompressible (\vdn) and compressible (\vdnp) saturation speeds derived in this work, 
and the range of speeds \vK\ predicted near-Earth from the radial gradient of the solar wind's kinetic energy flux \citep{Wind:SWE:Wk} observed with Parker Solar Probe near-Sun observations \citep{Liu2021c} along with the range of speeds predict for solar wind from continuously open source regions if there was no Alfvén wave pressure accelerating the solar wind in transit \citep{Rivera2025}.
}
\end{table}
}

\NewDocumentCommand{\DescRow}{m m}{\multirow{#1}{35ex}{#2}}

\RenewDocumentCommand{\SpeedTable}{}{
\begin{table}
\centering
\caption{\label{tbl:speeds}
Key speeds highlighted in \cref{fig:vsw-hist}.
}
\vspace{-3ex}
\begin{tabular}{c c c}
\hline
\hline
Speed & Typical Value  & Description \\ 
 & $\left[\mathrm{km \, s^{-1}}\right]$ &  \\ 
\hline
$v_K$ & $557$ to $700$ & \DescRow{5}{The range of speeds predicted near-Earth from the radial gradient of the solar wind's kinetic energy flux\footnote{\citet{Wind:SWE:Wk}} observed with Parker Solar Probe near-Sun observations\footnote{\citet{Liu2021c}}.}  \\ 
& & \\
& & \\
& & \\
& & \\
$v_\mathrm{fast}$ & $622 \pm 59$ & \DescRow{2}{The fast wind peaks during solar minima\footnote{\citetOne}.} \\ 
& & \\
$v_\mathrm{IP} $ & $434$ to $554$ & \DescRow{8}{The range of speeds predicted for fast wind observed at 1 AU if Alfvén wave forcing had been absent during solar wind propagation, i.e. "wave-poor" solar wind from continuously open source regions in which the energy in the Alfvénic fluctuations is less than typical fast wind.\footnote{\citetOne}.} \\ 
& & \\
& & \\
& & \\
& & \\
& & \\
& & \\
& & \\
& & \\
\vdnp & $435$ to $508$ & \DescRow{2}{The range of compressible saturation speeds.} \\ 
& & \\
\vdn & $417$ to $436$ &\DescRow{2}{The range of incompressible saturation speeds.} \\ 
& & \\
\vsigma & $407$ to $439$ & \DescRow{2}{The range of saturation speeds across 15 quantiles in \xhel\footnote{\citetOne}.} \\ 
& & \\
$v_n$ & $409 \pm 15$ & \DescRow{3}{The range of speeds corresponding to the peak of \n[\He] as a function of \vsw\footnote{\citetOne}.} \\ 
& & \\
& & \\
$v_\mathrm{slow}$ & $355 \pm 44$ & \DescRow{2}{The slow wind peak during solar minima\footnote{\citetOne}.} \\ 
& & \\
$v_\mathrm{heavy}$ & $316$ to $340$ & \DescRow{2}{The range of heavy element abundance saturation speeds\footnote{\citet{ACE:SWICS:FStransition}}.} \\ 
& & \\
\hline
\hline
\end{tabular}
\vspace{-2.5ex}
\end{table}
} 

\newcommand{\TblByThresholds}{
\begin{table*}
\centering
\begin{tabular}{ccccccccc}
\hline\hline
{} & \multicolumn{4}{c}{Sat Slope} & \multicolumn{4}{c}{$v_0$} \\
{} & \multicolumn{4}{c}{$\left[\% \, \mathrm{km^{-1} \, s}\right]$} & \multicolumn{4}{c}{$\left[\kms\right]$} \\
\hline
{\xhel} &                                      0.085 &              0.1 &             0.15 &             None &               0.085 &             0.1 &            0.15 &           None \\
\hline
0.05 &                            $ 3.4 \pm 0.5 $ &  $ 3.4 \pm 0.7 $ &  $ 3.6 \pm 0.6 $ &  $ 3.9 \pm 0.7 $ &       $ 301 \pm 4 $ &   $ 301 \pm 4 $ &   $ 301 \pm 4 $ &  $ 304 \pm 4 $ \\
0.14 &                            $ 3.5 \pm 0.5 $ &  $ 2.9 \pm 0.6 $ &  $ 2.8 \pm 0.6 $ &  $ 4.2 \pm 0.4 $ &       $ 302 \pm 4 $ &   $ 302 \pm 4 $ &   $ 302 \pm 4 $ &  $ 305 \pm 3 $ \\
0.22 &                            $ 2.8 \pm 0.6 $ &  $ 2.7 \pm 0.6 $ &  $ 2.9 \pm 0.6 $ &  $ 3.8 \pm 0.5 $ &       $ 298 \pm 5 $ &   $ 298 \pm 4 $ &   $ 300 \pm 4 $ &  $ 303 \pm 3 $ \\
0.31 &                            $ 2.8 \pm 0.4 $ &  $ 2.7 \pm 0.4 $ &  $ 3.1 \pm 0.4 $ &  $ 3.7 \pm 0.4 $ &       $ 291 \pm 5 $ &   $ 293 \pm 4 $ &   $ 296 \pm 4 $ &  $ 302 \pm 4 $ \\
0.39 &                            $ 2.2 \pm 0.4 $ &  $ 2.4 \pm 0.4 $ &  $ 3.0 \pm 0.5 $ &  $ 3.5 \pm 0.5 $ &       $ 302 \pm 7 $ &   $ 297 \pm 4 $ &   $ 300 \pm 7 $ &  $ 298 \pm 4 $ \\
0.47 &                            $ 2.5 \pm 0.4 $ &  $ 2.7 \pm 0.3 $ &  $ 2.2 \pm 0.5 $ &  $ 3.3 \pm 0.4 $ &       $ 293 \pm 5 $ &   $ 296 \pm 4 $ &   $ 299 \pm 5 $ &  $ 303 \pm 4 $ \\
0.55 &                            $ 2.0 \pm 0.4 $ &  $ 1.8 \pm 0.4 $ &  $ 1.9 \pm 0.4 $ &  $ 2.6 \pm 0.4 $ &      $ 298 \pm 11 $ &   $ 296 \pm 5 $ &   $ 298 \pm 4 $ &  $ 302 \pm 4 $ \\
0.62 &                            $ 2.1 \pm 0.4 $ &  $ 2.3 \pm 0.4 $ &  $ 2.3 \pm 0.4 $ &  $ 2.6 \pm 0.4 $ &       $ 293 \pm 5 $ &   $ 293 \pm 5 $ &   $ 295 \pm 6 $ &  $ 297 \pm 5 $ \\
0.68 &                            $ 1.4 \pm 0.3 $ &  $ 1.3 \pm 0.3 $ &  $ 1.3 \pm 0.3 $ &  $ 1.9 \pm 0.3 $ &      $ 295 \pm 13 $ &  $ 285 \pm 13 $ &   $ 299 \pm 5 $ &  $ 302 \pm 4 $ \\
0.74 &                            $ 1.3 \pm 0.3 $ &  $ 1.5 \pm 0.3 $ &  $ 1.7 \pm 0.2 $ &  $ 1.7 \pm 0.2 $ &       $ 294 \pm 5 $ &   $ 295 \pm 4 $ &   $ 297 \pm 4 $ &  $ 299 \pm 4 $ \\
0.80 &                            $ 2.3 \pm 0.5 $ &  $ 1.3 \pm 0.2 $ &  $ 1.3 \pm 0.2 $ &  $ 1.8 \pm 0.2 $ &      $ 279 \pm 23 $ &  $ 280 \pm 13 $ &  $ 277 \pm 11 $ &  $ 302 \pm 3 $ \\
0.85 &                            $ 1.2 \pm 0.2 $ &  $ 1.2 \pm 0.2 $ &  $ 1.3 \pm 0.2 $ &  $ 1.6 \pm 0.2 $ &      $ 266 \pm 14 $ &   $ 296 \pm 4 $ &   $ 298 \pm 4 $ &  $ 297 \pm 4 $ \\
0.89 &                            $ 1.0 \pm 0.1 $ &  $ 0.9 \pm 0.2 $ &  $ 0.9 \pm 0.1 $ &  $ 1.1 \pm 0.1 $ &      $ 279 \pm 12 $ &  $ 282 \pm 16 $ &  $ 268 \pm 14 $ &  $ 287 \pm 4 $ \\
0.93 &                            $ 1.0 \pm 0.1 $ &  $ 0.9 \pm 0.1 $ &  $ 0.9 \pm 0.1 $ &  $ 1.1 \pm 0.1 $ &      $ 291 \pm 11 $ &  $ 293 \pm 10 $ &  $ 275 \pm 15 $ &  $ 296 \pm 3 $ \\
0.98 &                            $ 0.5 \pm 0.1 $ &  $ 0.5 \pm 0.1 $ &  $ 0.6 \pm 0.1 $ &  $ 0.8 \pm 0.1 $ &      $ 253 \pm 16 $ &  $ 253 \pm 16 $ &  $ 260 \pm 14 $ &  $ 287 \pm 4 $ \\
\hline\hline
{} & \multicolumn{4}{c}{$v_s$} & \multicolumn{4}{c}{$A_s$} \\
{} & \multicolumn{4}{c}{$\left[\kms\right]$} & \multicolumn{4}{c}{$\left[\%\right]$} \\
\hline
{\xhel} &               0.085 &            0.1 &           0.15 &           None &              0.085 &                0.1 &               0.15 &               None \\
\hline
0.05 &       $ 428 \pm 5 $ &  $ 429 \pm 6 $ &  $ 430 \pm 6 $ &  $ 433 \pm 6 $ &  $ 3.94 \pm 0.08 $ &   $ 3.95 \pm 0.1 $ &    $ 3.9 \pm 0.1 $ &  $ 3.87 \pm 0.11 $ \\
0.14 &       $ 423 \pm 5 $ &  $ 428 \pm 6 $ &  $ 431 \pm 6 $ &  $ 430 \pm 4 $ &  $ 3.88 \pm 0.09 $ &  $ 3.95 \pm 0.11 $ &  $ 3.97 \pm 0.11 $ &  $ 3.82 \pm 0.09 $ \\
0.22 &       $ 429 \pm 6 $ &  $ 430 \pm 6 $ &  $ 431 \pm 6 $ &  $ 431 \pm 5 $ &   $ 3.99 \pm 0.1 $ &   $ 3.98 \pm 0.1 $ &   $ 3.95 \pm 0.1 $ &  $ 3.85 \pm 0.09 $ \\
0.31 &       $ 432 \pm 5 $ &  $ 432 \pm 5 $ &  $ 431 \pm 5 $ &  $ 432 \pm 4 $ &  $ 4.03 \pm 0.08 $ &  $ 4.02 \pm 0.08 $ &  $ 3.97 \pm 0.08 $ &  $ 3.89 \pm 0.07 $ \\
0.39 &       $ 428 \pm 6 $ &  $ 429 \pm 5 $ &  $ 428 \pm 6 $ &  $ 432 \pm 5 $ &  $ 4.06 \pm 0.08 $ &  $ 4.03 \pm 0.08 $ &  $ 3.94 \pm 0.09 $ &  $ 3.88 \pm 0.09 $ \\
0.47 &       $ 427 \pm 5 $ &  $ 427 \pm 4 $ &  $ 431 \pm 5 $ &  $ 429 \pm 5 $ &  $ 4.02 \pm 0.09 $ &  $ 3.99 \pm 0.07 $ &  $ 4.08 \pm 0.11 $ &   $ 3.9 \pm 0.09 $ \\
0.55 &       $ 427 \pm 6 $ &  $ 430 \pm 4 $ &  $ 431 \pm 4 $ &  $ 431 \pm 4 $ &   $ 4.1 \pm 0.08 $ &  $ 4.14 \pm 0.08 $ &  $ 4.11 \pm 0.08 $ &  $ 4.01 \pm 0.08 $ \\
0.62 &       $ 425 \pm 5 $ &  $ 425 \pm 5 $ &  $ 427 \pm 6 $ &  $ 430 \pm 5 $ &  $ 4.09 \pm 0.08 $ &  $ 4.07 \pm 0.08 $ &  $ 4.05 \pm 0.09 $ &  $ 4.01 \pm 0.08 $ \\
0.68 &       $ 426 \pm 6 $ &  $ 429 \pm 7 $ &  $ 427 \pm 5 $ &  $ 429 \pm 5 $ &   $ 4.2 \pm 0.08 $ &  $ 4.21 \pm 0.07 $ &  $ 4.19 \pm 0.07 $ &   $ 4.1 \pm 0.07 $ \\
0.74 &       $ 420 \pm 5 $ &  $ 420 \pm 4 $ &  $ 421 \pm 4 $ &  $ 428 \pm 4 $ &  $ 4.21 \pm 0.07 $ &  $ 4.18 \pm 0.06 $ &  $ 4.14 \pm 0.06 $ &  $ 4.15 \pm 0.05 $ \\
0.80 &       $ 417 \pm 8 $ &  $ 422 \pm 5 $ &  $ 428 \pm 4 $ &  $ 421 \pm 3 $ &   $ 4.06 \pm 0.1 $ &  $ 4.19 \pm 0.05 $ &   $ 4.2 \pm 0.05 $ &  $ 4.08 \pm 0.04 $ \\
0.85 &       $ 421 \pm 5 $ &  $ 410 \pm 4 $ &  $ 411 \pm 4 $ &  $ 419 \pm 4 $ &  $ 4.21 \pm 0.04 $ &  $ 4.18 \pm 0.04 $ &  $ 4.16 \pm 0.05 $ &  $ 4.12 \pm 0.04 $ \\
0.89 &       $ 412 \pm 4 $ &  $ 413 \pm 6 $ &  $ 421 \pm 5 $ &  $ 420 \pm 3 $ &  $ 4.23 \pm 0.03 $ &  $ 4.23 \pm 0.04 $ &  $ 4.23 \pm 0.03 $ &  $ 4.18 \pm 0.03 $ \\
0.93 &       $ 403 \pm 4 $ &  $ 403 \pm 4 $ &  $ 412 \pm 6 $ &  $ 409 \pm 3 $ &   $ 4.2 \pm 0.03 $ &   $ 4.2 \pm 0.03 $ &  $ 4.19 \pm 0.03 $ &  $ 4.14 \pm 0.03 $ \\
0.98 &       $ 414 \pm 4 $ &  $ 414 \pm 4 $ &  $ 414 \pm 4 $ &  $ 410 \pm 3 $ &  $ 4.22 \pm 0.03 $ &  $ 4.21 \pm 0.03 $ &  $ 4.18 \pm 0.03 $ &  $ 4.13 \pm 0.03 $ \\
\hline\hline
\end{tabular}
\caption{\label{tbl:by-thresholds}
Saturation fit parameters for observations with \dnn_\Hy_[0.085][<], 0.1, 0.15
}
\end{table*}
}

\renewcommand{\TblByThresholds}{
\begin{table*}
\centering
\begin{tabular}{ccccccccccccc}
\hline\hline
{} & \multicolumn{4}{c}{Saturation Slope} & \multicolumn{4}{c}{Vanishing Speed} \\
{} & \multicolumn{4}{c}{\mfast} & \multicolumn{4}{c}{\vv} \\
{} & \multicolumn{4}{c}{$\left[\pten{-3} \% \, \mathrm{km^{-1} \, s}\right]$} & \multicolumn{4}{c}{$\left[\kms\right]$} \\
\hline
{\xhel} &                                      0.085 &              0.1 &             0.15 &            Comp. &               0.085 &             0.1 &            0.15 &           Comp. \\
\hline
0.05 &     $ 3.4 \pm 0.5 $ & $ 3.4 \pm 0.7 $ & $ 3.6 \pm 0.6 $ & $ 8.3 \pm 1.6 $ & $ 301 \pm 4 $ & $ 301 \pm 4 $ & $ 301 \pm 4 $ & $ 320 \pm 8 $ \\
0.14 &     $ 3.5 \pm 0.5 $ & $ 2.9 \pm 0.6 $ & $ 2.8 \pm 0.6 $ & $ 8.4 \pm 0.8 $ & $ 302 \pm 4 $ & $ 302 \pm 4 $ & $ 302 \pm 4 $ & $ 316 \pm 4 $ \\
0.22 &     $ 2.8 \pm 0.6 $ & $ 2.7 \pm 0.6 $ & $ 2.9 \pm 0.6 $ & $ 7.7 \pm 0.9 $ & $ 298 \pm 5 $ & $ 298 \pm 4 $ & $ 300 \pm 4 $ & $ 313 \pm 4 $ \\
0.31 &     $ 2.8 \pm 0.4 $ & $ 2.7 \pm 0.4 $ & $ 3.1 \pm 0.4 $ & $ 7.0 \pm 0.6 $ & $ 291 \pm 5 $ & $ 293 \pm 4 $ & $ 296 \pm 4 $ & $ 306 \pm 4 $ \\
0.39 &     $ 2.2 \pm 0.4 $ & $ 2.4 \pm 0.4 $ & $ 3.0 \pm 0.5 $ & $ 5.7 \pm 0.6 $ & $ 302 \pm 7 $ & $ 297 \pm 4 $ & $ 300 \pm 7 $ & $ 306 \pm 4 $ \\
0.47 &     $ 2.5 \pm 0.4 $ & $ 2.7 \pm 0.3 $ & $ 2.2 \pm 0.5 $ & $ 6.5 \pm 0.8 $ & $ 293 \pm 5 $ & $ 296 \pm 4 $ & $ 299 \pm 5 $ & $ 319 \pm 5 $ \\
0.55 &     $ 2.0 \pm 0.4 $ & $ 1.8 \pm 0.4 $ & $ 1.9 \pm 0.4 $ & $ 6.7 \pm 0.8 $ & $ 298 \pm 11 $ & $ 296 \pm 5 $ & $ 298 \pm 4 $ & $ 309 \pm 5 $ \\
0.62 &     $ 2.1 \pm 0.4 $ & $ 2.3 \pm 0.4 $ & $ 2.3 \pm 0.4 $ & $ 5.3 \pm 0.6 $ & $ 293 \pm 5 $ & $ 293 \pm 5 $ & $ 295 \pm 6 $ & $ 308 \pm 8 $ \\
0.68 &     $ 1.4 \pm 0.3 $ & $ 1.3 \pm 0.3 $ & $ 1.3 \pm 0.3 $ & $ 5.2 \pm 0.7 $ & $ 295 \pm 13 $ & $ 285 \pm 13 $ & $ 299 \pm 5 $ & $ 304 \pm 10 $ \\
0.74 &     $ 1.3 \pm 0.3 $ & $ 1.5 \pm 0.3 $ & $ 1.7 \pm 0.2 $ & $ 5.4 \pm 0.5 $ & $ 294 \pm 5 $ & $ 295 \pm 4 $ & $ 297 \pm 4 $ & $ 309 \pm 7 $ \\
0.8 &     $ 2.3 \pm 0.5 $ & $ 1.3 \pm 0.2 $ & $ 1.3 \pm 0.2 $ & $ 4.3 \pm 0.5 $ & $ 279 \pm 23 $ & $ 280 \pm 13 $ & $ 277 \pm 11 $ & $ 308 \pm 10 $ \\
0.85 &     $ 1.2 \pm 0.2 $ & $ 1.2 \pm 0.2 $ & $ 1.3 \pm 0.2 $ & $ 5.1 \pm 0.7 $ & $ 266 \pm 14 $ & $ 296 \pm 4 $ & $ 298 \pm 4 $ & $ 287 \pm 8 $ \\
0.89 &     $ 1.0 \pm 0.1 $ & $ 0.9 \pm 0.2 $ & $ 0.9 \pm 0.1 $ & $ 2.6 \pm 0.3 $ & $ 279 \pm 12 $ & $ 282 \pm 16 $ & $ 268 \pm 14 $ & $ 299 \pm 9 $ \\
0.93 &     $ 1.0 \pm 0.1 $ & $ 0.9 \pm 0.1 $ & $ 0.9 \pm 0.1 $ & $ 3.3 \pm 0.3 $ & $ 291 \pm 11 $ & $ 293 \pm 10 $ & $ 275 \pm 15 $ & $ 305 \pm 5 $ \\
0.98 &     $ 0.5 \pm 0.1 $ & $ 0.5 \pm 0.1 $ & $ 0.6 \pm 0.1 $ & $ 2.6 \pm 0.2 $ & $ 253 \pm 16 $ & $ 253 \pm 16 $ & $ 260 \pm 14 $ & $ 284 \pm 7 $ \\
\hline
Avg &     $1.2 \pm 0.1$ & $1.2 \pm 0.1$ & $1.1 \pm 0.1$ & $4.0 \pm 0.1$ & $420.0 \pm 2.0$ & $421.0 \pm 1.0$ & $424.0 \pm 1.0$ & $443.0 \pm 1.0$ \\
\hline\hline
{} & \multicolumn{4}{c}{Saturation Speed} & \multicolumn{4}{c}{Saturation Abundance} \\
{} & \multicolumn{4}{c}{\vs} & \multicolumn{4}{c}{\As} \\
{} & \multicolumn{4}{c}{$\left[\kms\right]$} & \multicolumn{4}{c}{$\left[\%\right]$} \\
\hline
{\xhel} &               0.085 &            0.1 &           0.15 &           Comp. &              0.085 &                0.1 &               0.15 &  Comp. \\
\hline
0.05 & $ 428 \pm 5 $ & $ 429 \pm 6 $ & $ 430 \pm 6 $ & $ 441 \pm 13 $ & $ 3.94 \pm 0.08 $ & $ 3.95 \pm 0.1 $ & $ 3.9 \pm 0.1 $ & $ 3.59 \pm 0.23 $ \\
0.14 & $ 423 \pm 5 $ & $ 428 \pm 6 $ & $ 431 \pm 6 $ & $ 436 \pm 7 $ & $ 3.88 \pm 0.09 $ & $ 3.95 \pm 0.11 $ & $ 3.97 \pm 0.11 $ & $ 3.5 \pm 0.14 $ \\
0.22 & $ 429 \pm 6 $ & $ 430 \pm 6 $ & $ 431 \pm 6 $ & $ 444 \pm 8 $ & $ 3.99 \pm 0.1 $ & $ 3.98 \pm 0.1 $ & $ 3.95 \pm 0.1 $ & $ 3.59 \pm 0.14 $ \\
0.31 & $ 432 \pm 5 $ & $ 432 \pm 5 $ & $ 431 \pm 5 $ & $ 456 \pm 6 $ & $ 4.03 \pm 0.08 $ & $ 4.02 \pm 0.08 $ & $ 3.97 \pm 0.08 $ & $ 3.74 \pm 0.1 $ \\
0.39 & $ 428 \pm 6 $ & $ 429 \pm 5 $ & $ 428 \pm 6 $ & $ 458 \pm 6 $ & $ 4.06 \pm 0.08 $ & $ 4.03 \pm 0.08 $ & $ 3.94 \pm 0.09 $ & $ 3.8 \pm 0.1 $ \\
0.47 & $ 427 \pm 5 $ & $ 427 \pm 4 $ & $ 431 \pm 5 $ & $ 439 \pm 7 $ & $ 4.02 \pm 0.09 $ & $ 3.99 \pm 0.07 $ & $ 4.08 \pm 0.11 $ & $ 3.56 \pm 0.13 $ \\
0.55 & $ 427 \pm 6 $ & $ 430 \pm 4 $ & $ 431 \pm 4 $ & $ 441 \pm 8 $ & $ 4.1 \pm 0.08 $ & $ 4.14 \pm 0.08 $ & $ 4.11 \pm 0.08 $ & $ 3.53 \pm 0.12 $ \\
0.62 & $ 425 \pm 5 $ & $ 425 \pm 5 $ & $ 427 \pm 6 $ & $ 450 \pm 8 $ & $ 4.09 \pm 0.08 $ & $ 4.07 \pm 0.08 $ & $ 4.05 \pm 0.09 $ & $ 3.68 \pm 0.11 $ \\
0.68 & $ 426 \pm 6 $ & $ 429 \pm 7 $ & $ 427 \pm 5 $ & $ 450 \pm 9 $ & $ 4.2 \pm 0.08 $ & $ 4.21 \pm 0.07 $ & $ 4.19 \pm 0.07 $ & $ 3.7 \pm 0.12 $ \\
0.74 & $ 420 \pm 5 $ & $ 420 \pm 4 $ & $ 421 \pm 4 $ & $ 444 \pm 7 $ & $ 4.21 \pm 0.07 $ & $ 4.18 \pm 0.06 $ & $ 4.14 \pm 0.06 $ & $ 3.62 \pm 0.1 $ \\
0.8 & $ 417 \pm 8 $ & $ 422 \pm 5 $ & $ 428 \pm 4 $ & $ 444 \pm 8 $ & $ 4.06 \pm 0.1 $ & $ 4.19 \pm 0.05 $ & $ 4.2 \pm 0.05 $ & $ 3.69 \pm 0.11 $ \\
0.85 & $ 421 \pm 5 $ & $ 410 \pm 4 $ & $ 411 \pm 4 $ & $ 439 \pm 9 $ & $ 4.21 \pm 0.04 $ & $ 4.18 \pm 0.04 $ & $ 4.16 \pm 0.05 $ & $ 3.55 \pm 0.13 $ \\
0.89 & $ 412 \pm 4 $ & $ 413 \pm 6 $ & $ 421 \pm 5 $ & $ 452 \pm 7 $ & $ 4.23 \pm 0.03 $ & $ 4.23 \pm 0.04 $ & $ 4.23 \pm 0.03 $ & $ 3.89 \pm 0.08 $ \\
0.93 & $ 403 \pm 4 $ & $ 403 \pm 4 $ & $ 412 \pm 6 $ & $ 428 \pm 4 $ & $ 4.2 \pm 0.03 $ & $ 4.2 \pm 0.03 $ & $ 4.19 \pm 0.03 $ & $ 3.66 \pm 0.06 $ \\
0.98 & $ 414 \pm 4 $ & $ 414 \pm 4 $ & $ 414 \pm 4 $ & $ 433 \pm 6 $ & $ 4.22 \pm 0.03 $ & $ 4.21 \pm 0.03 $ & $ 4.18 \pm 0.03 $ & $ 3.65 \pm 0.05 $ \\
\hline
Avg & $420.0 \pm 1.0$ & $421.0 \pm 1.0$ & $424.0 \pm 1.0$ & $443.0 \pm 2.0$ & $4.17 \pm 0.01$ & $4.17 \pm 0.01$ & $4.16 \pm 0.01$ & $3.68 \pm 0.03$ \\
\hline\hline
\end{tabular}
\caption{\label{tbl:by-thresholds}
Saturation fit parameters for incompressible observations with \dnn_\Hy_[0.085][<], 0.1, 0.15 along with compressible \dnn_\Hy_[0.15][>], under \emph{Comp.}\ columns.
The bottom row is the weighted average for each column.
}
\end{table*}
}

\newcommand{\TblTypicalAhe}{
\begin{table}
\centering
\begin{tabular}{cc}
\hline\hline
Type & Typical Abundance \\
\hline
Slow Wind\footnote{\citet{Aellig:Ahe,Wind:SWE:Ahe:phase,Wind:SWE:ahe:shutoff,Kasper:Ahe,ACE:SWICS:SSN,Song2022,Wind:SWE:ahe:xhel,Yogesh:Ahe}} & 1 - 5\% \\
Fast Wind\footnote{\citet{Wind:SWE:ahe:xhel,Wind:SWE:ahe:dnn,Yogesh:Ahe,Song2022}} & 5\%  \\
ICMEs\footnote{\citet{Starkey2024,Khokhlachev2022,Song2022}} & Highly variable, up to $>10\%$  \\
SIRs\footnote{\citet{Yogesh:ahe:SIR,Durovcova:SIR}} & 2-3\% (Slow Wind) to 6-8\% (Fast Wind) \\
\hline\hline
\end{tabular}
\caption{The typical helium abundance at L1 in different types of solar wind and transient events including ICMEs and stream interaction regions (SIRs).}
\label{tbl:ahe:typical}
\end{table}
}

\graphicspath{{./}{figures/}}
\DeclareGraphicsExtensions{.pdf,.png,.jpg,.eps,}

\received{\today}
\submitjournal{ApJ}

\usepackage{lipsum}
\NewDocumentCommand{\BlindText}{O{6}}{\todo{Remove blind text. This is here to help figures render nicely.}
\textcolor{white}{\lipsum*[1-#1]}}

\NewDocumentCommand{\gradAhe}{o}{\ensuremath{
\IfNoValueTF{#1}{\grad_\vsw_[\ahe]}{\grad_\vsw_[\ahe][#1]}}}

\newcommand{\BibOne}{Wind:SWE:ahe:xhel}
\defcitealias{\BibOne}{Paper 1}

\newcommand{\citetOne}{\citetalias{\BibOne}}
\newcommand{\citepOne}{\citepalias{\BibOne}}

\newcommand{\satpoint}{\ensuremath{\left(\vs,\As\right)}}

\begin{document}

\title{
On the Regulation of the Solar Wind Helium Abundance by the Hydrogen Compressibility
}

\shorttitle{Helium Abundance and Hydorgen Compressibility}
\shortauthors{Alterman and D'Amicis}

\newcommand{\Goddard}{
\affiliation{Heliophysics Science Division \\
NASA Goddard Space Flight Center \\
8800 Greenbelt, RD\\
Greenbelt, MD 20771, USA}
}

\newcommand{\IAPS}{
\affiliation{INAF - Institute for Space Astrophysics and Planetology\\
Via Fosso del Cavaliere, 100\\
00133 Rome, Italy}
}

\correspondingauthor{B.\ L.\ Alterman}
\email{b.l.alterman@nasa.gov}

\author[0000-0001-6673-3432]{B.\ L.\ Alterman}
\email{b.l.alterman@nasa.gov}
\Goddard

\author[0000-0003-2647-117X]{Raffaella D'Amicis}
\email{raffaella.damicis@inaf.it}
\IAPS

\begin{abstract}

Traditionally, fast solar wind is considered to originate in solar source regions that are continuously open to the heliosphere.
In contrast, slow solar wind is considered to originate in source regions that are only intermittently open to the heliosphere.
In fast wind, the gradient of the solar wind helium abundance (\ahe) with increasing solar wind speed (\vsw) is approximately 0 and \ahe\ is fixed at $\sim50\%$ of the photospheric value.
In slow wind, this gradient is large and \ahe\ is highly variable and \ahe\ doesn't exceed this $\sim50\%$ value.
Although the normalized cross helicity in fast wind is typically observed to approach 1, this is not universally true and \citet{\BibOne} show that \grad_\vsw_[\ahe] in fast wind unexpectedly increases with decreasing \xhel.
We show that these large \grad_\vsw_[\ahe] are due to the presence of compressive fluctuations in fast wind.
Accounting for the solar wind's compressibility (\dnn), there exists two subsets of enhanced \ahe\ in excess of typical fast wind values.
The subset corresponding to large solar wind compressibility is likely from neither continuously nor intermittently open sources.
The portion of the solar wind speed distribution over which these fluctuations are most significant corresponds to the range of Alfvén wave-poor solar wind from continuously open source regions, which is likely analogous to the Alfvénic slow wind.
Mapping the results of this work to \citet{\BibOne} and vice versa shows that, in any given \dnn\ quantile, \xhel[0.65][\lesssim], an upper bound on non-Alfvénic cross helicity.
Similarly, \dnn[0.15][\lesssim] in any given \xhel\ quantile, an upper bound on incompressible solar wind fluctuations.
We conclude that \dnn\ is essential for characterizing the solar wind helium abundance and possibly regulating it.

\end{abstract}

\keywords{Solar wind (1534), Fast solar wind (1872), Slow solar wind (1873), Abundance ratios (11), Chemical abundances (224), Alfven waves (23), Magnetohydrodynamics (1964)}

\section{Introduction \label{sec:intro}} 

Based on the bimodal distribution of speeds observed near Earth in the ecliptic plane during solar minima, the solar wind has be classified into fast and slow speeds.
Fast wind is typically associated with sources on the Sun that are continuously open to the heliosphere and have radial magnetic field lines.
Slow wind is typically tied to sources that are intermittently open to the heliosphere.
Coronal holes (CHs) are the prototypical example of the former \citep{Phillips1994,Geiss1995}.
The latter include solar sources like helmet streamers, pseudostreamers, and the boundaries between pseudostreamers and CHs \citep{Fisk1999,Subramanian2010,Antiochos2011,Crooker2012,Abbo2016,Antonucci2005}.
These intermittently open sources are also referred to as ``magnetically closed''.
The differences between these sources lead to distinct \emph{in situ} signatures.
For example, fast solar wind from magnetically open regions displays enhanced helium- and heavy ion-to-hydrogen abundances \citep{ACE:SWICS:FStransition,Zurbuchen2016,ACE:SWICS:AUX,Pilleri2015} and temperature ratios \citep{Kasper2008,Kasper2017,Tracy2016,Stakhiv2016}.
The speed at which helium and heavy ions stream along the magnetic field in fast wind from magnetically open sources is also faster than the protons and constrained to a fraction of the Alfvén speed that depends on the ion species \citep{Wind:SWE:bimax,Stakhiv2016,Alterman2018,Berger2011,Klein2021,Verniero2020,Verniero2022,Durovcova2019}.
Moreover, the fast wind carries a chemical composition and charge-state ratios that are different from those observed in the slow wind, which reflects the differences in the magnetically open and closed topology of their sources \citep{vonSteiger2000,Geiss1995,Geiss1995b,Zhao:InSituComposition:Sources,Zhao2022,Xu2014,Fu2017,Fu2015,Brooks2015}.
Of these properties, \citet{\BibOne} used the chemical composition and the Alfvénic content of solar wind fluctuations to statistically tie \emph{in situ} observations to their solar source regions.
We refer to \citet{\BibOne} as \citetOne.

\plotAheVsw
In \citetOne, the chemical composition is characterized by the helium abundance, given by \cref{eq:ahe}
\eqAhe*
where \n[\Hy] and \n[\He] are the ionized hydrogen and fully ionized helium number densities.
\cref{tbl:ahe:typical} gives typical values for \ahe\ in various types of solar wind and transient events.
\cref{fig:vsw-ahe} is a simplified version of Figure 3 in \citetOne, plotting \ahe\ as a function of \vsw.
Here, the columns are normalized to their maximum values so that slow wind, which is observed in the ecliptic plane more frequently than fast wind, does not obscure the overall trend.
Reducing the data in each column to a mean and standard deviation yields the solid blue line.
Because this figure reproduces a result in \citetOne\, the standard deviations, which fall at approximately the 0.6 color level, are omitted for visual clarity.
The authors then fit the trend of these central values in each column to produce the bi-linear fit in the dash-dotted line.
The bi-linear function is given by \cref{eq:two-line}
\twolinefcn
where $A(v)$ is the abundance as a function of speed.
The slopes and x-intercepts of the two lines are $m_i$ and $v_i$ for subscripts 1 and 2.
Nominally, we consider line 1 to be the steeper gradient (slow wind).
The speed at which the two lines intersect is 
\eqVsat
The corresponding abundance is \As.

\TblTypicalAhe
\citetOne\ identifies the point in the $\left(\vsw,\ahe\right)$-plane at which the gradient of \ahe\ as a function of \vsw\ decreases as $\satpoint = \left(\kms[433], 4.19\%\right)$.
\As[4.19 \pm 0.05] is $49 \pm 2\%$ of the photospheric \ahe\ \citep{Asplund2021}.
At speeds \vsw*[\vs][<], the gradient of \ahe\ as a function of \vsw\ (\grad_\vsw_[\ahe]) is linear and strongly increasing.
In contrast, \grad_\vsw_[\ahe] is close to zero at speeds \vsw*[\vs][>].
Because this gradient is approximately flat at speeds \vsw*[\vs][>], the authors refer to this as the ``saturation point’’.
This change in the gradient statistically indicates the transition from slow to fast solar wind.
We will refer to $\left(\vsw,\ahe\right) < \satpoint$ as ``below saturation'' and $\left(\vsw,\ahe\right) > \satpoint$ as ``above saturation''.

\citetOne\ quantifies the Alfvénic content of the fluctuations with the normalized cross helicity \citep{Tu1995,LR:turbulence,Woodham2018}
\eqXhel*
where the quantities on the right hand side $e^\pm = \frac{1}{2}\langle \left(z^\pm\right)^2\rangle$ are the energies associated with the Elsässer variables.
The Elsässer variables are given by the sum and difference of the velocity and magnetic field fluctuations in Alfvén units $\Bz^\pm = \Bv \pm \frac{1}{\sqrt{\mu_0\rho}}\Bb$ for plasma velocity fluctuation \Bv, magnetic field fluctuation \Bb, and solar wind mass density $\rho$ \citep{ElsasserVariables,Tu1989,Grappin1991}.
We calculate these fluctuations as $X - \langle X \rangle$ where $\langle X \rangle$ is a 1hr rolling mean, a typical Alfvénic timescale at \au[1] \citep{Tu1989,DAmicis2015,DAmicis2022}.
The normalized cross helicity approaches $\pm1$ as the dominant mode in a plasma observation is closer to a pure Alfvén mode.
Typically, \xhel*\ is a signed quantity where the sign indicates the direction (towards or away from the Sun) a given Alfvén wave propagates.
However, this does not account for local rotations in the plasma associated, e.g., with switchbacks \citep{McManus2020}.
As they are not concerned with the direction of the fluctuations, \citetOne\ takes the absolute value.

The solar wind can be described as originating in an isothermal corona, evolving polytropically through the inner heliosphere, and in the case of fast wind, it continues to accelerate in transit due to forcing by Alfvén wave pressure gradients \citep{Rivera2024}.
\citet{Wind:SWE:Wk} shows that, in the absence of Alfvén wave forcing, solar wind from continuously open source regions would only be accelerated to ``intermediate'' speeds for which the source region is ambiguous, not the fastest, non-transient speeds typical of fast wind.
This suggests that the energy in the Alfvénic fluctuations at these intermediate speeds is less than the energy in the fluctuations carried in faster speed wind.
We refer to this as ``wave-poor'' fast wind and it is likely synonymous with the Alfvénic slow wind \citep{Marsch1981,DAmicis2011a}.
This suggests the Alfvénic content of the solar wind is essential for understanding its evolution through interplanetary space.

\plotVswXhel
\cref{fig:vsw:xhel} is a contour plot of \func{\xhel}{\vsw} using Wind/FC observations from L1, recreating the relevant details from Figure 4 in \citetOne.
For the same reason as in \cref{fig:vsw-ahe}, the columns in the underlying histogram have been normalized to their maximum value.
These contours are smoothed with a $1\sigma$ Gaussian kernel for visual clarity.
This figure shows that $\xhel \rightarrow 1$ as \vsw\ increases, indicating that the dominant wave mode observed in the solar wind is more likely to be Alfvénic as \vsw\ increases.

\citetOne\ summarizes the observations that have characterized the observed \emph{in situ} relationship between \ahe\ and \vsw\ that relate these observations to source regions.
\citetOne\ also summarizes the \emph{in situ} observations that relate \xhel\ to source regions.
Relevant to this work, \ahe\ carries information about solar wind source regions based on processes that happen in the chromosphere and transition region \citep{Lie-Svendsen2001,Lie-Svendsen2002,Endeve2005,Lie-Svendsen2003,Hansteen1997}, below the solar wind’s sonic critical point (\rc).
In contrast, \xhel\ carries information about solar wind source regions above \rc\ and below or near the Alfvén radius (\rA), above which the solar wind magnetically disconnects from the Sun in the sense that modifications of the magnetic field driven by motion on the Sun's surface, solar rotation, etc. can no longer propagate along a given field line to reach the plasma and modify its state.
Combining the observations of \xhel, \ahe, and \vsw, \citetOne\ analyze how the gradient of \ahe\ as a function of \vsw\ and the saturation point change with increasing \xhel.

Repeating their analysis for 15 quantiles in \xhel,
\citetOne\ shows that the saturation speed (\vs) and abundance (\As) are divided into three intervals of \xhel: \textit{Low}, \textit{Mid}, and \textit{High}.
Across these intervals, \vs\ decreases and \As\ increases.
In other words, $\min\left(\vs\right)$ associated with the most Alfvénic wind (highest \xhel) and largest \As\ is less than $\max\left(\vs\right)$ in the least Alfvénic wind (lowest \xhel) with the smallest \As.

\citetOne\ also analyzes the gradients of \ahe\ with \vsw\ above and below saturation.
The authors show that \grad_\vsw_[\ahe] is independent of \xhel\ for \vsw*[\vs][<], suggesting the process responsible for the highly variable \ahe\ in solar wind from magnetically closed sources is independent of \xhel\ and therefore independent of Alfvénic processes near the source region.
Analogous analysis with ACE/SWICS \citep{ACE:SWICS:FStransition} reveals a similar trend in which the gradients of heavy ion abundances below their respective saturation speeds are also indistinguishable and the process responsible for their variability does not preferentially couple to any particular ion(s).
In contrast, \grad_\vsw_[\ahe] increases with decreasing \xhel\ for \vsw*[\vs][>], suggesting that the Alfvénic content of solar wind observed above saturation is inversely related to these gradients.
However, \citetOne\ leaves this difference in \grad_\vsw_[\ahe] above and below saturation unexplained.

In summary, \citetOne\ makes the following observations.
\begin{compactenum}
    \item \ahe\ is set below \rc.
    \item \ahe\ is larger in solar wind originating in open sources than in solar wind originating in intermittently open sources. 
    \item \vsw\ increases during solar wind propagation above \rA.
    \item The in-transit acceleration mechanisms depend on source region \citep{Rivera2024,Rivera2025}. 
    \item \vs\ and \As\ are anti-correlated across across \xhel.
    \item \grad_\vsw_[\ahe] is independent of \xhel\ at \vsw*[\vs][<] and increases with decreasing \xhel\ for \vsw*[\vs][>].
\end{compactenum}
Based on these observations, \citetOne\ infers that the minimum speed of solar wind from continuously open sources is less than the maximum intermittently open source regions.
Such an inference is consistent with the Alfvénic slow wind originating from the low speed extension of plasma that would otherwise be categorized as fast wind from coronal holes \citep{DAmicis2015,Wang1994a,Wang2019,Panasenco2019,Panasenco2020,Panasenco2013,Panasenco2019,Chitta2025}.

\plotCategorizationCartoon
This analysis leads \citetOne\ to propose that \func{\vsw}{\xhel,\ahe} suggests a categorization scheme for solar wind observations at \au[1].
\cref{fig:xhel-ahe-vsw:cartoon} is a slightly modified version of their Figure 11 that illustrates it.
There are two contours of \vsw\ plotted in the $\left(\vsw,\ahe\right)$-plane that separate the observations into four distinct regions.
There is a helium-poor region without a preferred \xhel\ across the bottom of the plane that \citetOne\ suggests is solar wind from intermittently open sources.
On the right side of the plane, there is an Alfvénic and helium rich region that \citetOne\ suggests is solar wind from continuously open source regions.
In the middle of the plane is a large region where solar wind observations from both sources mix across a narrow range of speeds.
The range of \vsw\ in this region of the plane is consistent with the observed range of coronal mass ejection (CME) speeds \citep{LR:CME:obs} and \citetOne\ hypothesizes that the the speed enhancement in the top left corner is due to the presence of transients \citep{Good2022,Yogesh:ICME,Song2022,Scolini2024a}.
A detailed analysis of this is the subject of on going work.
The authors then use this information to characterize significant processes in different regions of the distribution of solar wind speeds observed near Earth during solar minima.

In Alfvénic fluctuations, by definition, $\delta \! \abs{\BB}$ and $\delta n$ tend towards zero.
This is often referred to as ``weak compressibility’’ \citep{Bruno2001,Matteini2015}.
Intuitively, one expects that compressibility would decrease with \vsw\ given \xhel\ increases with it.
The solar wind's compressibility can be calculated from time series observations collected by a single spacecraft as
\eqdnn
where \n[\Hy] is the instantaneous hydrogen density and $\langle \n[\Hy] \rangle$ is a rolling mean.
The fluctuations $\delta \n[\Hy]$ are normalized to the observed \n[\Hy] to quantify the amplitude of these fluctuations with respect to the observed density.
We calculate $\langle \n[\Hy] \rangle$ on the same 1hr time scale used for calculating \xhel* and take the absolute value because we are concerned with the magnitude of the fluctuations.
Compressive fluctuations, which can be defined as \dnn[0.1][>] \citep{Tu1994,Cuesta2023}, are most common in non-Alfvénic solar wind.

\plotVswDnn
\cref{fig:vsw:dnn} is a contour plot of the PDF of \func{\dnn}{\vsw} using Wind/FC observations.
As in \cref{fig:vsw:xhel}, the frequency of observation in each column in the underlying 2D histograms has been normalized to its maximum value and contours are smoothed with a $1\sigma$ Gaussian kernel for visual clarity.
\cref{fig:vsw:xhel,fig:vsw:dnn} show that, although compressive fluctuations (\dnn[0.1][>]) are less common than non-compressive fluctuations (\dnn[0.1][<]), they are present across all \vsw\ and \xhel.
\cref{fig:vsw:xhel:dnn} is a contour plot of \func{\dnn}{\vsw,\xhel}.
We calculate the average \dnn\ as the logarithmic mean.
It has been smoothed with a $1\sigma$ Gaussian kernel for visual clarity.
The solid black line indicates \dnn[0.1].
The solid purple line indicates the 0.7 contour from \cref{fig:vsw:xhel}, indicating that \xhel\ tends towards 1 as \vsw\ increases above \vsw[433][>].
This figure shows that, at these speeds, even as \xhel[1][\rightarrow], compressive fluctuations occupy a significant fraction of plane and that compressive fluctuations become significant at high speeds with low cross helicity.
In other words, the intuitive anti-correlation between \dnn\ and \xhel\ is most prominent at high speeds and virtually absent at low speeds.
This work shows that the absence of the anti-correlation between \dnn\ and \xhel\ at lower speeds is likely is because the range of \vsw\ that correspond to incompressible, Alfvénic solar wind overlaps with the range corresponding to compressible, non-Alfvénic solar wind.

Compressive fluctuations are associated with slow modes and pressure balanced structures (PBS).
Under magnetohydrodynamics (MHD), these two phenomena can be considered as different limits that both lead to $k_\perp/k_\parallel \rightarrow \infty$ \citep{Hollweg2014}.
PBS can be described as the $k_\parallel \rightarrow 0$ limit, leading to structures in which the total thermal plus magnetic pressure is constant and that are convected along with the solar wind \citep{Tu1994,LR:turbulence}.
Although the total thermal plus magnetic pressure is constant in PBS, both quantities can fluctuate \citep{LR:turbulence,Tu1994,Marsch1990a} and the fluctuations in temperature likely maintain the pressure balance \citep{Tu1994}.
\citet{Marsch1990a} argue these convected structures are related to the source of the solar wind.
Specifically, \citet{LR:turbulence} suggest that they are, ``related to the fine ray-like structures or plumes associated with the underlying chromospheric network and interpreted as signatures of the interplanetary flowtubes.''
The more recently identified periodic density structures (PDS) may be related to these PBS \citep{Kepko2024a,Viall2015,Viall2009}.

\plotVswXhelDnn
Slow mode waves can be derived as the $k_\perp \rightarrow \infty$ limit while $k_\parallel \rightarrow \infty$ is held finite \citep{Hollweg2014}.
The are more difficult to study than PBS because they are typically Landau damped \citep{Vasquez1999,Cuesta2023}.
MHD predictions for slow mode waves agree with solar wind observations better than their kinetic counterparts \citep{Verscharen2017a} and compressive fluctuations may help maintain the stability of fully ionized helium that is differentially streaming along the magnetic field faster than ionized hydrogen \citep{Zhu2023}.
Using Parker Solar Probe (Probe) observations below \Rs[139.64], \citet{Adhikari2020} show that the solar wind compressibility stays roughly between 0.1 and 0.2 and is consistent with nearly incompressible MHD (NI-MHD).
NI-MHD is an extension of MHD in which compressible fluctuations are generated by the nonlinear interaction between Alfvén waves propagating along the mean magnetic field and incompressible fluctuations propagating perpendicular to it \citep{Zank1992}.

In this paper, we analyze the increase of \grad_\vsw_[\ahe] with decreasing \xhel\ for solar wind with speeds \vsw*[\vs][>], which is unexplained in \citetOne.
We show that they are related to the solar wind's compressibility.
We also show that there are two distinct populations of enhanced \ahe*[\As][>], one Alfvénic and one compressible, and infer that the compressible subset with enhanced \ahe*[\As][>] is from neither of the traditional sources of solar wind that are intermittently or continuously open to the heliosphere.
Mapping the results of this work to \citet{\BibOne} and vice versa shows that, in any given \dnn\ quantile, \xhel[0.65][\lesssim], an upper bound on non-Alfvénic cross helicity.
Similarly, \dnn[0.15][\lesssim] in any given \xhel\ quantile, an upper bound on incompressible solar wind fluctuations.
We conclude that the hydrogen compressibility is essential for characterizing and possibly regulating the solar wind helium abundance.

\section{Observations \label{sec:obs}} 

We use solar wind measurements provided by the Wind Solar Wind Experiment (SWE) Faraday cups (FCs) \citep{Wind:SWE} and Magnetic field Investigation \citep[MFI,][]{Wind:MFI:A,Wind:MFI:B}.
Our data selection follows \citetalias{\BibOne}.

\section{Analysis\label{sec:analysis}}

\plotVswAheXhel
\subsection{Combining the Solar Wind Speed (\vsw), Helium Abundance (\ahe), and Normalized Cross Helicity (\xhel) \label{sec:analysis:vsw-ahe-xhel}}

We begin with a semi-qualitative characterization of the relationship between \xhel, \ahe, and \vsw.
\cref{fig:vsw-ahe-xhel} plots the former as a function of the other two.
The contours are smoothed with a $1\sigma$ Gaussian filter for visual clarity.
The white area in the bottom right is where the contours are sufficiently noisy that the smoothing fails.
The contour at \xhel[0.6] is highlighted with a thick black line.
The dash-dotted lines are fits to $\ahe\left(\vsw\right)$ in 15 quantiles of \xhel\ derived in \citetalias[Figure 4]{\BibOne}.
The corresponding saturation points \satpoint\ across all 15 \xhel\ quantiles are plotted in black and are located at \xhel[0.6][\gtrsim], just beyond the highlighted contour.
In \citetalias{\BibOne} the saturation speed (\vs) and saturation abundance (\As) are each divided into \emph{Low}, \emph{Mid}, and \emph{High} ranges.
The three \vs\ and three \As\ ranges are not identical.

Overall, \xhel\ increases with increasing \vsw.
For \vsw*[\vs][\gtrsim], this increase becomes increasingly independent of \ahe, especially for \ahe[4.19].
Visually inspecting the underlying 2D histogram, we have qualitatively determined by eye that the slope of these contours larger and smaller than \ahe[4.19] are roughly symmetric with values of $\pm 3\%/\kms[200]$, and are independent of \xhel.
These slopes are not shown for visual clarity.
In other words, at \ahe\ larger and smaller than 4.19\%, a larger \vsw\ is required to observe an increase in \xhel\ and, as \vsw\ and \xhel\ increase, the range of \ahe\ on which higher \xhel\ are observed decreases.

The fits of \func{\ahe}{\vsw} for each \xhel\ quantile are derived and originally presented in \citetalias[][Figures 4 and 5]{\BibOne}.
\cref{fig:vsw-ahe-xhel} plots these fits as a black shaded region with the lines corresponding to the lowest (blue) and highest (red) \xhel\ plotted as dash-dotted lines.
For \vsw*[\vs][>], the of \func{\ahe}{\vsw} for low \xhel\ becomes larger than the \func{\ahe}{\vsw} fits for high \xhel\ solar wind over the range of speeds $\kms[530] < \vsw*[560][<]$.
This range of speeds corresponds to the region of \cref{fig:vsw-ahe-xhel} in which the average \xhel\ crosses the 0.7 contour, which \citetalias{\BibOne} shows, corresponding to the \xhel\ in the non-Alfvénic subset as identified with \As\ and in which \As\ reaches its largest value.
Over this \kms[30] range, the blue and red dash-dotted lines are replaced with solid black lines.
This suggests the paradox that \ahe\ in non-Alfvénic solar wind is greater than \ahe\ in Alfvénic solar wind in the region of the plane where the solar wind is, on average, Alfvénic.

\plotVswAheDnn
\subsection{\vsw, \ahe, and the Solar Wind Compressibility (\dnn) \label{sec:analysis:vsw-ahe-dnn}}

Figures 4 \& 5 of \citetalias{\BibOne} show that \grad_\vsw_[\ahe] for speeds \vsw*[\vs][>] increases with decreasing \xhel.
\cref{fig:vsw-ahe-xhel} shows that \xhel\ decreases for increasing \ahe\ smaller and larger than \ahe*[\As] and slower wind.
\cref{fig:vsw-ahe-dnn} shows contour plots of the \dnn\ as a function of \vsw\ and \ahe.
Again, contours are smoothed with a $1\sigma$ filter for clarity.
The solid black lines are the 1D fits of Gaussians to \ahe\ PDFs in each \vsw\ column: the central line is the mean from the fit; the upper and lower lines indicate the corresponding $1\sigma$ widths.
The contour at \dnn[0.09] is at the boundary between yellow and blue; the \dnn[0.085] contour is between the two lightest blue regions.
The saturation points and fits to \func{\ahe}{\vsw} are plotted in the same manner as \cref{fig:vsw-ahe-xhel}.
The black dotted rectangle indicates the region enlarged in \cref{fig:vsw-ahe-dnn:zoom}.
For speeds \vsw*[\vs][>], low compressibility \dnn[0.09][\lesssim] is roughly limited to a band in \ahe\ that is narrower than the solid black lines corresponding to the $1\sigma$ fits to distributions of \ahe\ in each \vsw\ bin.

\plotVswAheDnnZoom
\cref{fig:vsw-ahe-dnn:zoom} enlarges the region surrounding the saturation point \satpoint\ and the crossover between \func{\ahe}{\vsw} at \vsw*[\vs][>] for low and high \xhel\ in \cref{fig:vsw-ahe-dnn}.
Several observations stand out:
\begin{compactenum}
    \item A significant fraction of these saturation points are confined to the light blue region in which \dnn[0.085] to 0.090 and \As[4][\geq], which corresponds to the \emph{Mid} and \emph{High} ranges of \As\ defined in \citetalias{\BibOne}.
    \item The contour of \dnn[0.09] changes at approximately $\left(\As,\vsw\right) = \left(4\%, \kms[400]\right)$.
    For lower speeds, this contour of constant \dnn\ follows increasing \ahe\ and \vsw.
    At higher speeds, this contour is independent of \As\ and follows increasing \vsw\ and then follows decreasing \As\ as \vsw\ increases above \vsw[515][\approx].
\item There are two regions corresponding to \dnn[0.085][\leq], one corresponding to \vsw[470][\lesssim] and the other corresponding to \vsw[530][\gtrsim].
    The region where \func{\ahe}{\vsw} for low \xhel\ becomes larger than \func{\ahe}{\vsw} for high \xhel\ begins at \vsw[530] and crosses this \dnn[0.085] contour.
    The location of this intersection between the crossover region and \dnn[0.085] is also where \func{\xhel}{\ahe,\vsw}[0.7] in \cref{fig:vsw-ahe-xhel}.
    \item The contour \dnn[0.09] traces out an increase in both \ahe\ and \vsw\ for \vsw*[\vs][<].
    For \vsw*[\vs][>], the \dnn[0.09] contour traces out constant or very slowly decreasing values of \ahe\ with increasing \vsw.
\end{compactenum}
These observations suggest that, in addition to \ahe's dependence on \vsw\ and \xhel, it also depends on \dnn\ and that \dnn\ may cross a value separating compressible and incompressible fluctuations in a region of the \func{\relax}{\vsw,\ahe}-plane that \cref{fig:vsw-ahe-xhel} shows the apparent paradox described above.

\subsection{Saturation Fits \label{sec:analysis:sat-fits}}

\plotSaturationFits
To characterize how the change in gradient \gradAhe\ depends on the compressibility \dnn, we apply \citetalias{\BibOne}'s bilinear fitting method and repeat the analysis of \cref{fig:vsw-ahe} for 15 quantiles of \dnn.
\cref{fig:sat-fits} plots the resulting fits of \func{\ahe}{\vsw} in each \dnn\ quantile, which is indicated by the color bar.
In \citetalias{\BibOne}, Figure 4 is the equivalent analysis for 15 quantiles in \xhel.
\cref{tbl:dnn} summarizes these saturation points.
Quantiles corresponding to \dnn[0.15][>] are plotted in pink.
The green line plots the saturation points \satpoint, connected in \dnn\ order.
The insert zooms in to the region immediately surrounding the saturation points.
Broadly, there are two groups of quantiles: an incompressible subset (\dnn[0.15][<]) and a compressible subset (\dnn[0.15][>]).
Below the saturation point -- i.e. with $\left(\vsw,\ahe\right) < \satpoint$ -- \gradAhe\ is roughly consistent across all \dnn\ quantiles.
Above the saturation point -- i.e. with $\left(\vsw,\ahe\right) > \satpoint$ -- \gradAhe\ is several times larger for the compressible quantiles (pink lines) than for the incompressible quantiles.

\plotSaturationFitsScaled
\cref{fig:sat-fits:scaled} highlights the difference in \gradAhe\ above and below the saturation point by scaling the fits in \cref{fig:sat-fits} to their \satpoint, which are plotted in green. 
Figure 5 in \citetOne\ is the equivalent analysis for 15 quantiles in \xhel.
This figure shows that the \gradAhe\ are nearly indistinguishable below the saturation points.
Above the saturation point, fits from the incompressible quantiles occupy a very narrow region.
In contrast, \gradAhe\ above the saturation points are similar when \dnn[0.15][<] and markedly larger when \dnn[0.15][>].

\TblDnn
The insert in \cref{fig:sat-fits} shows that, similar to the gradients \gradAhe, the saturation points \satpoint\ themselves also fall into incompressible and compressible subsets. 
The incompressible subset is approximately a line for which \As\ decreases with increasing \vs, while the compressible subset does.
To characterize the dependence of the saturation points on \dnn\ without the parametric dependence in \cref{fig:sat-fits}, \cref{fig:sat-dn} plots (a) the saturation speed \vs\ and (b) the saturation abundance \As\  as a function of \dnn.
The color scale matches \cref{fig:sat-fits,fig:sat-fits:scaled} and will be used going forward in all figures for which it is appropriate to facilitate comparison between plots.
Additionally, markers for the two \dnn\ subsets differ: incompressible quantiles are plotted with circles and compressible quantiles are plotted with squares.
Markers corresponding to \dnn[0.085] (see \cref{fig:xhel-ahe-dnn}) are indicated by their blue instead of black edge.
The blue horizontal bands indicate points corresponding to the identified ranges of \xhel\ in Figure 6 of \citetOne, the analogous analysis performed as a function of \xhel.
From this figure we make the following observations about the saturation speed \vs.
\begin{compactenum}
\item The saturation speed is approximately constant for incompressible quantiles with \dnn[0.1][\leq].
The weighted average of values in this range are \vs[428 \pm 1.4], which corresponds to the \emph{Low \xhel} subset defined in \citetOne.
\item \vs\ increases slightly across the two largest incompressible \dnn\ quantiles between \dnn[0.1] and \dnn[0.15].
\item For compressible quantiles with \dnn[0.15][>], the maximum saturation speed increases to \vs[502 \pm 7] and average saturation speed increases markedly to \vs[461 \pm 3.4].
\item The saturation speed belonging to the incompressible subset with the largest \dnn\ is larger than the smallest \vs\ in the compressible subset.
\end{compactenum}
For the saturation abundance \As, we make the following observations.
\begin{compactenum}
\item The weighted average of \As\ in the incompressible subset is $4.2 \pm 0.02\%$.
\item In the compressible subset, \As
    \begin{compactenum} 
        \item reaches a local minimum of \As[3.9 \pm 0.1] at \dnn[0.252] and 
        \item a maximum value of \As[4.13 \pm 0.1] at \dnn[0.649].
    \end{compactenum}
\item Incompressible \As\ is larger than, or including error bars, within the range of \As\ corresponding to \emph{High} \xhel\ in \citetOne.
\item Compressible \As\ corresponding to the largest \dnn\ falls within the range of \As\ for \emph{High} \xhel\ in \citetOne, while compressible \As\ corresponding to the other two \dnn[0.15][>] quantiles fall within the \As\ range corresponding to \emph{Low} \xhel\ in \citetOne.
\end{compactenum}

\plotSaturationCompressibility

To determine why non-Alfvénic \ahe\ is greater than Alfvénic \ahe\ for speeds \vsw[545][\gtrsim], \cref{fig:sat-slopes} plots the gradients \gradAhe\ for \vsw*[\vs][>] (i.e.~above the saturation point) as a function of \dnn.
The style matches \cref{fig:sat-dn}.
The marker indicating the slope at \dnn[0.085] is marked with a light blue circle for comparison with \cref{fig:xhel-ahe-dnn}.
This ``saturation slope'' (\mfast) is approximately constant and independent of \dnn\ in the incompressible subset with an average value $0.84 \pm 0.06 \; \pten{-3} \% \, \mathrm{km^{-1} \, s}$.
For the compressible subset, the saturation slope increases markedly and monotonically to a maximum of $2.7 \pm 0.2 \; \pten{-3} \% \, \mathrm{km^{-1} \, s}$ at \dnn[0.649].
This analysis suggests that \dnn[0.15] may separate compressible and incompressible fluctuations in the solar wind, incompressible solar wind may have consistent behavior above saturation, and compressible fluctuations may introduce the apparent paradox in \cref{fig:vsw-ahe-xhel}.

\plotSaturationSlope

\subsection{Excluding Compressive Solar Wind from \func{\ahe}{\vsw,\xhel} \label{sec:analysis:dnn-cut}}

\plotCompareThresholds
To verify the significance of compressive solar wind in the saturation analysis of \citetOne, we repeat that analysis selecting incompressible solar wind with \dnn_\Hy_[0.15][<], 0.1, and 0.085.
The first two are chosen based on \cref{sec:analysis:vsw-ahe-dnn}.
The latter is the contour of \dnn\ in \cref{fig:vsw-ahe-dnn:zoom} across which \ahe\ in low \xhel\ quantiles becomes larger than \ahe\ in high \xhel\ quantiles.
We also select compressive solar wind with \dnn_\Hy_[0.15][>].
\cref{fig:compare-thresholds} plots the resulting saturation 
\begin{inparaenum}[(a)]
\item slopes, 
\item speed, and 
\item abundances.
\end{inparaenum}
The incompressible observations with  \dnn_\Hy_[0.085][<], 0.1, and 0.15 are presented in dash-dotted red, green, and orange lines, respectively.
The number of dots in the line increases with the threshold from 1 to 3.
The compressible subset with \dnn_\Hy_[0.15][>] is plotted with a dotted purple line.
The fit parameters from all the data irrespective of threshold is plotted in a solid blue line.
Shaded regions corresponding to the uncertainties for \dnn_\Hy_[0.1][<], which is representative of all three incompressible subsets, and all \dnn_\Hy_ are shown.
For visual clarity, the lines and uncertainties are smoothed with a Savitzky-Golay filter \citep{SavgolFilter,SavgolFilter:Comment}.
The fit parameters themselves are given in \cref{tbl:by-thresholds}.
The solid gray lines are the fit parameters derived from \cref{fig:vsw-ahe} across all observations.

\TblByThresholds
In panel (a), we observe that all saturation slopes decrease with increasing \xhel.
Comparing the compressible subset to all the data or any of the incompressible subsets show that the slopes in the compressible subset are a factor of 2 to 5x larger across \xhel.
Comparison of the lines for any of the incompressible subsets along with all the data also shows that selecting data below any of the \dnn\ thresholds reduce \mfast\ and this change is most significant for \xhel[0.4][<].
Comparing these trends to the gray line, we also observe that \mfast\ in \cref{fig:vsw-ahe} is closest to Alfvénic \mfast\ incompressible solar wind or all the observations, irrespective of compressibility.

In panels (b), compressible \vs\ is less than 10\% larger than any of the incompressible subsets of all the data.
However, it also typically exceeds \vs\ derived in \cref{fig:vsw-ahe} (gray line).
In contrast, \vs\ for the incompressible subsets along with all the data are nearly indistinguishable and well below \vs\ indicated by the gray line for \xhel[0.75][>].

Panel (c) shows that compressible \As\ is at most 16\% smaller than all other saturation abundances.
The mean values of \As\ for the incompressible subsets (green, red, and orange lines) are larger than the mean value of \As\ across all data (blue line), their mutual uncertainties overlap.
For non-Alfvénic solar wind with \xhel[0.6][<], all incompressible subsets along with all the data irrespective of compressibility have \As[4.19][<], which is below \As\ in \cref{fig:vsw-ahe} (gray line), but \As\ in incompressible Alfvénic solar wind with \xhel[0.683][\geq] is (excluding \xhel[0.74] for \dnn_\Hy_[0.85][<]) within 1\% of \As[4.19] from \cref{fig:vsw-ahe}.
That \As\ in compressible solar wind is larger than \As\ in incompressible solar wind suggests that the increase in \mfast\ with decreasing \dnn\ in \cref{fig:sat-fits,fig:sat-fits:scaled} drives \As\ to smaller values and \vs\ to larger values.
This suggests that there is a subset of solar wind with an enhanced \ahe\ that is related to compressive fluctuations and that obscures the transition in our observations from slow to fast wind.
This analysis also confirms that \dnn[0.15] is a threshold across which the solar wind transitions from compressible to incompressible.

\subsection{Visualizing the Speed Ranges on the Bimodal Distribution of \vsw\ Observed During Solar Minima\label{sec:analysis:vsw-hist}}

\plotVswHist
The classification of solar wind into slow and fast subsets is based, in part, on the bimodal nature of the distribution of speeds observed at \au[1] during solar minima.
Although this classification is known to be limited in both precision and accuracy, it relies solely on observations of \vsw, which most spacecraft provide.
\citet{Wind:SWE:Wk} quantify the location and width of these peaks by fitting the slow wind distribution with \vsw[400][<] and the fast wind distribution with \vsw[600][>] each to a Gaussian.
The speeds are \vslow[355 \pm 44] and \vfast[622 \pm 59].
\citet{Wind:SWE:Wk} show that the range of near-Earth \vsw\ predicted from near-Sun observations of the solar wind's kinetic energy flux \citep{Liu2021c} is \vK*[557] to \kms[700].
The same work shows that the range of speeds to which solar wind from continuously open source regions would be accelerated in the absence of Alfvén wave forcing in transit, i.e. ``wave-poor'' fast wind in which the energy in the Alfvénic fluctuations is less than typical fast wind, is \vIP*[435] to \kms[554].
Figure 1 of \citet{Wind:SWE:Wk} contextualizes these observations by highlighting the relevant segments of the \vsw\ probability distribution function (PDF) for observations collected during solar minima.
\citetOne\ determines that \vsigma*[407] to \kms[439] is the range of speeds over which the solar wind transitions from its slow to fast state when characterized by \gradAhe\ across 15 \xhel\ quantiles.
\citetOne\ also shows that \func{\n[\He\!]}{\vsw} peaks at speeds \vn[409 \pm 15].
Figure 8 of \citetOne\ contextualizes \vsigma\ and \vn\ in the same way as \citet{Wind:SWE:Wk} using the PDF of \vsw\ observations collected during solar minima.
The range of speeds over which heavy ion abundances saturate is \vheavy*[316] to \kms[340] \citep{ACE:SWICS:FStransition}.

\SpeedTable

\cref{fig:vsw-hist} synthesizes three versions of this histogram with the above speeds (\citeauthor{Wind:SWE:Wk} \citeyear{Wind:SWE:Wk}; \citetOne; \citeauthor{ACE:SWICS:FStransition} \citeyear{ACE:SWICS:FStransition}) and adds the range of speeds at which the slow/fast wind transition occurs across the 15 \dnn\ quantiles derived in this paper.
\cref{tbl:speeds} summarizes these speeds.
The PDF is plotted in black.
Each speed range is highlighted with a color with line thickness chosen so that overlapping segments are all visible.
Because this figure identifies nine overlapping speed intervals, each segment is also plotted below the PDF and labeled here.
The vertical location of these labeled segments is chosen so that the labels do not overlap, but are not significant otherwise.
The transition between slow and fast wind for incompressible solar wind (\dnn_\Hy_[0.15][<]) is labeled \vdn*[417] to \kms[436].
The transition for compressible solar wind (\dnn_\Hy_[0.15][>]) is labeled \vdnp*[435] to \kms[508].

\cref{fig:vsw-hist} shows that the speeds are organized in the following increasing order.
\begin{compactenum}
\item The slow wind peak \vslow\ covers the lowest range of speeds.
\item The range of speeds at which the gradients of heavy ion abundances transition from slow to fast wind (\vheavy) is within the slower portion of the range \vslow, but does not exceed the center of the slow wind peak \citep{ACE:SWICS:FStransition}.
\item The range of speeds \vn\ at which the helium number density peaks as a function of \vsw\ derived in \citetOne\ overlaps with the fast end of \vslow.
\item The saturation speed \vsigma\ derived across \xhel\ quantiles overlaps with \vn, but the slowest \vsigma\ is faster than the fastest \vslow.
\item The incompressible saturation speed \vdn\ overlaps with both the fast end of the \vn\ range and the majority of the \vsigma\ range.
\item The range of compressible saturation speeds \vdnp\ overlaps with \vn\ by \kms[4].
By construction, the slowest \vdnp\ is faster than the fastest \vdn.
\item The range of speeds \vIP\ corresponding to Alfvén ``wave-poor'' fast wind, which is likely synonymous with Alfvénic slow wind, effectively spans the range of speeds from \vdn\ to the lower range of \vfast\ and \vK.
\item The range of near-Earth speeds derived from the near-Sun radial gradients of the solar wind's total and kinetic energy fluxes (\vK) corresponds to and exceeds the fast wind peak (\vfast).
\end{compactenum}
Although not plotted, the range of speeds in \cref{fig:vsw-ahe-xhel} over which non-Alfvénic \ahe\ (\xhel[0.7][<]) exceed Alfvénic \ahe\ (\xhel[0.7][>]) is \kms[530-560], which corresponds to the gap between \vdnp\ and the lower bounds of \vK\ and \vfast.
This analysis suggests that compressive fluctuations are significant in the \vdnp\ region of the \vsw\ PDF, where the power in Alfvénic fluctuations is reduced in comparison to typical fast wind \citep{Wind:SWE:Wk} (which may be related to the Alfvénic slow wind), and which corresponds to the intermediate range of speeds over which the literature specifies ad hoc speeds for transitioning from slow to fast wind, with the implied change in solar source region.

\plotXhelDnnAhe*
\subsection{Contextualizing Observed \vsw\ and \ahe\ in the $\left(\xhel,\dnn\right)$ Plane \label{sec:analysis:context}}

Alfvénic and compressive fluctuations are mutually exclusive.
As such, solar wind is broadly considered to be incompressible and incompressible solar wind is broadly considered to be Alfvénic, but these relationships do not imply that incompressible and Alfvénic solar wind are interchangeable nor that compressible and non-Alfvénic solar wind are interchangeable without carefully considering data selection and solar wind conditions.
\cref{fig:xhel-dnn-ahe} presents contour plots of
\begin{inparaenum}[(a)]
\item mean,
\item 90\% quantile, and
\item 10\% quantile of \func{\ahe}{\xhel,\dnn} along with
\item the variability of \ahe.
\end{inparaenum}
This variability is referenced to the mean value and calculated as 
\begin{equation}\label{eq:variability}
100 \times \left(\frac{90\% \, \mathrm{quantile} - 10\% \, \mathrm{quantile}}{\mathrm{mean}} - 1\right).
\end{equation}
A variability $>1$ indicates that the distance between the 90\% and 10\% quantiles is larger than the mean value.
A variability $<1$ indicates that the distance between the quantiles is less than the mean value.
Visually, this is how much the difference between panels (b) and (c) exceeds the values in panel (a), normalized to that panel.
All contours are smoothed by a $1\sigma$ Gaussian kernel for visual clarity.
The black contours are \ahe[3.82] derived in panel (a), where 3.82\% is the smallest \As\ in both this work and \citetOne.

In panel (a), we observe two distinct populations of \ahe*[\As][\geq].
A low compressibility subset with high cross helicity occupies the right side of the figure.
We will refer to this as the helium-rich, Alfvénic subset.
A high compressibility subset without a preferred cross helicity occupies the top of the figure.
We refer this as the helium-rich, compressive subset.
There is also a helium-poor subset with \ahe[\As][<] that occupies the majority of the plane outside of these two helium-rich regions.
Comparing panels (b) and (c), we observe that the 90\% quantile of this high compressibility subset exceeds \ahe[7.5], while the 10\% quantile only exceeds \ahe[1.4] in Alfvénic wind with \xhel[0.7][\gtrsim].
In contrast, the 90\% quantile of the low compressibility subset has, effectively, \ahe[6][<] and the 10\% quantile exceeds \ahe[1.4] across the range of \dnn\ in this subset.
The variability in panel (d) quantifies this comparison of the 90\% and 10\% quantiles.
It shows that in this incompressible subset, \ahe, for which the average is above saturation (\ahe*[\As][>]), the variability is within approximately 20\% of the mean value.
In contrast, the variability in the compressible subset exceeds 70\%.

\plotXhelDnnVsw
\cref{fig:xhel-dnn-vsw} is an abridged presentation of the analogous analysis of \vsw.
\begin{inparaenum}[{Panel} (a)]
\item plots mean \vsw\ and
\item plots the variability of \vsw\ calculated in the same manner as \cref{fig:xhel-dnn-ahe}.
\end{inparaenum}
Both panels include contours of mean \vsw*[433] and \kms[450].
\vs[450] is the largest \vs\ in this work for \dnn[0.15][<].
\vs[433] is the least Alfvénic and fastest \vs\ in \citetOne\ and within a few \kms\ of incompressible \vs\ calculated in this work.
Comparing panel (a) in \cref{fig:xhel-dnn-ahe,fig:xhel-dnn-vsw}, we see that \vsw*[\vs][>] corresponds to the incompressible, Alfvénic, and helium-rich subset of \ahe\ in \cref{fig:xhel-dnn-ahe}.
Panel (a) also shows that mean \vsw\ only exceeds the largest (incompressible) \vs\ in the incompressible subset.
Panel (b) shows that mean \vsw\ in the Alfvénic subset is more variable than in the compressible subset.
Comparing panels (b) and (d) in \cref{fig:xhel-dnn-ahe} to panel (b) in \cref{fig:xhel-dnn-vsw}, we observe that that the largest \ahe\ coincide with the most variable \ahe\ in the compressible subset where the variability in mean \vsw\ is not as large as in the Alfvénic subset.
This implies that the compressibility is driving the changes in \ahe, not \vsw\ itself.
Comparing panel (a) and (b) here, we also observe the expected anti-correlation between \xhel\ and \dnn\ along contours of constant average \vsw, even though \vsw\ may be highly variable at any given point across most of the plane.
This suggests that a wide range of \vsw\ can be observed in both incompressible, Alfvénic solar wind and compressible, non-Alfvénic solar wind and these ranges of speeds are largely overlapping, which may explain the apparent paradox presented in \cref{sec:analysis:vsw-ahe-xhel,sec:analysis:vsw-ahe-dnn}.

\subsection{Contextualizing Observed \dnn\ in the $\left(\xhel,\ahe\right)$ Parameter Space \label{sec:analysis:context}}

Broadly, slow and fast solar wind are accelerated in different types of solar wind source regions.
Fast wind emerges from solar wind source regions like CHs with magnetic fields that are continuously open to the heliosphere.
Slow wind emerges from solar wind source regions that are only intermittently open to the heliosphere, e.g. pseudostreamers, helmet streamers, etc.
These latter regions are sometimes colloquially referred to as magnetically ``closed''.
As reviewed in \sect{intro}, a variety of observations show that this classification scheme is overly general.
Per \citetOne\ and as illustrated in \cref{fig:xhel-ahe-vsw:cartoon}, classifying \emph{in situ} solar wind observations as in the $\left(\xhel,\ahe\right)$-plane space may separate solar wind by the magnetic topology source region of the solar source region from which it observes.
Generally, helium-poor solar wind without a preferred \xhel\ corresponds to solar wind from magnetically closed regions.
In contrast, helium-rich and Alfvénic solar wind originates in magnetically open source regions.

Solar wind from magnetically closed regions is often slower than solar wind from magnetically open regions.
However, this is not always the case.
The Alfvénic slow wind (ASW) is an emerging class of solar wind that is most similar to fast wind in all respects except its speed, which is more typical of slow wind \citep{DAmicis2021a,DAmicis2021,DAmicis2018,Damicis2016,DAmicis2015,Yardley2024,Rivera2025}.
\citet{Wind:SWE:Wk} suggests that ASW may be wave-poor solar wind from continuously open source regions in which the energy in the Alfvénic fluctuations is reduced in comparison to typical fast wind..
Consistent with these works, \citetOne\ argues that ASW is the slow speed extension of solar wind originating in magnetically open sources and this range of speeds is slower than the maximum speed that solar wind from magnetically closed source can obtain.
To compare our observations of \dnn\ with the results of \citetOne, we now contextualize the compressibility in the $\left(\xhel,\ahe\right)$ parameter space.

\plotXhelAheDnnMultiPanel

\cref{fig:xhel-ahe-dnn} plots 
\begin{inparaenum}[(a)]
\item mean and 
\item the 75\% quantile 
\end{inparaenum} 
of \dnn\ as a function of \xhel\ and \ahe.
Solid black contours in panel (a) indicate \dnn[0.085], 0.1, and 0.15, the three thresholds for separating incompressible and compressible solar wind that we have identified in this work.
The dash-dotted black line indicates where the 75\% quantile of \dnn[0.15] in panel (b).
In panel (b), the solid contours indicate where the 75\% quantiles of \dnn[0.1] and 0.15.
Solar wind with \xhel\ greater than that indicated by the dash-dotted blue line is in the region of the plane identified in \citetOne\ as originating in magnetically open source regions.
Here, we observe that solar wind carrying compressive fluctuations largely occupies the top left corner of the plane where the minimum \ahe\ along the \dnn[0.15] contour is more than 7\% (in panel a) and 6\% (in panel b), both larger than \As[4.34], the largest \As\ in this work and \citetOne\ accounting for uncertainty on \As.
On the right side of panel (b), the 75\% quantile of the solar wind compressibility in the central portion of the open field region \citepalias[bounded on the left by the blue dash-dotted line,][]{\BibOne} does not exceed \dnn[0.1] and the majority of this plane does not exceed \dnn[0.15], the largest threshold we identified for the separation between incompressible and compressible solar wind based only on \func{\grad_\vsw_[\ahe]}{\vsw*[\vs][>]}.
In panel (a), mean \dnn does not exceed 0.085 in a slightly larger region of the plane.
Based on the source region identification in \citetOne, this analysis suggests that solar wind is from continuously open sources is incompressible and solar wind from intermittently open source regions has a compressibility that approaches the \dnn[0.15] threshold.
Forthcoming work shows that transients are located at a compressibility \dnn[0.15][>] in panel (a) \citep{Wind:SWE:ahe:ICMEs}.

\subsection{Mapping Between \dnn\ and \xhel \label{sec:appendix:mapping}}

\TblDnnMapping

We have shown an enhancement in helium abundance \ahe*[\As][>] in two different, non-overlapping regions of the $\left(\xhel,\dnn\right)$-plane: one compressible and one Alfvénic. 
While the Alfvénic subset of enhanced \ahe\ is limited to \dnn[0.2][<], the compressible subset is observed across all \xhel.
Despite the physical definition of Alfvénic fluctuations, compressible solar wind is not interchangeable with non-Alfvénic solar wind.
Rather both processes must be considered because statistical aggregation across a range of solar wind conditions mixes these two types of plasma.
This is why our analysis shows that the saturation fits across \dnn\ or \xhel\ quantiles averages over variability in the other quantity.
The comparison between \cref{fig:vsw:xhel,fig:vsw:dnn,fig:vsw:xhel:dnn} support this inference.
To characterize the impact of this on our determination of the saturation points, we have mapped each \dnn\ quantile to \xhel\ using both \vs\ and \As.
We refer to this as ``mapped \xhel''.
We  similarly mapped each \xhel\ quantile from \citetOne\ to \dnn\ using both \vs\ and \As\ derived in that work.
Here, we derive and present those mappings.
We then discuss their implications.
\cref{tbl:dnn:mapping} summarizes mapped \xhel\ in each \dnn\ quantile using \As\ and \vs.
Similarly, \cref{tbl:xhel} summarizes mapped \dnn\ in each \xhel\ quantile.
These mappings and their implications serve to confirm the results presented in the main portion of this work.

\TblXhel

\subsubsection{Mapping Quantiles of \dnn\ to \xhel\ with \As\ \label{sec:appendix:mapping:dnn2xhel:As}}

\plotLinesAheXhelDnn
\cref{fig:lines:ahe-xhel-dnn} plots the average \xhel\ as a function of \ahe\ in 0.2\% wide bins for each \dnn\ quantile.
With this figure, we can map \dnn\ quantiles to \xhel\ using \As.
The shaded regions indicate the uncertainty, calculated as the standard error of the mean (SEM).
Lines and uncertainties are smoothed with a $1\sigma$ Gaussian filter for visual clarity.
The \dnn\ quantiles are given by the color bar and the line styles vary across quantiles.
The line for the quantile in which \dnn[0.085] is over plotted on a black line to highlight it.
As in prior figures, quantiles with \dnn[0.15][>] are plotted in pink.
The green line indicates \xhel\ at \As\ in each \dnn\ quantile.
The inset axis enlarges the region surrounding \As.
We make the following observations in \cref{fig:lines:ahe-xhel-dnn}.
\begin{compactenum}
\item \xhel\ increases with \ahe\ in all but the most compressive \dnn\ quantile, reaching a maximum at $\gtrsim \As$ and then decreasing as \ahe\ increases.
\item For incompressible quantiles (\dnn[0.15][<]), we observe the following:
    \begin{compactenum}
    \item \func{\xhel}{\ahe,\dnn} is independent of \dnn\ for \ahe*[\As][<].
\item As \ahe\ increases to values greater than \As, \xhel\ depends on \dnn\ and the spread in \xhel\ increases with increasing \dnn.
    \item The dependence of average \xhel\ on \ahe\ for \ahe*[\As][>] decreases with decreasing \dnn.
\end{compactenum}
\item For compressible quantiles (\dnn[0.15][>]), we observe the following:
    \begin{compactenum}
    \item The overall dependence of \xhel\ on \ahe\ decreases with decreasing \dnn.
    \item For the most compressible solar wind (largest \dnn), average \xhel\ is just smaller than 0.5 over the range \ahe*[1][\approx] to $\sim 6\%$. 
    For \ahe[6][>], even \xhel\ in the most compressible solar wind decreases with increasing \ahe.
    \end{compactenum}
\item The gradients of \func{\xhel}{\ahe,\dnn} are steeper for \ahe*[\As][<] than for \ahe*[\As][>].
\end{compactenum}
In other words, at low \ahe, all \dnn\ map to \xhel[0.5][\sim] and the possible values of \xhel\ at any given \func{\relax}{\ahe, \dnn} increases as \ahe\ increases.

\subsubsection{Mapping Quantiles of \dnn\ to \xhel\ with \vs\ \label{sec:appendix:mapping:dnn2xhel:vs}}

\plotLinesVswXhelDnn
\cref{fig:lines:vsw-xhel-dnn} allows us to apply the same method in \cref{fig:lines:ahe-xhel-dnn} to mapping \dnn\ quantiles to \xhel\ using \vs.
Here, we use \vsw\ bins that are \kms[10] wide.
The figure style matches \cref{fig:lines:ahe-xhel-dnn}.
Here, the green line indicates \vs\ and the inset axis highlights the region surrounding it.
We make the following observations.
\begin{compactenum}
    \item The dependence of average \xhel\ on \vsw\ becomes weaker as \dnn\ increases.
    \item For incompressible solar wind with \dnn[0.15][<], there is a strong, monotonic increase of \xhel\ with increasing \vsw\ up to \kms[\sim750].
    The value of average \xhel\ in this range of speeds only weakly depends on \dnn\ for any given \vsw.
    At speeds above \kms[\sim750], \xhel\ decreases with increasing \vsw.
    Due to the low frequency at which solar wind speeds of this range are observed, we do not draw an inference about \xhel\ in these fastest speeds.
    \item Incompressible solar wind with \dnn[0.15][>], the dependence of \xhel\ on \dnn\ for any given \vsw\ is markedly stronger, while \xhel\ strongly depends on \vsw\ in the compressible quantile with the smallest \dnn.
    For the solar wind in the most compressive \dnn\ quantile, \xhel\ is almost independent of speed.
\end{compactenum}
Similar to the \As\ mapping in \cref{fig:lines:ahe-xhel-dnn}, \dnn\ in slow wind maps to \xhel[0.5][\sim], while it maps to a large range \xhel\ in fast wind and the range increases with \dnn.

\subsubsection{Mapping Quantiles of \xhel\ to \dnn\ with \vs\ \label{sec:appendix:mapping:xhel2dnn:vs}}

\plotLinesVswDnnXhel
To fully compare the results in \citetOne\ across \xhel\ quantiles and those presented in this paper across \dnn\ quantiles, \cref{fig:lines:vsw-dnn-xhel,fig:lines:ahe-dnn-xhel} apply the technique in the previous section generate the inverse mapping from \xhel\ in each of 15 quantiles at the saturation point to \dnn.
The insert in each figure zooms in on the boxed region surrounding the appropriate saturation quantity (\vs\ or \As).

\plotLinesAheDnnXhel
 \cref{fig:lines:vsw-dnn-xhel} plots \func{\dnn}{\vsw} for each \xhel\ quantile.
 From it, we make the following observations.
\begin{compactenum}
    \item In Alfvénic wind with \xhel[0.7][\gtrsim], the compressibility is approximately constant with \dnn[0.09][<] for all but the fastest speeds with \vsw[750][\lesssim], above which \dnn\ increases by increasing amounts as \xhel\ increases.
    \dnn\ increases markedly above 0.3 in the fastest wind.
\item The insert shows that \vs\ is approximately independent of \dnn\ when \dnn[0.09][>].
    \item In the slowest wind, \dnn[0.1][<] across \xhel.
\end{compactenum}
That \dnn\ increases with increasing \vsw[560][>], the magnitude of the increase is larger with in non-Alfvénic solar wind, and \vsw*[\vs][>] is observed across all \dnn\ in \cref{fig:xhel-dnn-vsw}, reinforces the observation made throughout this paper that \vsw\ does not uniquely map to either non-Alfvénic, compressible or Alfvénic, incompressible solar wind.
Rather there is a wide range of speeds for both types of plasmas.

\subsubsection{Mapping Quantiles of \xhel\ to \dnn\ with \As\ \label{sec:appendix:mapping:xhel2dnn:As}}

\plotSaturationCompressibilityMapped
\cref{fig:lines:ahe-dnn-xhel} plots \func{\dnn}{\ahe} for each \xhel\ quantile.
The range of \dnn\ values in this figure is half as large as the range in \cref{fig:lines:vsw-dnn-xhel} so that the variability in the plot is readable.
The green markers indicate the compressibility \dnn\ at \As\ in each \xhel\ quantile, which is labeled with \As\ on the plot.
The insert axis zooms in on the indicated region surrounding \As.
From this figure, we make the following observations.
\begin{compactenum}
    \item As \ahe\ approaches 0, the compressibility \dnn[0.1][\sim] is approximately constant across \xhel.
\item For abundances \ahe[5][\lesssim], \dnn\ decreases with increasing \ahe\ in Alfvénic wind and is approximately constant for non-Alfvénic wind. 
\item For abundances \ahe[5][\gtrsim],  the minimum compressibility is 
        \begin{compactenum}
            \item is \dnn[0.067] , 
            \item reaches this minimum at \ahe[5.3][\approx], and
            \item \dnn\ is smallest for the largest \xhel,
        \end{compactenum}
    \item At \ahe[5.3][\gtrsim], \dnn\ increases across all \xhel\ and the gradient of \dnn\ as a function of \ahe\ increases with decreasing \xhel\ such that \dnn\ takes on its largest range of values at the largest \ahe.
    In particular, 
        \begin{compactenum}
            \item In Alfvénic wind with large \xhel, \dnn\ tends towards the same 10\% observed with vanishingly small \ahe.
            \item In non-Alfvénic wind with small \xhel, the compressibility tends towards \dnn[0.20][>]. the maximum \xhel\ in Alfvénic subset of enhanced \ahe\ (\cref{fig:xhel-dnn-ahe}).
        \end{compactenum}
\end{compactenum}
Similar to the observations in \cref{fig:lines:vsw-dnn-xhel}, we observe a larger gradient of \dnn\ across \xhel\ as \ahe\ increases.
Although the maximum \dnn\ in the largest \ahe\ is $\sim2x$ smaller than the maximum \dnn\ in the fastest \vsw, the anti-correlation between \dnn\ and \xhel\ is still recovered as \ahe\ increases.

\subsection{Mapping between the dependence of \satpoint\ and the saturation slope on \dnn\ and \xhel \label{sec:appendix:map-at-point}}

\plotSaturationSlopeMapped
Using the relationships derived in \cref{sec:appendix:mapping:dnn2xhel:As,sec:appendix:mapping:dnn2xhel:vs,sec:appendix:mapping:xhel2dnn:vs,sec:appendix:mapping:xhel2dnn:As}, we can map the dependence of the saturation point \satpoint\ along with the saturation slope from \dnn\ to \xhel.
While much of this can be observed combining the mappings with observations in \sect{analysis}, it is presented here for completeness.
\cref{fig:sat-dn:mapped} plots the dependence of the saturation point derived in \dnn\ quantiles as a function of mapped \xhel.
Panel (a) plots \func{\vs}{\xhel}.
Here, we derive \xhel\ from \cref{fig:lines:vsw-xhel-dnn} and calculate the mean \xhel\ in each \dnn\ quantile for the \vsw\ bin in which \vs\ falls.
This is what ``\vsw\ Bins'' next to the panel label indicates.
Panel (b) plots \func{\As}{\xhel}, applying the same mapping method using \cref{fig:lines:ahe-xhel-dnn} to determine the mean \xhel\ in each \dnn\ quantile for the \ahe\ bin in which \As\ falls.
The figure style follows \cref{fig:sat-dn}, including the \xhel\ ranges from \citetOne\ highlighted with blue bars.
The major change in style is that markers are unfilled, with the exception of the \dnn[0.085] quantile, which is filled and has a black edge.
We make the following observations in Panel (a) regarding the mapping using \vs.
\begin{compactenum}
	\item Incompressible solar wind mapped to \xhel\ with \vs\ has \xhel[0.6][\approx].
	\item Compressible solar wind mapped to \xhel\ with \vs\ has \xhel[0.06][<].
\end{compactenum}
In Panel (b), we observe the following about the mapping using \As.
\begin{compactenum}
	\item Incompressible solar wind mapped to \xhel\ using \As\ has \xhel[0.65][\approx].
	\item Compressible solar wind happed to \xhel\ using \As\ has \xhel[0.6][\lesssim].
\end{compactenum}
Comparing Panels (a) and (b), we observe that the minimum mapped \xhel\ is approximately the same when mapped \xhel\ is derived for \vs\ and for \As.
In contrast, the maximum mapped \xhel\ is larger when derived for \As\ than when derived for \vs.

\plotSaturationCompressibilityMappedOld
\cref{fig:sat-slopes:mapped} applies the same mapping techniques in \cref{fig:sat-dn:mapped} to the saturation slope in \cref{fig:sat-slopes} using both (a) \vs\ and (b) \As.
The figure style matches \cref{fig:sat-dn:mapped}.
As expected, compressible slopes mapped to \xhel\ from \dnn\ with \vs\ have \xhel[0.575][<], while incompressible slopes $\sim 0.001 \% \, km^{-1} \, s$ are fixed at a constant \xhel[0.06].
Similarly, compressible slopes mapped to \xhel\ from \dnn\ with \As\ have \xhel[0.6][\lesssim], while incompressible slopes map to \xhel[0.65][\approx].

Broadly, \cref{fig:sat-dn:mapped,fig:sat-slopes:mapped} imply that less compressible solar wind is more Alfvénic it is and the less compressible it is, the less \ahe\ departs from \As.
However, it also confirms that the average cross helicity or Alfvénicity in any incompressible \dnn\ quantile tends towards \xhel[0.6][\sim], which is the value in \cref{fig:vsw-ahe-xhel} at which we observe saturation, and does not reflect the full range of Alfvénic solar wind.

\cref{fig:sat-dn:old} plots
\begin{inparaenum}[(a)] 
\item the saturation speed \vs\ and
\item the saturation abundance \As\ 
\end{inparaenum}
as a function of mapped \dnn.
The general style matches \cref{fig:sat-dn:mapped,fig:sat-slopes:mapped}.
The exception is that \dnn\ and \xhel\ are swapped.
The color bar now indicates \xhel\ and \vs\ and \As\ are now plotted as a function of mapped \dnn.
The blue bands are distinct ranges of \xhel\ defined in \citetOne.
We observe the following observations about \vs.
\begin{compactenum}
    \item The mid and high \xhel\ subsets defined in \citetOne\ are grouped tightly into mapped \dnn\ corresponding to \dnn[0.085][\lesssim].
    \item The low \xhel\ range of \vs\ extends to higher mapped \dnn.
\end{compactenum}
From Panel (b), we observe the following about \As.
\begin{compactenum}
    \item The dependence of \As\ on mapped \dnn\ is roughly linear.
    Fitting the observations of \As\ as a function of mapped \dnn\ with a line yields a slope of $-8.6 \pm 0.5 \%/\#$ and a y-intercept of \As[4.76 \pm 0.04], i.e. \As\ in perfectly incompressible solar wind.
\item Again, the mid \xhel\ subset is tightly clustered.
    The low and high \xhel\ subsets are less clustered.
\end{compactenum}
Comparing mapped \dnn\ derived in both panels, we observe the following.
\begin{compactenum}
    \item Neither \vs\ nor \As\ are observed at a mapped \dnn\ corresponding to incompressible \dnn[0.15][>].
    \item The minimum and maximum mapped \dnn\ is smaller when derived for \As\ than when derived for \vs, but does not reach the lowest \dnn[0.01][\approx] observed in \cref{fig:sat-dn,fig:sat-slopes}.
\end{compactenum}
This mapping shows that incompressible \dnn\ is dominant across \xhel, but this value depends how the solar wind observations are aggregated.

\section{Discussion \label{sec:disc}}

\plotAfastAs

The saturation abundance \As\ is $49 \pm 2\%$ of the photospheric abundance.
\citet[Figure 1]{ACE:SWICS:SSN} show that the the helium abundance above saturation oscillates around $51 \pm 3\%$ of the photospheric helium abundance across solar activity, while the helium abundance below saturation only reaches a maximum of this value during solar maxima.
These observations suggest that this factor of 2 depletion of solar wind \ahe\ from its photospheric value is robust across solar activity as the heliographic latitude of open and closed source regions evolve with the solar cycle.

\citetOne\ shows that \grad_\vsw_[\ahe] for speeds \vsw*[\vs][>] increase with decreasing \xhel.
\cref{fig:ahe-enh} plots \func{\ahe}{\vsw[800]}, the helium abundance in each \xhel\ quantile from \citetOne\ at the fastest analyzed speed (\vsw[800]) normalized to \As\ in that quantile.
In the most Alfvénic solar wind, \ahe\ is enhanced by $\sim10\%$ with respect to \As.
In the least Alfvénic solar wind, \ahe\ is enhanced by $\sim40\%$.
The enhancement of \ahe\ above \As\ decreases linearly with increasing \xhel.
The dash-dotted line is a fit to the trend with a slope of $-0.4 \pm 0.02$ and a y-intercept of $1.46 \pm 0.02$, indicating that \ahe\ in the fastest wind would be 46\% larger than \As\ in perfectly non-Alfvénic and therefore compressible solar wind.
As \vsw*[\vs][>] is nominally from magnetically open sources and such solar wind is typically Alfvénic and incompressible, this seems to suggest that \ahe\ is not limited to $\sim50\%$ of its photospheric value, contradicting \citet{ACE:SWICS:SSN}.

We have combined the helium abundance with the normalized cross helicity and solar wind compressibility to address this apparent contradiction.
We observe that \dnn[0.85][\lesssim] in the \func{\relax}{\vsw,\ahe}-plane where \ahe\ observations are most common.
This suggests that \dnn\ may regulate \ahe\ in the solar wind.
To verify this inference, we have performed the following analysis.
\begin{compactenum}
\item Comparing \func{\dnn}{\vsw,\ahe} with the saturation fits derived in \citetOne.
\item Analyzing helium saturation as a function of \vsw\ across 15 \dnn\ quantiles.
\item Repeating \citetOne's analysis selecting for \dnn[0.085][<], 0.1, and 0.15 along with \dnn[0.15][>].
\item Characterizing \vsw\ and \ahe\ in the $\left(\xhel,\dnn\right)$-plane.
\item Characterizing \dnn\ in the $\left(\xhel,\ahe\right)$-plane, which \citetOne\ uses to identify solar wind observations predominantly from magnetically open and closed source regions.
\item Mapping saturation points derived in quantiles of \dnn\ or \xhel\ to the other quantity.
\end{compactenum}
We have also visualized the saturation speeds derived here with other key speeds derived by \citetOne, \citet{ACE:SWICS:FStransition}, and \citet{Wind:SWE:Wk} in the bimodal distribution of speeds observed near-Earth during solar minima.

\subsection{\func{\dnn}{\vsw,\ahe} and the Saturation Fits in \citetOne}
Comparing \func{\dnn}{\vsw,\ahe} with the saturation fits derived in \citetOne\ reveals the following.
\begin{compactenum}
    \item The central values and standard deviation of \ahe\ across \vsw\ are roughly limited to the incompressible region of the $\left(\ahe,\vsw\right)$-plane where \dnn[0.1][<].
    This is below the range of compressibilities observed by Probe below \Rs[139.64] \citet{Adhikari2020}.
\item Above saturation, \ahe\ in non-Alfvénic solar wind exceeds \ahe\ in Alfvénic solar wind over the range of speeds \vsw*[530] to \kms[560], which corresponds to the same region where \func{\xhel}{\ahe,\vsw}[0.7] in \cref{fig:vsw-ahe-xhel}.
    \xhel[0.68] is the lower bound on the Alfvénic subset defined in \citetOne\ based on \As.
    This speed range is below \vfast\ defined in \citetOne, where the frequency of speed observations has dropped significantly from the slow wind peak and where solar wind observed at \au[1] predominantly originates in open field regions, and \vK.
    \vfast\ is the peak of the fast wind distribution and \vK\ is the speed predicted at \au[1] from the radial scaling of the kinetic energy flux observed with Probe near-Sun observations \citep{Liu2021c}.
    This range of speeds also falls within the range \vIP, which \citet{Wind:SWE:Wk} identifies as solar wind from source regions that typically generate fast wind, but is Alfvén wave-poor in that the energy in the Alfvénic fluctuations is reduced in comparison to typical fast wind and is therefore not accelerated to the fastest, non-transient speeds observed near Earth.
    \item The solar wind's compressibility increases with increasing \ahe\ and \vsw, which are observed less frequently at \au[1] than slower speeds and lower \ahe.
\end{compactenum}
These suggest that the increase of \grad_\vsw_[\ahe] with decreasing \xhel\ at speeds above saturation (\vsw*[\vs][>]) are related to the presence of compressive fluctuations, which are a significant fraction of the observations in this region of parameter space where observations are less frequent.

\subsection{\func{\ahe}{\vsw} Saturation Fits Across \dnn\ Quantiles and Repeating \citetOne's Analysis Across \xhel\ Quantiles for Incompressible and Compressible Subsets}
Analyzing helium saturation as a function of \vsw\ across 15 quantiles of \dnn\ shows that the large \grad_\vsw_[\ahe] above saturation are limited to compressible solar wind with \dnn[0.15][>], while \grad_\vsw_[\ahe] in incompressible solar wind is virtually indistinguishable.
Repeating the analysis from \citetOne\ for incompressible subsets of our observations with \dnn[0.085][<], 0.1, 0.15; the compressible subset with \dnn[0.15][>]; and the results in \citetOne\ derived without regard to \dnn\ shows that \vs, \As, and \mfast\ for any of the incompressible thresholds are similar and are also similar to the values derived irrespective of compressibility.
This suggests that the overall behavior of \func{\ahe}{\vsw} across \xhel\ is dominated by incompressible solar wind and that compressive fluctuations across \xhel\ drive the large \grad_\vsw_[\ahe] for speeds \vsw*[\vs][>].
It also suggests that the incompressible results are not sensitive to a threshold chosen below \dnn[0.015].

Comparing \vs\ and \As\ derived across \xhel\ and \dnn\ quantiles reveals
\begin{compactenum}
    \item \vs[428 \pm 1.4] in incompressible solar wind doesn't exceed the largest \vs[430]\ in \citetOne.
    This largest \vs\ in \citetOne\ corresponds to non-Alfvénic solar wind.
\item \vs\ in compressive solar wind reaches speeds 17\% larger than non-Alfvénic \vs[430] from \citetOne.
    \item Excluding the largest \dnn\ quantile, compressible \As\ is within the range of non-Alfvénic \As\ derived in \citetOne.
    \item Incompressible \As\ only drop to Alfvénic \As\ at the largest incompressible \dnn.
    \item The saturation slopes are fixed at $\sim \pten{-3}$ for incompressible \dnn\ and jumps by $> 6\times$ in the most compressible \dnn.
\end{compactenum}
This substantiates the broad observation in \cref{fig:vsw:xhel,fig:vsw:dnn} that compressive fluctuations are present across \xhel\ when \xhel\ is plotted as a function of \vsw.
We infer that these compressive fluctuations impact the saturation points derived in \citetOne\ and lead to the apparent paradox in \cref{sec:analysis:vsw-ahe-xhel,sec:analysis:vsw-ahe-dnn}.

\subsection{\ahe\ and \vsw\ in the $\left(\xhel,\dnn\right)$-Plane}
Characterizing \ahe\ in the $\left(\xhel,\dnn\right)$-plane shows that:
\begin{compactenum}
    \item Helium-rich solar wind with \ahe*[\As][>] can be both compressible and incompressible.
    \item \ahe*[\As][>] in the the compressible subset reaches \ahe\ in excess of 7.5\% and is highly variable.
    This subset does not have a preferred \xhel.
    \item Excluding a trivial region, \ahe*[\As][>] in the incompressible subset does not exceed 6\% and displays very little variability.
    This subset has \dnn[0.2][\lesssim].
    While this threshold is larger than any defined previously, only a small fraction of the enhanced \ahe\ in this Alfvénic region of the plane reach this compressibility.
    As such, averages in \dnn\ or \xhel\ quantiles likely obscure this small fraction of the observations.
\end{compactenum}
The presence of \ahe*[\As][>] that is highly variable across \xhel\ at high \dnn\ further implies that a non-trivial subset of solar wind has an enhanced \ahe\ due to compressive fluctuations.
From this, we infer that the enhancement of \grad_\vsw_[\ahe] for speeds \vsw*[\vs][>] is due to compressive fluctuations in Alfvénic solar wind.
This is supported by \cref{fig:vsw:dnn,fig:vsw:xhel:dnn}, which show that \dnn[0.1][>] is present across speeds \vsw[300][\gtrsim] and compressive fluctuations become more significant at higher \vsw\ with low \xhel.

Characterizing \vsw\ in the $\left(\xhel,\dnn\right)$-plane shows that Alfvénic solar wind has faster average speeds and smaller variability than compressive solar wind across \xhel.
In particular,
\begin{compactenum}
    \item the average \vsw\ in the incompressible subset exceeds the largest \vs\ observed in the incompressible \dnn[0.143] quantile.
    \item \vs[430] from the non-Alfvénic solar wind in \citetOne, which is slower than \vsw\ in the incompressible subset, falls within the range of \vsw\ in the compressive subset.
    \item The variability of \vsw\ in the incompressible subset is larger than the variability in the incompressible subset.
    \item The expected anti-correlation between \dnn\ and \xhel\ is recovered across contours of constant \vsw, even though \vsw\ is highly variable, suggesting that a wide range of \vsw\ are observed in incompressible, Alfvénic and compressible, non-Alfvénic solar wind and these speed ranges significantly overlap.
\end{compactenum}
Together, these imply that the compressibility is driving changes in \ahe, not changes in \vsw\ itself.
They also confirm that the overlap in \vsw\ corresponding to incompressible, Alfvénic and compressible, non-Alfvénic solar wind leads to the apparent paradox in \cref{sec:analysis:vsw-ahe-xhel,sec:analysis:vsw-ahe-dnn}. 

The comparison between \ahe\ and \vsw\ in the $\left(\xhel,\dnn\right)$-plane suggests that the enhancement in \ahe\ is not limited to fast solar wind, typically considered to have originated in continuously open field source regions.
Rather, there also exists a subset of solar wind with an enhanced and highly variable \ahe*[\As][>] where the compressibility is large, the average \vsw\ does not exceed \vs\ for the compressible subset (\vdnp\ in \cref{fig:vsw-hist}), and \vsw\ is less variable than in the Alfvénic subset.
This suggests that enhanced \ahe*[\As][>] is not limited to solar wind from continuously open sources on the Sun, in agreement with \citet{ACE:SWICS:SSN}.

\subsection{\dnn\ in the $\left(\xhel,\ahe\right)$-Plane}
\citetOne\ shows that solar wind from closed and open source regions can be separated into different regions of the $\left(\xhel,\ahe\right)$-plane.
Characterizing \dnn\ in this plane shows that:
\begin{compactenum}
	\item compressible solar wind with \dnn[0.15][>] is limited to non-Alfvénic, helium-rich solar wind,
	\item incompressible solar wind with \dnn[0.85][<] is limited to $2.5\% \lesssim \ahe\ \lesssim 7.5\%$,
	\item and this incompressible range corresponds to the region of the $\left(\xhel,\ahe\right)$-plane with solar wind from magnetically open source regions.
\end{compactenum}
In other words, solar wind is least compressible in the region of the plane that \citetOne\ identifies as solar wind from open sources like coronal holes and compressive solar wind is limited the region of the plane where \ahe\ is more than $1.4\times$ larger than the largest \As.
From this, we infer that the compressive subset is of unique origin with a source region not traditionally considered to be continuously or intermittently magnetically open.
Ongoing work suggest that this subset of observations corresponds to transients.

\subsection{The Bimodal Distribution of Speeds Observed Near Earth During Solar Minima}
\citetOne\ visualizes the key speeds derived in that work on the distribution of \vsw\ observed during solar minima.
To this distribution, \cref{fig:vsw-hist} adds the speeds \vdn, \vdnp, \vheavy\ \citep{ACE:SWICS:FStransition}, \vIP\ \citep{Wind:SWE:Wk}, and \vK\ \citep{Wind:SWE:Wk}.
\begin{compactenum}
    \item The range of incompressible \vs\ (\vdn) falls within the full range of \vsigma\ and the faster portion of \vn.
    \vsigma\ is the range of saturation speeds in \citetOne\ and \vn\ is the range of speeds over which \func{\n[\He]}{\vsw} reaches a local maximum.
    This local maximum is indicative of how \He's role in solar wind at the sonic point changes between continuously and intermittently open source regions (\citetOne; \citeauthor{Lie-Svendsen2001} \citeyear{Lie-Svendsen2001}; \citeauthor{Lie-Svendsen2002} \citeyear{Lie-Svendsen2002}; \citeauthor{Endeve2005} \citeyear{Endeve2005}; \citeauthor{Lie-Svendsen2003} \citeyear{Lie-Svendsen2003}; \citeauthor{Hansteen1997} \citeyear{Hansteen1997}).
    \item The range of compressible \vs\ (\vdnp) marginally overlaps the fastest portion of \vsigma\ and is within the range of speeds \vIP, which corresponds to Alfvén wave-poor solar wind from continuously open source regions with reduced energy in the Alfvénic fluctuations.
    \item That \vheavy\ coincides with the slow portion of \vslow\ speed range, effectively at speeds slower than the peak of the \vsw\ distribution, suggests that the processes that lead to the observed helium saturation speeds (which necessarily increases during transit through interplanetary space due to in-transit acceleration) do not impact heavier elements in the same manner \citep{ACE:SWICS:FStransition}.
    \item The range of speeds \vIP, overlaps with \vsigma, \vdn, and \vdnp, reaching a maximum value just greater than the minimum \vfast\ and \vK.
    \item The range of speeds over which non-Alfvénic \ahe\ above saturation (\vsw*[\vs][>]) becomes larger than Alfvénic \ahe\ above saturation is \kms[530-560] (\cref{fig:vsw-ahe-dnn:zoom}), which sits within the \vIP\ range above \vdnp\ and below \vfast and \vK.
\end{compactenum}
This suggests that the transition in observations from slow to fast wind at \au[1] is dominated by incompressible and Alfvénic solar wind, but compressive solar wind is significant in intermediate speed range \vIP, where \emph{ad hoc} thresholds for the separation between slow and fast wind have often been set \citep{Schwenn2006,Fu2018} and likely corresponds to Alfvénic slow wind \citep{Marsch1981,DAmicis2011a}.

\subsection{Relevance for Source Regions, Pressure Balanced Structures, Slow Modes}
Fully ionized hydrogen constitutes approximately 95\% of the solar wind ions.
As such, it is typically assumed that the energy carried by helium and heavier elements can be neglected or can be treated as a fixed value.
For example, this is common practice in deriving of marginal stability thresholds \citep[e.g.][]{Verscharen2016a}.
In ambient solar wind, fully ionized helium constitutes around 4.19\% to 4.25\% of the ionized hydrogen density, corresponding to approximately 17\% of the solar wind mass density.

The solar wind's compressibility may also be related to properties of the solar wind's source region on the Sun \citep{Marsch1990a,LR:turbulence}.
We have shown that there are two distinct subsets of solar wind that are helium-rich and have distinct compressibilities.
We have also shown that there are, effectively, two sets of saturation points: a compressible and an incompressible set.
The incompressible subset corresponds to Alfvénic solar wind.
The compressible subset does not have a preferred Alfvénicity.
The two subsets of helium-rich solar wind may point to two distinct sources of helium-rich solar wind at the Sun, each leading to solar wind with distinct compressibilities.
\citet{Wind:SWE:ahe:ICMEs} shows that the incompressible subset corresponds to ambient solar wind and ICMEs drive the increased compressibility in the compressible subset.

NI-MHD suggests that compressible fluctuations are generated by the nonlinear interaction between Alfvén waves propagating along the magnetic field and incompressible fluctuations propagating perpendicular to it \citep{Zank1992}.
PBS and slow modes in the solar wind can both be described in the limit of $k_\perp/k_\parallel \rightarrow \infty$ \citep{Hollweg2014}.
MHD predictions for slow modes in the solar wind agree better with observations than their kinetic counter parts \citep{Verscharen2017a} and compressive fluctuations may help maintain the stability of differentially streaming helium by modifying the relevant marginal stability conditions \citep{Zhu2023}.
PBS may be related to the more recently identified periodic density structures (PDS) \citep{Kepko2024a,Viall2015,Viall2009}, which are considered to be signatures of the solar wind's source regions.
The overall regulation of \func{\ahe}{\vsw} by \dnn\ and the narrower ranges in \vs\ and \As\ when compressive fluctuations are excluded suggest that solar wind helium may play a role in the local evolution of the solar wind in transit that is related to the solar wind's compressibility and that the compressive fluctuations may be driven by a process related to the solar wind's sources.

\subsection{Addressing the Hidden Variable(s)}

The mapping between the saturation analysis as a function of \dnn\ quantiles to \xhel\ reveals the following.
\begin{compactenum}
    \item The dependence of \xhel\ on \ahe\ and \vsw\ is weakest when \dnn\ is largest.
    \item Below saturation, \func{\xhel}{\ahe} is independent of \dnn\ in incompressible solar wind and, above saturation, the dependence increases with decreasing \dnn.
    \item In incompressible solar wind, \func{\xhel}{\vsw} is independent of \dnn, but strongly dependent on \dnn\ in compressible solar wind.
    \item Incompressible \vs\ and \As\ each occupy small ranges of mapped \xhel\ centered around $\sim0.6$ and $\sim0.65$, respectively, while compressible \vs\ and \As\ cover a wide range of \xhel.
    \item Only the compressible saturation slopes for \dnn[0.15][>] depart from these two clusters and reach \xhel\ smaller than them
\end{compactenum}
From these observations, we infer that there is a critical transition in the range \xhel[0.6] to 0.65.

The mapping between the saturation analysis in \xhel\ quantiles \citepOne\ to \dnn\ reveals the following.
\begin{compactenum}
    \item The minimum average \dnn\ across \xhel\ occurs in the most Alfvénic wind.
    \item At the fastest speeds, the largest compressibility is in the least Alfvénic wind.
    \item The compressibility in the most Alfvénic wind is independent of speed.
    \item The compressibility varies with \ahe\ across all \xhel\ quantiles and reaches a local minimum at \ahe*[\As][\gtrsim].
    \item As with \vsw, the range of \dnn\ is largest at the largest \ahe.
    However, the maximum \dnn\ in the fastest speeds is approximately twice as large as the maximum \dnn\ in the largest \ahe, suggesting that there is a narrower range of \dnn\ within each \xhel\ quantile at any \vsw\ than at any \ahe.
\end{compactenum}
From these observations, we infer that the transition from slow to fast solar wind at \au[1] is dominated by incompressible solar wind and that compressive fluctuations may lead to large variations in observations of fast wind.

In short, the empirical functions derived to map between \dnn\ and \xhel\ using saturation points reveals that the anti-correlation between \xhel\ and \dnn\ is absent in slow or helium-poor solar wind.
As \vsw\ or \ahe\ increase, the expected anti-correlation is recovered and it is most prominent in fast or helium-rich wind.
These observations further support the inference there is not a unique mapping from a given \vsw\ to compressible, non-Alfvénic solar wind or incompressible, Alfvénic solar wind.

The mappings between saturation points as a function of \dnn\ and \xhel\ also reveals the dominant value of the other quantity.
Across all \dnn\ quantiles, only incompressible solar wind has a \xhel[0.6][\geq], which is the \xhel\ contour in \cref{fig:vsw-ahe-xhel} at which we observe the saturation points derived in \citetOne, and the typical cross helicity is larger when \dnn\ is mapped to \xhel\ using \As\ than when using \vs.
The saturation points for compressible solar wind map to non-Alfvénic \xhel[0.6][<].
Across all \xhel\ quantiles, \dnn\ does not exceed 0.15, the largest incompressible threshold derived above.
In particular, the dominant \dnn\ across \xhel\ is $< 0.1$ in Alfvénic solar wind.
The \dnn\ derived in each \xhel\ quantile does not exceed 0.11 using \As\ and does not exceed 0.132 using \xhel.
These observations suggest that the enhanced \ahe*[\As][>] subsets in \cref{fig:xhel-dnn-ahe} and the enhanced \vsw*[\vs][>] regions in \cref{fig:xhel-dnn-vsw} do show that averaging in \dnn\ quantiles or \xhel\ quantiles does select for a wide range of values in the other quantity, impacting our determination of the saturation points and fits.
Given the observations in \cref{fig:vsw:xhel,fig:vsw:dnn,fig:vsw:xhel:dnn}, we infer that the saturation point is highly sensitive to both \xhel\ and \dnn\ and neglecting either leaves the analysis sensitive to a confounding variable.
We also infer that aggregating solar wind observations without considering both the compressibility and the cross helicity mixes compressive and Alfvénic fluctuations, creating the apparent paradox in \cref{sec:analysis:vsw-ahe-xhel,sec:analysis:vsw-ahe-dnn}.

\section{Conclusion \label{sec:conclusion}}

\citetOne\ shows that the solar wind can be separated into fast and slow based on the gradients of the helium abundance with solar wind speed, \grad_\vsw_[\ahe].
This gradient changes at the saturation speed (\vs), which has a characteristic saturation abundance (\As).
Observations of this transition between slow and fast wind depends on \xhel.
\citetOne\ suggests that this transition can be used to classify classify solar wind by the magnetic topology of its source region using the $\left(\xhel,\ahe\right)$-plane.
\citetOne\ also infers that the minimum \vsw\ from source regions with continuously open magnetic topologies is slower than the maximum speed of solar wind originating in intermittently open source regions, which are sometimes referred to as ``closed''.
This observation is consistent with Alfvénic slow wind as the low speed extension of solar wind that would otherwise be categorized as fast \citep{DAmicis2015,Wang1994a,Wang2019,Panasenco2019,Panasenco2020,Panasenco2013,Panasenco2019}.
However, \citetOne\ shows that the gradient of \ahe\ with \vsw\ at speeds \vsw*[\vs][>] increase with decreasing \xhel.
\cref{fig:ahe-enh} shows that this enhancement in \ahe\ at fast speeds leads to a helium abundance far in excess of 50\% the photospheric value, which is a consistent upper bound on \ahe\ across the solar cycle \citep{ACE:SWICS:SSN}.

Analyzing \ahe\ as a function of \vsw\ and \dnn\ and comparing these results to the analysis in \citetOne\ of \ahe\ as a function of \vsw\ and \xhel, we have shown that the dependence of \grad_\vsw_[\ahe] on \xhel\ at speeds \vsw*[\vs][>] is due to the presence of compressive fluctuations in Alfvénic solar wind with high (\vsw*[\vs][>]) speeds.
We have also shown that there exists two subsets of solar wind with enhanced \ahe*[\As][>].
In the Alfvénic subset, \ahe\ does not vary.
In the compressible subset, \ahe\ varies by up to 80\% in excess of the mean value.
The average \vsw\ in this compressible subset is less variable than \ahe\ in the Alfvénic subset.
Comparing these two subsets of observations to the source region identification in \citetOne, we have shown that the incompressible subset likely corresponds to solar wind from magnetically open sources, but the compressible subset likely originated in neither open nor closed source regions.
This does not exclude that these compressive fluctuations are generated in transit.

Visualizing the range of speeds over which \grad_\vsw_[\ahe] changes in the compressible subset (\vdnp) on the distribution of speeds observed during solar minima at \au[1] shows the significance of compressible fluctuations on the separation of observations from open and closed sources.
In particular, \vdnp\ falls within the range of \emph{ad hoc} speeds typically used to separate fast and slow wind.
This range of speeds (\vIP) corresponds to wave-poor solar wind from continuously open source regions in which the energy in the Alfvénic fluctuations is reduced in comparison to typical fast wind \citep{Wind:SWE:Wk} and is likely analogous to the Alfvénic slow wind \citep{Marsch1981,DAmicis2011a}.
Given the enhancement of \ahe*[\As][>] due to compressive fluctuations in non-Alfvénic wind occurs at speeds \vsw*[\vs][>], this suggests that solar wind with compressive fluctuations from neither open nor closed sources is significant at these intermediate speeds.

The mappings between saturation points as a function of \dnn\ and \xhel\ reveals the dominant value of the other quantity.
Across all \dnn\ quantiles, only incompressible solar wind has a \xhel[0.6][\geq], which is the \xhel\ contour in \cref{fig:vsw-ahe-xhel} at which we observe the saturation points derived in \citetOne, and the typical cross helicity is larger when \dnn\ is mapped to \xhel\ using \As\ than when using \vs.
The saturation points for compressible solar wind map to non-Alfvénic \xhel[0.6][<].
Across all \xhel\ quantiles, \dnn\ does not exceed 0.15, the largest incompressible threshold derived above.
In particular, the dominant \dnn\ across \xhel\ is $< 0.1$ in Alfvénic solar wind.
The \dnn\ derived in each \xhel\ quantile does not exceed 0.11 using \As\ and does not exceed 0.132 using \xhel.
These observations suggest that the enhanced \ahe*[\As][>] subsets in \cref{fig:xhel-dnn-ahe} and the enhanced \vsw*[\vs][>] regions in \cref{fig:xhel-dnn-vsw} do show that averaging in \dnn\ quantiles or \xhel\ quantiles selects for a wide range of values in the other quantity, impacting our determination of the saturation points and fits.
Given the observations in \cref{fig:vsw:xhel,fig:vsw:dnn,fig:vsw:xhel:dnn}, we infer that the saturation point is highly sensitive to both \xhel\ and \dnn\ and neglecting either leaves the analysis sensitive to a confounding variable.

From these observations, we conclude that a given \vsw\ does not uniquely map to either Alfvénic, incompressible or non-Alfvénic, compressible solar wind.
Rather, the presence of enhanced compressive fluctuations across \vsw\ obscures the transition between fast and slow wind in our observations and confounds the mapping from fast/slow wind to open/closed source regions.
We further conclude that the hydrogen compressibility is essential for characterizing and possibly regulating the solar wind helium abundance.
This inference may be consistent with the role of pressure balanced structures (PBS), periodic density structures (PDS), or slow modes in the solar wind, all of which may also be related to solar wind source regions.
Future work will determine how significant the role of transients is in this characterization of the role of compressibility in the solar wind and its relationship to the helium abundance.
Combined with \citetOne, this work also suggests that a combination of \ahe, \xhel, and \dnn\ may be sufficient to map the solar wind to its source region when composition observations are unavailable.

\begin{acknowledgments}
The authors thank the referee for their helpful and supportive feedback.
 The authors are grateful to Mihailo M. Martinović and Kristopher G. Klein for valuable discussions.
 The authors also thank Nicholeen Viall and Brent Randol for useful discussions about periodic density structures.
 The authors thank Yogesh for discussions about the typical helium abundance in SIRs.
Wind Faraday cup data are obtained from CDAWeb.
The authors acknowledge Justin C. Kasper for the development and Michael L. Stevens for the delivery of this data product.
B.L.A. is funded by grants 80NSSC22K0645 (LWS/TM) and 80NSSC22K1011 (LWS) along with Parker Solar Probe and Solar Orbiter funding at NASA Goddard Space Flight Center.

\end{acknowledgments}

\software{
IPython \citep{Perez2007}, 
Jupyter \citep{Kluyver2016}, 
Matplotlib \citep{Hunter2007}, 
Numpy \citep{Harris2020,VanderWalt2011}, 
SciPy \citep{Jones2001,scipy},
Pandas \citep{Mckinney2010,McKinney2011,Mckinney2013}, 
Python \citep{Millman2011,Oliphant2007},
Mathematica \citep{Mathematica:14.0}
}

\bibliography{Zotero.bib}{}

\begin{thebibliography}{}
\expandafter\ifx\csname natexlab\endcsname\relax\def\natexlab#1{#1}\fi
\providecommand{\url}[1]{\href{#1}{#1}}
\providecommand{\dodoi}[1]{doi:~\href{http://doi.org/#1}{\nolinkurl{#1}}}
\providecommand{\doeprint}[1]{\href{http://ascl.net/#1}{\nolinkurl{http://ascl.net/#1}}}
\providecommand{\doarXiv}[1]{\href{https://arxiv.org/abs/#1}{\nolinkurl{https://arxiv.org/abs/#1}}}

\bibitem[{Abbo {et~al.}(2016)Abbo, Ofman, Antiochos, Hansteen, Harra, Ko,
  Lapenta, Li, Riley, Strachan, von Steiger, \& Wang}]{Abbo2016}
Abbo, L., Ofman, L., Antiochos, S.~K., {et~al.} 2016, Space Science Reviews,
  201, 55, \dodoi{10.1007/s11214-016-0264-1}

\bibitem[{Adhikari {et~al.}(2020)Adhikari, Zank, Zhao, Kasper, Korreck,
  Stevens, Case, Whittlesey, Larson, Livi, \& Klein}]{Adhikari2020}
Adhikari, L., Zank, G.~P., Zhao, L.~L., {et~al.} 2020, The Astrophysical
  Journal Supplement Series, 246, 38, \dodoi{10.3847/1538-4365/ab5852}

\bibitem[{Aellig {et~al.}(2001)Aellig, Lazarus, \& Steinberg}]{Aellig:Ahe}
Aellig, M.~R., Lazarus, A.~J., \& Steinberg, J.~T. 2001, Geophysical Research
  Letters, 28, 2767, \dodoi{10.1029/2000GL012771}

\bibitem[{Alterman(2025)}]{Wind:SWE:Wk}
Alterman, B. 2025, Astrophysical Journal Letters

\bibitem[{Alterman \& D'Amicis(2025{\natexlab{a}})}]{Wind:SWE:ahe:dnn}
Alterman, B.~L., \& D'Amicis, R. 2025{\natexlab{a}}, Astrophysical Journal
  Letters (in prep)

\bibitem[{Alterman \& D'Amicis(2025{\natexlab{b}})}]{Wind:SWE:ahe:ICMEs}
---. 2025{\natexlab{b}}, in prep

\bibitem[{Alterman \& D’Amicis(2025)}]{Wind:SWE:ahe:xhel}
Alterman, B.~L., \& D’Amicis, R. 2025, The Astrophysical Journal Letters,
  982, L40, \dodoi{10.3847/2041-8213/adb48e}

\bibitem[{Alterman \& Kasper(2019)}]{Wind:SWE:Ahe:phase}
Alterman, B.~L., \& Kasper, J.~C. 2019, The Astrophysical Journal, 879, L6,
  \dodoi{10.3847/2041-8213/ab2391}

\bibitem[{Alterman {et~al.}(2021)Alterman, Kasper, Leamon, \&
  McIntosh}]{Wind:SWE:ahe:shutoff}
Alterman, B.~L., Kasper, J.~C., Leamon, R.~J., \& McIntosh, S.~W. 2021, Solar
  Physics, 296, 67, \dodoi{10.1007/s11207-021-01801-9}

\bibitem[{Alterman {et~al.}(2018)Alterman, Kasper, Stevens, \&
  Koval}]{Alterman2018}
Alterman, B.~L., Kasper, J.~C., Stevens, M., \& Koval, A. 2018, The
  Astrophysical Journal, 864, 112, \dodoi{10.3847/1538-4357/aad23f}

\bibitem[{Alterman {et~al.}(2025)Alterman, Rivera, Raines, Lepri, \&
  D'Amicis}]{ACE:SWICS:SSN}
Alterman, B.~L., Rivera, Y.~J., Raines, J.~M., Lepri, S.~T., \& D'Amicis, R.
  2025, Astronomy \& Astrophysics (in review)

\bibitem[{Alterman {et~al.}(2024)Alterman, {Y. J. Rivera}, Lepri, \&
  Raines}]{ACE:SWICS:FStransition}
Alterman, B.~L., {Y. J. Rivera}, Lepri, S.~T., \& Raines, J.~M. 2024, Astronomy
  \& Astrophysics, \dodoi{10.1051/0004-6361/202451550}

\bibitem[{Antiochos {et~al.}(2011)Antiochos, Mikic, Titov, Lionello, \&
  Linker}]{Antiochos2011}
Antiochos, S.~K., Mikic, Z., Titov, V.~S., Lionello, R., \& Linker, J.~A. 2011,
  The Astrophysical Journal, 112, \dodoi{10.1088/0004-637X/731/2/112}

\bibitem[{Antonucci {et~al.}(2005)Antonucci, Abbo, \& Dodero}]{Antonucci2005}
Antonucci, E., Abbo, L., \& Dodero, M.~A. 2005, Astronomy \& Astrophysics, 435,
  699, \dodoi{10.1051/0004-6361:20047126}

\bibitem[{Asplund {et~al.}(2021)Asplund, Amarsi, \& Grevesse}]{Asplund2021}
Asplund, M., Amarsi, A.~M., \& Grevesse, N. 2021, Astronomy \& Astrophysics,
  653, A141, \dodoi{10.1051/0004-6361/202140445}

\bibitem[{Berger {et~al.}(2011)Berger, Wimmer-Schweingruber, \&
  Gloeckler}]{Berger2011}
Berger, L., Wimmer-Schweingruber, R.~F., \& Gloeckler, G. 2011, Physical Review
  Letters, 106, 151103, \dodoi{10.1103/PhysRevLett.106.151103}

\bibitem[{Brooks {et~al.}(2015)Brooks, Ugarte-Urra, \& Warren}]{Brooks2015}
Brooks, D.~H., Ugarte-Urra, I., \& Warren, H.~P. 2015, Nature Communications,
  6, \dodoi{10.1038/ncomms6947}

\bibitem[{Bruno \& Carbone(2013)}]{LR:turbulence}
Bruno, R., \& Carbone, V. 2013, Living Reviews in Solar Physics, 10, 1,
  \dodoi{10.12942/lrsp-2013-2}

\bibitem[{Bruno {et~al.}(2001)Bruno, Carbone, Veltri, Pietropaolo, \&
  Bavassano}]{Bruno2001}
Bruno, R., Carbone, V., Veltri, P., Pietropaolo, E., \& Bavassano, B. 2001,
  Planetary and Space Science, 49, 1201, \dodoi{10.1016/S0032-0633(01)00061-7}

\bibitem[{Chitta {et~al.}(2025)Chitta, Huang, D’Amicis, Calchetti, Zhukov,
  Kraaikamp, Verbeeck, Aznar~Cuadrado, Hirzberger, Berghmans, Horbury, Solanki,
  Owen, Harra, Peter, Schühle, Teriaca, Louarn, Livi, Giunta, Hassler, \&
  Wang}]{Chitta2025}
Chitta, L.~P., Huang, Z., D’Amicis, R., {et~al.} 2025, Astronomy \&
  Astrophysics, 694, A71, \dodoi{10.1051/0004-6361/202452737}

\bibitem[{Crooker {et~al.}(2012)Crooker, Antiochos, Zhao, \&
  Neugebauer}]{Crooker2012}
Crooker, N.~U., Antiochos, S.~K., Zhao, X., \& Neugebauer, M. 2012, Journal of
  Geophysical Research: Space Physics, 117, n/a, \dodoi{10.1029/2011JA017236}

\bibitem[{Cuesta {et~al.}(2023)Cuesta, Chhiber, Fu, Du, Yang, Pecora,
  Matthaeus, Li, Steinberg, Guo, Gan, Conrad, \& Swanson}]{Cuesta2023}
Cuesta, M.~E., Chhiber, R., Fu, X., {et~al.} 2023, The Astrophysical Journal
  Letters, 949, L19, \dodoi{10.3847/2041-8213/acd4c2}

\bibitem[{D'Amicis \& Bruno(2015)}]{DAmicis2015}
D'Amicis, R., \& Bruno, R. 2015, Astrophysical Journal, 805, 1,
  \dodoi{10.1088/0004-637X/805/1/84}

\bibitem[{D'Amicis {et~al.}(2016)D'Amicis, Bruno, \& Matteini}]{Damicis2016}
D'Amicis, R., Bruno, R., \& Matteini, L. 2016, AIP Conference Proceedings,
  1720, \dodoi{10.1063/1.4943813}

\bibitem[{D'Amicis {et~al.}(2018)D'Amicis, Matteini, \& Bruno}]{DAmicis2018}
D'Amicis, R., Matteini, L., \& Bruno, R. 2018, Monthly Notices of the Royal
  Astronomical Society, 14, 1, \dodoi{10.1093/mnras/sty3329}

\bibitem[{D’Amicis {et~al.}(2021{\natexlab{a}})D’Amicis, Alielden, Perrone,
  Bruno, Telloni, Raines, Lepri, \& Zhao}]{DAmicis2021}
D’Amicis, R., Alielden, K., Perrone, D., {et~al.} 2021{\natexlab{a}},
  Astronomy \& Astrophysics, 654, A111, \dodoi{10.1051/0004-6361/202140600}

\bibitem[{D’Amicis {et~al.}(2011)D’Amicis, Bruno, \&
  Bavassano}]{DAmicis2011a}
D’Amicis, R., Bruno, R., \& Bavassano, B. 2011, Journal of Atmospheric and
  Solar-Terrestrial Physics, 73, 653, \dodoi{10.1016/j.jastp.2011.01.012}

\bibitem[{D’Amicis {et~al.}(2021{\natexlab{b}})D’Amicis, Perrone, Bruno, \&
  Velli}]{DAmicis2021a}
D’Amicis, R., Perrone, D., Bruno, R., \& Velli, M. 2021{\natexlab{b}},
  Journal of Geophysical Research: Space Physics, 126,
  \dodoi{10.1029/2020JA028996}

\bibitem[{D’Amicis {et~al.}(2022)D’Amicis, Perrone, Velli, Sorriso-Valvo,
  Telloni, Bruno, \& De~Marco}]{DAmicis2022}
D’Amicis, R., Perrone, D., Velli, M., {et~al.} 2022, Universe, 8, 352,
  \dodoi{10.3390/universe8070352}

\bibitem[{Elsasser(1950)}]{ElsasserVariables}
Elsasser, W.~M. 1950, Physical Review, 79, 183, \dodoi{10.1103/PhysRev.79.183}

\bibitem[{Endeve {et~al.}(2005)Endeve, Lie‐Svendsen, Hansteen, \&
  Leer}]{Endeve2005}
Endeve, E., Lie‐Svendsen, O., Hansteen, V.~H., \& Leer, E. 2005, The
  Astrophysical Journal, 624, 402, \dodoi{10.1086/428938}

\bibitem[{Fisk {et~al.}(1999)Fisk, Zurbuchen, \& Schwadron}]{Fisk1999}
Fisk, L.~A., Zurbuchen, T.~H., \& Schwadron, N.~A. 1999, The Astrophysical
  Journal, 521, 868, \dodoi{10.1086/307556}

\bibitem[{Fu {et~al.}(2015)Fu, Li, Li, Huang, Mou, Jiao, \& Xia}]{Fu2015}
Fu, H., Li, B., Li, X., {et~al.} 2015, Solar Physics, 290, 1399,
  \dodoi{10.1007/s11207-015-0689-9}

\bibitem[{Fu {et~al.}(2018)Fu, Madjarska, Li, Xia, \& Huang}]{Fu2018}
Fu, H., Madjarska, M.~S., Li, B., Xia, L., \& Huang, Z. 2018, Monthly Notices
  of the Royal Astronomical Society, 478, 1884, \dodoi{10.1093/mnras/sty1211}

\bibitem[{Fu {et~al.}(2017)Fu, Madjarska, Xia, Li, Huang, \& Wangguan}]{Fu2017}
Fu, H., Madjarska, M.~S., Xia, L., {et~al.} 2017, The Astrophysical Journal,
  836, 169, \dodoi{10.3847/1538-4357/aa5cba}

\bibitem[{Geiss {et~al.}(1995{\natexlab{a}})Geiss, Gloeckler, \& von
  Steiger}]{Geiss1995b}
Geiss, J., Gloeckler, G., \& von Steiger, R. 1995{\natexlab{a}}, Space Science
  Reviews, 72, 49

\bibitem[{Geiss {et~al.}(1995{\natexlab{b}})Geiss, Gloeckler, Von~Steiger,
  Balsiger, Fisk, Galvin, Ipavich, Livi, McKenzie, Ogilvie, Et, \&
  Wilken}]{Geiss1995}
Geiss, J., Gloeckler, G., Von~Steiger, R., {et~al.} 1995{\natexlab{b}},
  Science, 268, 1033, \dodoi{10.1126/science.7754380}

\bibitem[{Good {et~al.}(2022)Good, Hatakka, Ala-Lahti, Soljento, Osmane, \&
  Kilpua}]{Good2022}
Good, S.~W., Hatakka, L.~M., Ala-Lahti, M., {et~al.} 2022, Monthly Notices of
  the Royal Astronomical Society, 514, 2425, \dodoi{10.1093/mnras/stac1388}

\bibitem[{Grappin {et~al.}(1991)Grappin, Velli, \& Mangeney}]{Grappin1991}
Grappin, R., Velli, M., \& Mangeney, A. 1991, Annales Geophysicae, 9, 416

\bibitem[{Hansteen {et~al.}(1997)Hansteen, Leer, \& Holzer}]{Hansteen1997}
Hansteen, V.~H., Leer, E., \& Holzer, T.~E. 1997, The Astrophysical Journal,
  482, 498, \dodoi{10.1086/304111}

\bibitem[{Harris {et~al.}(2020)Harris, Millman, van~der Walt, Gommers,
  Virtanen, Cournapeau, Wieser, Taylor, Berg, Smith, Kern, Picus, Hoyer, van
  Kerkwijk, Brett, Haldane, del Río, Wiebe, Peterson, Gérard-Marchant,
  Sheppard, Reddy, Weckesser, Abbasi, Gohlke, \& Oliphant}]{Harris2020}
Harris, C.~R., Millman, K.~J., van~der Walt, S.~J., {et~al.} 2020, Nature, 585,
  357, \dodoi{10.1038/s41586-020-2649-2}

\bibitem[{Hollweg {et~al.}(2014)Hollweg, Verscharen, \& Chandran}]{Hollweg2014}
Hollweg, J.~V., Verscharen, D., \& Chandran, B. D.~G. 2014, The Astrophysical
  Journal, 788, 35, \dodoi{10.1088/0004-637X/788/1/35}

\bibitem[{Hunter(2007)}]{Hunter2007}
Hunter, J.~D. 2007, Computing in Science \& Engineering, 9, 90,
  \dodoi{10.1109/MCSE.2007.55}

\bibitem[{Jones {et~al.}(2001)Jones, Oliphant, Peterson, \&
  {others}}]{Jones2001}
Jones, E., Oliphant, T.~E., Peterson, P., \& {others}. 2001, \{{SciPy}\}:
  {Open} source scientific tools for \{{Python}\}.
\newblock \url{http://www.scipy.org/}

\bibitem[{Kasper {et~al.}(2008)Kasper, Lazarus, \& Gary}]{Kasper2008}
Kasper, J.~C., Lazarus, A.~J., \& Gary, S.~P. 2008, Physical Review Letters,
  101, 261103, \dodoi{10.1103/PhysRevLett.101.261103}

\bibitem[{Kasper {et~al.}(2006)Kasper, Lazarus, Steinberg, Ogilvie, \&
  Szabo}]{Wind:SWE:bimax}
Kasper, J.~C., Lazarus, A.~J., Steinberg, J.~T., Ogilvie, K.~W., \& Szabo, A.
  2006, Journal of Geophysical Research, 111, A03105,
  \dodoi{10.1029/2005JA011442}

\bibitem[{Kasper {et~al.}(2007)Kasper, Stevens, Lazarus, Steinberg, \&
  Ogilvie}]{Kasper:Ahe}
Kasper, J.~C., Stevens, M., Lazarus, A.~J., Steinberg, J.~T., \& Ogilvie, K.~W.
  2007, The Astrophysical Journal, 660, 901, \dodoi{10.1086/510842}

\bibitem[{Kasper {et~al.}(2017)Kasper, Klein, Weber, Maksimovic, Zaslavsky,
  Bale, Maruca, Stevens, \& Case}]{Kasper2017}
Kasper, J.~C., Klein, K.~G., Weber, T., {et~al.} 2017, The Astrophysical
  Journal, 849, 126, \dodoi{10.3847/1538-4357/aa84b1}

\bibitem[{Kepko {et~al.}(2024)Kepko, Viall, \& DiMatteo}]{Kepko2024a}
Kepko, L., Viall, N.~M., \& DiMatteo, S. 2024, Journal of Geophysical Research:
  Space Physics, 129, e2023JA031403, \dodoi{10.1029/2023JA031403}

\bibitem[{Khokhlachev {et~al.}(2022)Khokhlachev, Yermolaev, Lodkina,
  Riazantseva, \& Rakhmanova}]{Khokhlachev2022}
Khokhlachev, A.~A., Yermolaev, Y.~I., Lodkina, I.~G., Riazantseva, M.~O., \&
  Rakhmanova, L.~S. 2022, Cosmic Research, 60, 67,
  \dodoi{10.1134/S0010952522020046}

\bibitem[{Klein {et~al.}(2021)Klein, Verniero, Alterman, Bale, Case, Kasper,
  Korreck, Larson, Lichko, Livi, McManus, Martinović, Rahmati, Stevens, \&
  Whittlesey}]{Klein2021}
Klein, K.~G., Verniero, J.~L., Alterman, B.~L., {et~al.} 2021, The
  Astrophysical Journal, 909, 7, \dodoi{10.3847/1538-4357/abd7a0}

\bibitem[{Kluyver {et~al.}(2016)Kluyver, Ragan-Kelley, Pérez, Granger,
  Bussonnier, Frederic, Kelley, Hamrick, Grout, Corlay, Ivanov, Avila, Abdalla,
  \& Willing}]{Kluyver2016}
Kluyver, T., Ragan-Kelley, B., Pérez, F., {et~al.} 2016, in Positioning and
  {Power} in {Academic} {Publishing}: {Players}, {Agents} and {Agendas}, ed.
  F.~Loizides \& B.~Schmidt (IOS Press), 87--90

\bibitem[{Koval \& Szabo(2013)}]{Wind:MFI:B}
Koval, A., \& Szabo, A. 2013, in {AIP} {Conference} {Proceedings}, Vol. 1539,
  211--214, \dodoi{10.1063/1.4811025}

\bibitem[{Lepping {et~al.}(1995)Lepping, Acũna, Burlaga, Farrell, Slavin,
  Schatten, Mariani, Ness, Neubauer, Whang, Byrnes, Kennon, Panetta, Scheifele,
  \& Worley}]{Wind:MFI:A}
Lepping, R.~P., Acũna, M.~H., Burlaga, L.~F., {et~al.} 1995, Space Science
  Reviews, 71, 207, \dodoi{10.1007/BF00751330}

\bibitem[{Lepri {et~al.}(2013)Lepri, Landi, \& Zurbuchen}]{ACE:SWICS:AUX}
Lepri, S.~T., Landi, E., \& Zurbuchen, T.~H. 2013, The Astrophysical Journal,
  768, 94, \dodoi{10.1088/0004-637X/768/1/94}

\bibitem[{Lie-Svendsen {et~al.}(2001)Lie-Svendsen, Leer, \&
  Hansteen}]{Lie-Svendsen2001}
Lie-Svendsen, O., Leer, E., \& Hansteen, V.~H. 2001, Journal of Geophysical
  Research: Space Physics, 106, 8217, \dodoi{10.1029/2000JA000409}

\bibitem[{Lie‐Svendsen {et~al.}(2003)Lie‐Svendsen, Hansteen, \&
  Leer}]{Lie-Svendsen2003}
Lie‐Svendsen, O., Hansteen, V.~H., \& Leer, E. 2003, The Astrophysical
  Journal, 596, 621, \dodoi{10.1086/377640}

\bibitem[{Lie‐Svendsen {et~al.}(2002)Lie‐Svendsen, Hansteen, Leer, \&
  Holzer}]{Lie-Svendsen2002}
Lie‐Svendsen, O., Hansteen, V.~H., Leer, E., \& Holzer, T.~E. 2002, The
  Astrophysical Journal, 566, 562, \dodoi{10.1086/337990}

\bibitem[{Liu {et~al.}(2021)Liu, Issautier, Meyer-Vernet, Moncuquet,
  Maksimovic, Halekas, Huang, Griton, Bale, Bonnell, Case, Goetz, Harvey,
  Kasper, MacDowall, Malaspina, Pulupa, \& Stevens}]{Liu2021c}
Liu, M., Issautier, K., Meyer-Vernet, N., {et~al.} 2021, Astronomy \&
  Astrophysics, 650, A14, \dodoi{10.1051/0004-6361/202039615}

\bibitem[{Marsch {et~al.}(1981)Marsch, Mühlhäuser, Rosenbauer, Schwenn, \&
  Denskat}]{Marsch1981}
Marsch, E., Mühlhäuser, K.-H., Rosenbauer, H., Schwenn, R., \& Denskat, K.~U.
  1981, Journal of Geophysical Research, 86, 9199,
  \dodoi{10.1029/JA086iA11p09199}

\bibitem[{Marsch \& Tu(1990)}]{Marsch1990a}
Marsch, E., \& Tu, C. 1990, Journal of Geophysical Research: Space Physics, 95,
  11945, \dodoi{10.1029/JA095iA08p11945}

\bibitem[{Matteini {et~al.}(2015)Matteini, Horbury, Pantellini, Velli, \&
  Schwartz}]{Matteini2015}
Matteini, L., Horbury, T.~S., Pantellini, F., Velli, M., \& Schwartz, S.~J.
  2015, The Astrophysical Journal, 802, 11, \dodoi{10.1088/0004-637X/802/1/11}

\bibitem[{Mckinney(2010)}]{Mckinney2010}
Mckinney, W. 2010, in Proceedings of the 9th {Python} in {Science}
  {Conference}, ed. S.~van~der Walt \& J.~Millman, 51 -- 56

\bibitem[{McKinney(2011)}]{McKinney2011}
McKinney, W. 2011, Python for High Performance and Scientific Computing, 1

\bibitem[{Mckinney(2013)}]{Mckinney2013}
Mckinney, W. 2013, Python for {Data} {Analysis} (O'Reilly),
  \dodoi{10.1145/1985441.1985476}

\bibitem[{McManus {et~al.}(2020)McManus, Bowen, Mallet, Chen, Chandran, Bale,
  Larson, Dudok De~Wit, Kasper, Stevens, Whittlesey, Livi, Korreck, Goetz,
  Harvey, Pulupa, MacDowall, Malaspina, Case, \& Bonnell}]{McManus2020}
McManus, M.~D., Bowen, T.~A., Mallet, A., {et~al.} 2020, The Astrophysical
  Journal Supplement Series, 246, 67, \dodoi{10.3847/1538-4365/ab6dce}

\bibitem[{Millman \& Aivazis(2011)}]{Millman2011}
Millman, K.~J., \& Aivazis, M. 2011, Computing in Science \& Engineering, 13,
  9, \dodoi{10.1109/MCSE.2011.36}

\bibitem[{Ogilvie {et~al.}(1995)Ogilvie, Chornay, Fritzenreiter, Hunsaker,
  Keller, Lobell, Miller, Scudder, Sittler, Torbert, Bodet, Needell, Lazarus,
  Steinberg, Tappan, Mavretic, \& Gergin}]{Wind:SWE}
Ogilvie, K.~W., Chornay, D.~J., Fritzenreiter, R.~J., {et~al.} 1995, Space
  Science Reviews, 71, 55, \dodoi{10.1007/BF00751326}

\bibitem[{Oliphant(2007)}]{Oliphant2007}
Oliphant, T.~E. 2007, Computing in Science \& Engineering, 9, 10,
  \dodoi{10.1109/MCSE.2007.58}

\bibitem[{Panasenco \& Velli(2013)}]{Panasenco2013}
Panasenco, O., \& Velli, M. 2013, in Solar {Wind} 13, Big Island, Hawaii,
  50--53, \dodoi{10.1063/1.4810987}

\bibitem[{Panasenco {et~al.}(2019)Panasenco, Velli, \&
  Panasenco}]{Panasenco2019}
Panasenco, O., Velli, M., \& Panasenco, A. 2019, The Astrophysical Journal,
  873, 25, \dodoi{10.3847/1538-4357/ab017c}

\bibitem[{Panasenco {et~al.}(2020)Panasenco, Velli, D’Amicis, Shi, Réville,
  Bale, Badman, Kasper, Korreck, Bonnell, Wit, Goetz, Harvey, MacDowall,
  Malaspina, Pulupa, Case, Larson, Livi, Stevens, \&
  Whittlesey}]{Panasenco2020}
Panasenco, O., Velli, M., D’Amicis, R., {et~al.} 2020, The Astrophysical
  Journal Supplement Series, 246, 54, \dodoi{10.3847/1538-4365/ab61f4}

\bibitem[{Perez \& Granger(2007)}]{Perez2007}
Perez, F., \& Granger, B.~E. 2007, Computing in Science \& Engineering, 9, 21,
  \dodoi{10.1109/MCSE.2007.53}

\bibitem[{Phillips {et~al.}(1994)Phillips, Balogh, Bame, Goldstein, Gosling,
  Hoeksema, McComas, Neugebauer, Sheeley, \& Wang}]{Phillips1994}
Phillips, J.~L., Balogh, A., Bame, S.~J., {et~al.} 1994, Geophysical Research
  Letters, 21, 1105, \dodoi{10.1029/94GL01065}

\bibitem[{Pilleri {et~al.}(2015)Pilleri, Reisenfeld, Zurbuchen, Lepri, Shearer,
  Gilbert, Steiger, \& Wiens}]{Pilleri2015}
Pilleri, P., Reisenfeld, D.~B., Zurbuchen, T.~H., {et~al.} 2015, The
  Astrophysical Journal, 812, 1, \dodoi{10.1088/0004-637X/812/1/1}

\bibitem[{Rivera {et~al.}(2024)Rivera, Badman, Stevens, Verniero, Stawarz, Shi,
  Raines, Paulson, Owen, Niembro, Louarn, Livi, Lepri, Kasper, Horbury,
  Halekas, Dewey, De~Marco, \& Bale}]{Rivera2024}
Rivera, Y.~J., Badman, S.~T., Stevens, M.~L., {et~al.} 2024, Science, 385, 962,
  \dodoi{10.1126/science.adk6953}

\bibitem[{Rivera {et~al.}(2025)Rivera, Badman, Verniero, Varesano, Stevens,
  Stawarz, Reeves, Raines, Raymond, Owen, Livi, Lepri, Landi, Halekas, Ervin,
  Dewey, De~Marco, D’Amicis, Dakeyo, Bale, \& Alterman}]{Rivera2025}
Rivera, Y.~J., Badman, S.~T., Verniero, J.~L., {et~al.} 2025, The Astrophysical
  Journal, 980, 70, \dodoi{10.3847/1538-4357/ada699}

\bibitem[{Savitzky \& Golay(1964)}]{SavgolFilter}
Savitzky, A., \& Golay, M. J.~E. 1964, Analytical Chemistry, 36, 1627,
  \dodoi{10.1021/ac60214a047}

\bibitem[{Schwenn(2006)}]{Schwenn2006}
Schwenn, R. 2006, Space Science Reviews, 124, 51,
  \dodoi{10.1007/s11214-006-9099-5}

\bibitem[{Scolini {et~al.}(2024)Scolini, Lugaz, Winslow, Farrugia, Magyar, \&
  Bacchini}]{Scolini2024a}
Scolini, C., Lugaz, N., Winslow, R.~M., {et~al.} 2024, The Astrophysical
  Journal, 961, 135, \dodoi{10.3847/1538-4357/ad0ed1}

\bibitem[{Song {et~al.}(2022)Song, Cheng, Li, Zhang, \& Chen}]{Song2022}
Song, H., Cheng, X., Li, L., Zhang, J., \& Chen, Y. 2022, The Astrophysical
  Journal, 925, 137, \dodoi{10.3847/1538-4357/ac3bbf}

\bibitem[{Stakhiv {et~al.}(2016)Stakhiv, Lepri, Landi, Tracy, \&
  Zurbuchen}]{Stakhiv2016}
Stakhiv, M.~O., Lepri, S.~T., Landi, E., Tracy, P.~J., \& Zurbuchen, T.~H.
  2016, The Astrophysical Journal, 829, 117,
  \dodoi{10.3847/0004-637X/829/2/117}

\bibitem[{Starkey {et~al.}(2024)Starkey, Fuselier, \& Dayeh}]{Starkey2024}
Starkey, M.~J., Fuselier, S.~A., \& Dayeh, M.~A. 2024, Journal of Geophysical
  Research: Space Physics, 129, e2024JA033099, \dodoi{10.1029/2024JA033099}

\bibitem[{Steinier {et~al.}(1972)Steinier, Termonia, \&
  Deltour}]{SavgolFilter:Comment}
Steinier, J., Termonia, Y., \& Deltour, J. 1972, Analytical Chemistry, 44,
  1906, \dodoi{10.1021/ac60319a045}

\bibitem[{Subramanian {et~al.}(2010)Subramanian, Madjarska, \&
  Doyle}]{Subramanian2010}
Subramanian, S., Madjarska, M.~S., \& Doyle, J.~G. 2010, Astronomy and
  Astrophysics, 516, A50, \dodoi{10.1051/0004-6361/200913624}

\bibitem[{Tracy {et~al.}(2016)Tracy, Kasper, Raines, Shearer, Gilbert, \&
  Zurbuchen}]{Tracy2016}
Tracy, P.~J., Kasper, J.~C., Raines, J.~M., {et~al.} 2016, Physical Review
  Letters, 255101, 255101, \dodoi{10.1103/PhysRevLett.116.255101}

\bibitem[{Tu \& Marsch(1994)}]{Tu1994}
Tu, C.-Y., \& Marsch, E. 1994, Journal of Geophysical Research: Space Physics,
  99, 21481, \dodoi{10.1029/94JA00843}

\bibitem[{Tu \& Marsch(1995)}]{Tu1995}
Tu, C.~Y., \& Marsch, E. 1995, Space Science Reviews, 73, 1,
  \dodoi{10.1007/BF00748891}

\bibitem[{Tu {et~al.}(1989)Tu, Marsch, \& Thieme}]{Tu1989}
Tu, C.-Y., Marsch, E., \& Thieme, K.~M. 1989, Journal of Geophysical Research,
  94, 11739, \dodoi{10.1029/ja094ia09p11739}

\bibitem[{van~der Walt {et~al.}(2011)van~der Walt, Colbert, \&
  Varoquaux}]{VanderWalt2011}
van~der Walt, S., Colbert, S.~C., \& Varoquaux, G. 2011, Computing in Science
  \& Engineering, 13, 22, \dodoi{10.1109/MCSE.2011.37}

\bibitem[{Vasquez \& Hollweg(1999)}]{Vasquez1999}
Vasquez, B.~J., \& Hollweg, J.~V. 1999, Journal of Geophysical Research: Space
  Physics, 104, 4681, \dodoi{10.1029/1998JA900090}

\bibitem[{Verniero {et~al.}(2020)Verniero, Larson, Livi, Rahmati, McManus,
  Pyakurel, Klein, Bowen, Bonnell, Alterman, Whittlesey, Malaspina, Bale,
  Kasper, Case, Goetz, Harvey, Korreck, MacDowall, Pulupa, Stevens, \&
  de~Wit}]{Verniero2020}
Verniero, J.~L., Larson, D.~E., Livi, R., {et~al.} 2020, The Astrophysical
  Journal Supplement Series, 248, 5, \dodoi{10.3847/1538-4365/ab86af}

\bibitem[{Verniero {et~al.}(2022)Verniero, Chandran, Larson, Paulson, Alterman,
  Badman, Bale, Bonnell, Bowen, de~Wit, Kasper, Klein, Lichko, Livi, McManus,
  Rahmati, Verscharen, Walters, \& Whittlesey}]{Verniero2022}
Verniero, J.~L., Chandran, B. D.~G., Larson, D.~E., {et~al.} 2022, The
  Astrophysical Journal, 924, 112, \dodoi{10.3847/1538-4357/ac36d5}

\bibitem[{Verscharen {et~al.}(2016)Verscharen, Chandran, Klein, \&
  Quataert}]{Verscharen2016a}
Verscharen, D., Chandran, B. D.~G., Klein, K.~G., \& Quataert, E. 2016, The
  Astrophysical Journal, 831, 128, \dodoi{10.3847/0004-637X/831/2/128}

\bibitem[{Verscharen {et~al.}(2017)Verscharen, Chen, \&
  Wicks}]{Verscharen2017a}
Verscharen, D., Chen, C. H.~K., \& Wicks, R.~T. 2017, The Astrophysical
  Journal, 840, 106, \dodoi{10.3847/1538-4357/aa6a56}

\bibitem[{Viall {et~al.}(2009)Viall, Spence, \& Kasper}]{Viall2009}
Viall, N.~M., Spence, H.~E., \& Kasper, J.~C. 2009, Geophysical Research
  Letters, 36, 1, \dodoi{10.1029/2009GL041191}

\bibitem[{Viall \& Vourlidas(2015)}]{Viall2015}
Viall, N.~M., \& Vourlidas, A. 2015, The Astrophysical Journal, 807, 176,
  \dodoi{10.1088/0004-637X/807/2/176}

\bibitem[{Virtanen {et~al.}(2020)Virtanen, Gommers, Oliphant, Haberland, Reddy,
  Cournapeau, Burovski, Peterson, Weckesser, Bright, Van Der~Walt, Brett,
  Wilson, Millman, Mayorov, Nelson, Jones, Kern, Larson, Carey, Polat, Feng,
  Moore, VanderPlas, Laxalde, Perktold, Cimrman, Henriksen, Quintero, Harris,
  Archibald, Ribeiro, Pedregosa, Van~Mulbregt, {SciPy 1.0 Contributors},
  Vijaykumar, Bardelli, Rothberg, Hilboll, Kloeckner, Scopatz, Lee, Rokem,
  Woods, Fulton, Masson, Haggstrom, Fitzgerald, Nicholson, Hagen, Pasechnik,
  Olivetti, Martin, Wieser, Silva, Lenders, Wilhelm, Young, Price, Ingold,
  Allen, Lee, Audren, Probst, Dietrich, Silterra, Webber, Slavič, Nothman,
  Buchner, Kulick, Schonberger, De~Miranda~Cardoso, Reimer, Harrington,
  Rodríguez, Nunez-Iglesias, Kuczynski, Tritz, Thoma, Newville, Kummerer,
  Bolingbroke, Tartre, Pak, Smith, Nowaczyk, Shebanov, Pavlyk, Brodtkorb, Lee,
  McGibbon, Feldbauer, Lewis, Tygier, Sievert, Vigna, Peterson, More, Pudlik,
  Oshima, Pingel, Robitaille, Spura, Jones, Cera, Leslie, Zito, Krauss,
  Upadhyay, Halchenko, \& Vazquez-Baeza}]{scipy}
Virtanen, P., Gommers, R., Oliphant, T.~E., {et~al.} 2020, Nature Methods, 17,
  261, \dodoi{10.1038/s41592-019-0686-2}

\bibitem[{von Steiger {et~al.}(2000)von Steiger, Schwadron, Fisk, Geiss,
  Gloeckler, Hefti, Wilken, Wimmer-Schweingruber, \&
  Zurbuchen}]{vonSteiger2000}
von Steiger, R., Schwadron, N.~A., Fisk, L.~A., {et~al.} 2000, Journal of
  Geophysical Research: Space Physics, 105, 27217, \dodoi{10.1029/1999JA000358}

\bibitem[{Wang(1994)}]{Wang1994a}
Wang, Y.-M. 1994, The Astrophysical Journal, 437, L67, \dodoi{10.1086/187684}

\bibitem[{Wang \& Ko(2019)}]{Wang2019}
Wang, Y.-M., \& Ko, Y.-K. 2019, The Astrophysical Journal, 880, 146,
  \dodoi{10.3847/1538-4357/ab2add}

\bibitem[{Webb \& Howard(2012)}]{LR:CME:obs}
Webb, D.~F., \& Howard, T.~A. 2012, Living Reviews in Solar Physics, 9,
  \dodoi{10.12942/lrsp-2012-3}

\bibitem[{{Wolfram Research, Inc.}(2024)}]{Mathematica:14.0}
{Wolfram Research, Inc.} 2024,  Champaign, Illinois: Wolfram Research, Inc.
\newblock \url{https://www.wolfram.com/mathematica}

\bibitem[{Woodham {et~al.}(2018)Woodham, Wicks, Verscharen, \&
  Owen}]{Woodham2018}
Woodham, L.~D., Wicks, R.~T., Verscharen, D., \& Owen, C.~J. 2018, The
  Astrophysical Journal, 856, 49, \dodoi{10.3847/1538-4357/aab03d}

\bibitem[{Xu \& Borovsky(2015)}]{Xu2014}
Xu, F., \& Borovsky, J. 2015, Journal of Geophysical Research: Space Physics,
  120, 70, \dodoi{10.1002/2014JA020412}

\bibitem[{Yardley {et~al.}(2024)Yardley, Brooks, D’Amicis, Owen, Long, Baker,
  Démoulin, Owens, Lockwood, Mihailescu, Coburn, Dewey, Müller, Suen,
  Ngampoopun, Louarn, Livi, Lepri, Fludra, Haberreiter, \&
  Schühle}]{Yardley2024}
Yardley, S.~L., Brooks, D.~H., D’Amicis, R., {et~al.} 2024, Nature Astronomy,
  \dodoi{10.1038/s41550-024-02278-9}

\bibitem[{{Yogesh} {et~al.}(2021){Yogesh}, Chakrabarty, \&
  Srivastava}]{Yogesh:Ahe}
{Yogesh}, Chakrabarty, D., \& Srivastava, N. 2021, Monthly Notices of the Royal
  Astronomical Society: Letters, 503, L17, \dodoi{10.1093/mnrasl/slab016}

\bibitem[{{Yogesh} {et~al.}(2022){Yogesh}, Chakrabarty, \&
  Srivastava}]{Yogesh:ICME}
---. 2022, Monthly Notices of the Royal Astronomical Society: Letters, 111,
  106, \dodoi{10.1093/mnrasl/slac044}

\bibitem[{{Yogesh} {et~al.}(2023){Yogesh}, Chakrabarty, \&
  Srivastava}]{Yogesh:ahe:SIR}
---. 2023, Monthly Notices of the Royal Astronomical Society, 526, L13,
  \dodoi{10.1093/mnrasl/slad112}

\bibitem[{Zank \& Matthaeus(1992)}]{Zank1992}
Zank, G.~P., \& Matthaeus, W.~H. 1992, Journal of Geophysical Research: Space
  Physics, 97, 17189, \dodoi{10.1029/92JA01734}

\bibitem[{Zhao {et~al.}(2022)Zhao, Landi, Lepri, \& Carpenter}]{Zhao2022}
Zhao, L., Landi, E., Lepri, S.~T., \& Carpenter, D. 2022, Universe, 8, 393,
  \dodoi{10.3390/universe8080393}

\bibitem[{Zhao {et~al.}(2017)Zhao, Landi, Lepri, Gilbert, Zurbuchen, Fisk, \&
  Raines}]{Zhao:InSituComposition:Sources}
Zhao, L., Landi, E., Lepri, S.~T., {et~al.} 2017, The Astrophysical Journal,
  846, 135, \dodoi{10.3847/1538-4357/aa850c}

\bibitem[{Zhu {et~al.}(2023)Zhu, Verscharen, He, Maruca, \& Owen}]{Zhu2023}
Zhu, X., Verscharen, D., He, J., Maruca, B.~A., \& Owen, C.~J. 2023, The
  Astrophysical Journal, 956, 66, \dodoi{10.3847/1538-4357/aced03}

\bibitem[{Zurbuchen {et~al.}(2016)Zurbuchen, Weberg, Von~Steiger, Mewaldt,
  Lepri, \& Antiochos}]{Zurbuchen2016}
Zurbuchen, T.~H., Weberg, M., Von~Steiger, R., {et~al.} 2016, The Astrophysical
  Journal, 826, 10, \dodoi{10.3847/0004-637X/826/1/10}

\bibitem[{Ďurovcová {et~al.}(2019{\natexlab{a}})Ďurovcová, Němeček, \&
  Šafránková}]{Durovcova:SIR}
Ďurovcová, T., Němeček, Z., \& Šafránková, J. 2019{\natexlab{a}}, The
  Astrophysical Journal, 873, 24, \dodoi{10.3847/1538-4357/ab01c8}

\bibitem[{Ďurovcová {et~al.}(2019{\natexlab{b}})Ďurovcová, Šafránková,
  \& Němeček}]{Durovcova2019}
Ďurovcová, T., Šafránková, J., \& Němeček, Z. 2019{\natexlab{b}}, Solar
  Physics, 294, 97, \dodoi{10.1007/s11207-019-1490-y}

\end{thebibliography}
\bibliographystyle{aasjournal}

\end{document}